\documentclass[prd,aps,twocolumn,a4paper,showkeys,nofootinbib,showpacs]{revtex4-2}

\usepackage[colorlinks=true,linkcolor=blue,linktocpage=true,anchorcolor=black,citecolor=blue,filecolor=black,menucolor=black,urlcolor=blue]{hyperref}

\usepackage{graphicx,psfrag}
\usepackage{mathrsfs}
\usepackage{amsmath,amsfonts,amssymb}
\usepackage{comment}
\usepackage{ulem}
\usepackage{commath}
\usepackage[top=2cm,bottom=2cm,left=1.4cm,right=1.4cm]{geometry}
\usepackage{multirow}

\usepackage[T1]{fontenc}

\newcommand{\be}{\begin{equation}}
\newcommand{\ee}{\end{equation}}
\newcommand{\bea}{\begin{eqnarray}}
\newcommand{\eea}{\end{eqnarray}}
\newcommand{\bel}{\begin{align}}
\newcommand{\eel}{\end{align}}

\def\e{{\rm e}}
\def\i{{\rm i}}

\def\GMc2{G M_{\odot} c^{-2}}

\def\F{{\cal F}}
\def\lm{{\ell m}}

\def\lm{{\ell m}}

\def\lm{{\ell m}}

\def\F{{\cal F}}

\def\TEOB{\texttt{TEOBResumS}}
\def\TEOBResumS{\texttt{TEOBResumS}}

\def\TEOBd{{\texttt{TEOBResumS-Dal\'i}}}
\newcommand\SEOBNRv[1]{\texttt{SEOBNRv{#1}}}

\def\Fphi{\hat{\mathcal{F}}_{\varphi}}

\def\pph{p_\varphi}
\def\prs{p_{r_*}}

\def\elm{\ellm}
\def\elm{{\ell m}}
\def\G12{\Tilde{G}_{ap}}

\DeclareSymbolFontAlphabet{\mathrsfs}{rsfs}
\DeclareMathAlphabet{\mathcal}{OMS}{cmsy}{m}{n}
\DeclareSymbolFontAlphabet{\mathrsfs}{rsfs}
\DeclareMathAlphabet\mathbfcal{OMS}{cmsy}{b}{n}

\usepackage{color}
\definecolor{cyan}{rgb}{0,0.9,0.9}
\definecolor{orange}{rgb}{0.9,0.5,0}
\definecolor{magenta}{rgb}{1,0,1}
\definecolor{purple}{rgb}{0.8,0.4,0.8}
\definecolor{gray}{rgb}{0.8242,0.8242,0.8242}
\definecolor{dodgerblue}{rgb}{0.12, 0.56, 1.0}

\begin{document}
        
\title{Effective-one-body waveform model for noncircularized, planar, coalescing black hole binaries II:
high accuracy by improving logarithmic terms in resummations}

\author{Alessandro \surname{Nagar}${}^{1,2}$}
\author{Danilo \surname{Chiaramello}${}^{1,3}$}
\author{Rossella \surname{Gamba}${}^{4,5}$}
\author{Simone \surname{Albanesi}${}^{1,6}$}
\author{Sebastiano \surname{Bernuzzi}${}^{6}$}
\author{Veronica \surname{Fantini}${}^{2}$}
\author{Mattia \surname{Panzeri}${}^{1,3}$}
\author{Piero \surname{Rettegno}${}^{1}$}

\affiliation{${}^1$INFN Sezione di Torino, Via P. Giuria 1, 10125 Torino, Italy}
\affiliation{${}^2$Institut des Hautes Etudes Scientifiques, 91440 Bures-sur-Yvette, France}
\affiliation{${}^3$Dipartimento di Fisica, Universit\`a di Torino, Via P. Giuria 1, 10125 Torino, Italy}
\affiliation{${}^4$Institute for Gravitation \& the Cosmos, The Pennsylvania State University, University Park PA 16802, USA}
\affiliation{${}^5$Department of Physics, University of California, Berkeley, CA 94720, USA}
\affiliation{${}^6$Theoretisch-Physikalisches Institut, Friedrich-Schiller-Universit{\"a}t Jena, 07743, Jena, Germany}  

\begin{abstract}
Effective-one-body (EOB) models are based on analytical building blocks that, mathematically, are truncated 
Taylor series with logarithms. These functions are usually resummed using Pad\'e approximants 
obtained first assuming that the logarithms are constant, and then replacing them back 
into the resulting rational functions. A recent study~\cite{Nagar:2024dzj} pointed out that this procedure 
introduces spurious logarithmic terms when the resummed functions are reexpanded. We therefore explore analytically more consistent 
resummation schemes. Here we update the \TEOBd{}  waveform model for spin-aligned, noncircularized 
coalescing black hole binaries by systematically implementing new 
(still Pad\'e based) resummations for all EOB functions (that is, the metric potentials $A, D$ and 
the residual waveform amplitude corrections $\rho_{\lm}$ up to $\ell=8$).
Once the model is informed by 50 Numerical Relativity simulations (with the usual two flexibility parameters,
one in the orbital and one in the spin sector), this new approach proves key in 
lowering the maximum EOB/NR unfaithfulness $\bar{\cal F}_{\rm EOBNR}^{\rm max}$ for the $\ell=m=2$ mode 
(with the Advanced LIGO noise in the total mass range $10-200M_{\odot}$) over 
530 spin-aligned waveforms of the Simulating eXtreme Spacetimes catalog. 
A median unfaithfulness equal to $3.09\times 10^{-4}$ is achieved, which is a marked improvement over the previous value, $1.06\times 10^{-3}$. 
The largest value, ${\rm Max}[\bar{\cal F}^{\rm max}_{\rm EOBNR}]= 6.80\times 10^{-3}$, is found for
an equal-mass, equal-spin simulation with dimensionless spins $\sim +0.998$;
only five configurations have $\bar{\cal F}^{\rm max}_{\rm EOBNR} > 5\times 10^{-3}$ 
(four of which equal-mass and with equal spins larger than $\sim +0.98$).
Results for eccentric binaries are similarly excellent (well below $10^{-2}$ and mostly around $10^{-3}$).
For scattering configurations, we find an unprecedented EOB/NR agreement ($\lesssim1\%$) for relatively small values
of the scattering angle, though differences progressively increase as the threshold of immediate capture
is approached. Our results thus prove that an improved treatment  of (apparently) minor analytical details 
is crucial to obtain  highly accurate waveform models for next generation detectors.
\end{abstract}

\maketitle

\section{Introduction}

Numerous works dedicated to the construction of waveform models for coalescing binary black holes (BBHs)
have recently called attention to the importance of apparently minor analytical details in the construction of 
a model~\cite{Nagar:2021xnh, Nagar:2023zxh, Nagar:2024dzj}.
In a recent paper~\cite{Nagar:2024dzj} (hereafter Paper~I), we highlighted the key role of radiation reaction 
(in analytical, effective-one-body (EOB) resummed form) in obtaining highly Numerical Relativity (NR)-faithful EOB waveforms for 
non-circularized, spin-aligned, coalescing black hole binaries. More precisely, we focused on two aspects: 
(i) on the one hand, we found evidence that it could be beneficial to tune the EOB fluxes to NR
data in order to decrease the EOB/NR unfaithfulness to the $10^{-4}$ level; (ii) on the other hand, we showed how a more careful 
resummation of the functions entering the EOB dynamics, and in particular the treatment of
logarithmic-dependent terms, might improve the model in its original form without the need of such additional NR-tuning
(see Appendix~A of Paper~I).
In this work, we build upon point (ii). Starting from the model of Paper~I, 
we systematically modify all resummed expression involving Pad\'e approximants, expressions 
that were previously obtained by treating the $\log(x)$ terms as constants. This results in new resummed expressions 
for the $(A,D)$ functions entering the conservative dynamics as well as for the residual amplitude corrections 
$\rho_\lm$ entering the waveform (and radiation reaction).
We then inform the model with 50 NR simulations and test its performance on a variety of configurations,
including quasi-circular, eccentric and scattering binaries.
The work presented here is part of a strong research effort aimed at generalizing the original 
\TEOBResumS{} model~\cite{Damour:2014sva,Nagar:2018zoe} for quasi-circular binaries
to more general orbital configurations, including extreme-mass-ratio-inspirals~\cite{Albertini:2022rfe,Albertini:2022dmc,Albertini:2023aol,Albertini:2024rrs} 
or the combined effect of orbital eccentricity and spin precession~\cite{Gamba:2024cvy}.

The paper is organized as follows: in Sec.~\ref{sec:resum} we review in detail the main elements in the dynamics
and the waveform model, before discussing the analytical changes we introduce in this work with respect
to Paper~I, namely the new resummed expressions that enter the conservative and nonconservative
part of the dynamics. In Sec.~\ref{sec:results} we NR-inform the new model and test its performance on bound
configurations, i.e. either quasi-circular or eccentric inspirals up to merger. A comparison with the 
{\tt SEOBNRv5HM} model~\cite{Pompili:2023tna} is also provided here. Scattering configurations are discussed
in Sec.~\ref{sec:scattering}, while concluding remarks are collected in Sec.~\ref{sec:end}.
The paper is complemented by a technical Appendix~\ref{app:derivs} that describes in detail the analytical (approximate)
procedure to compute time-derivatives of the phase-space variables needed to obtain the noncircular correction to
the waveform and radiation reaction, as originally proposed in Ref.~\cite{Chiaramello:2020ehz}. 
The paper is complemented by a few Appendices. Appendix~\ref{app:derivs} reports the
details of the implementation of the radiation reaction force and in particular the analytic computation
of time derivatives and their testing. In Appendix~\ref{app:newa6} we explore the impact of some of the
choices made in the calibration of the nonspinning sector of the model, while Appendix~\ref{sec:opt_dali}
discusses the performance of the model for eccentric configurations optimizing on initial data consistently 
with Ref.~\cite{Gamboa:2024hli}. Finally, Appendix~\ref{app:tables} lists numerical data used in the calibration phase.

We adopt the following notations and conventions. The black hole masses
are denoted as $(m_1,m_2)$, the mass ratio is $q=m_1/m_2\geq 1$, the total mass $M\equiv m_1+m_2$, 
the symmetric mass ratio $\nu\equiv m_1 m_2/M^2$
and the mass fractions $X_i\equiv m_i/M$ with $i=1,2$. 
The dimensionless spin magnitudes are $\chi_i\equiv S_i/m_1^2$ with $i=1,2$, 
and we indicate with $\tilde{a}_0\equiv \tilde{a}_1+\tilde{a}_2\equiv X_1\chi_1+X_2\chi_2$
the effective spin, usually called $\chi_{\rm eff}$ in the literature.
We also use the spin combinations $\tilde{a}_{12} = \tilde{a}_1 - \tilde{a}_2$, $\hat{S} = (S_1 + S_2)/M^2$, and
$\hat{S}_* = [(m_2/m_1) S_1 + (m_1/m_2) S_2]/M^2$.
Unless otherwise stated, we use geometric units with $G=c=1$.

\section{Effective-one-body model and resummations}
\label{sec:resum}
Our starting point is the model extensively discussed in Paper~I, that is in turn an update of the 
model of Ref.~\cite{Nagar:2021xnh}. To ease the discussion, we summarize in this Section the main
elements of the models for both the conservative dynamics and the waveform and radiation reaction
currently implemented in \TEOBd, before detailing the changes introduced in this work in each case.

\subsection{Conservative dynamics}
\label{sec:dynamics}
The main subjects of this section are the EOB potentials $A, D, Q$ that enter the Hamiltonian, which reads:
\begin{align}
	H_{\rm EOB} &= M \sqrt{1 + 2\nu (\hat{H}_{\rm eff} - 1)}  \ ,\\
	\hat{H}_{\rm eff} &= \dfrac{H_{\rm eff}}{\mu} = \hat{H}_{\rm eff}^{\rm orb} + \hat{H}_{\rm eff}^{\rm SO} \ , \\
	\hat{H}_{\rm eff}^{\rm orb} &= \sqrt{p_{r_*}^2 + A \left(1 + p_\varphi^2 u_c^2 + Q\right)} \ , \\
	\hat{H}_{\rm eff}^{\rm SO}  &= (G_S \hat{S} + G_{S_*} \hat{S}_*)p_\varphi = \tilde{G} p_\varphi \ .
\end{align}
We use rescaled phase-space variables, related to their physical counterparts by $r = R/M$, 
$p_{r_*} = P_{R_*}/\mu$, $p_\varphi = P_\varphi/(\mu M)$, $t = T/M$.
The centrifugal radius $r_c = u_c^{-1}$ incorporates even-in-spin effects, and is defined by:
\begin{equation}
r_c^2 = (r_c^{\rm LO})^2 \hat{r}_c^2 = \left[r^2 + \tilde{a}_0^2 \left(1 + \dfrac{2}{r}\right)\right] \left(1 + \dfrac{\delta a_{\rm NLO}^2}{r (r_c^{\rm LO})^2}\right) \  ,
\end{equation}
with the next-to-leading order (NLO) correction $\delta a_{\rm NLO}^2$ given in Eq.~(17) of Paper~I.
The conjugate momentum to the tortoise coordinate $r_*$ is $p_{r_*} = \sqrt{A/B} p_r$~\cite{Damour:2014sva},
with $D \equiv A B$. The gyro-gravitomagnetic functions $(G_S, G_{S_*})$ are used at next-to-next-to-leading
order (NNLO) following Ref.~\cite{Damour:2014sva}, with the addition of the next-to-next-to-next-to-leading order
(${\rm N^3LO}$) parameter $c_3 (\nu, \chi_{1,2})$ that is determined through NR calibration, see Eqs.~(42)-(43) in~\cite{Damour:2014sva}.

For spinning systems, the $A$ and $D$ functions are factorized as:
\begin{align*}
A &= \dfrac{1+2u_c}{1+2u} A_{\rm orb} (u_c) \\
D &= \dfrac{r^2}{r_c^2} D_{\rm orb} (u_c),
\end{align*}
where $u = r^{-1}$.
The PN-expanded potentials $A_{\rm orb}$ and $D_{\rm orb}$, as well as the $Q$ function, are 
the 5PN expressions of Eqs.~(2)-(3)-(5) of Ref.~\cite{Nagar:2021xnh}, respectively; they are used as
functions of $u_c$ rather than $u$ in the spinning case though~\cite{Damour:2014sva}.
As for $Q$, we only retain its local-in-time part (in the sense of Refs.~\cite{Bini:2019nra,Bini:2020wpo}), 
and keep it in Taylor-expanded form, without any resummation, as done in 
previous works~\cite{Nagar:2024dzj,Nagar:2021xnh}.
Following Paper~I, in $A$ we keep the function $a_6^c \left(\nu\right)$ as a free parameter to be informed
by NR simulations, while the yet uncalculated 5PN coefficient of the 
$D$ function is set to zero, $d^{\nu^2}_5=0$ (see Refs.~\cite{Bini:2019nra,Bini:2020wpo}).

In Paper~I, $A_{\rm orb}$ was resummed by means of a $(3,3)$ Pad\'e approximant, and $D_{\rm orb}$
by a $(3,2)$, in both cases treating the logarithms that appear at high PN order as constant coefficients
when calculating the resummed functions.
In this work, we implement and test a new resummation strategy for the potentials $A$ and $D$, according to the treatment 
of these logarithmic terms discussed in Appendix~A of Paper~I for the waveform modes. 
First, the rational part of each of $A_{\rm orb}$ and $D_{\rm orb}$ is separated out, 
leaving a remainder that is a polynomial multiplied by $\log u$: 
\begin{align}
	A_{\rm 5PN} =& 1-2u + 2\nu u^3 + \nu a_3 u^4 + \nu a_5^cu^5 + \nu a_6^c u^6 \nonumber\\
	   &+ \nu (a_5^{\log} u^5 + a_6^{\log} u^6)\log u \\ \nonumber 
	   \equiv& A_{\rm poly}(u)+A_{\log}(u) \log u \ , \\
	D_{\rm 5PN} =& 1 - 6\nu u^2 + \nu d_3 u^3 + \nu d_4^c u^4 + \nu d_5^c u^5 \nonumber\\
	 &+ \nu (d_4^{\log}u^4 + d_5^{\log}u^5)\log u \\ \nonumber 
	 \equiv& D_{\rm poly}(u)+D_{\log}(u)\log u \ .
\end{align}
Both the rational part and the polynomial $\log$ coefficient are then separately resummed,
and again the resulting functions are evaluated using the centrifugal inverse radius $u_c$
rather than $u$:
\begin{itemize}
\item[(i)] in $A$, $A_{\rm poly}$ is replaced by its $(3,3)$ Pad\'e approximant; 
in $A_{\rm log}$ we factor out $a_5^{\log} 
u^5$ and resum the remainder with a $(0,1)$ Pad\'e:
\begin{equation}
	\label{eq:Aorb_logresum}
	A_{\rm orb} (u_c) = P^3_3 [A_{\rm poly}] + a_5^{\rm log} u_c^5 P^0_1 \left[\dfrac{A_{\rm log}}{a_5^{\rm log} u_c^5}\right] \log(u_c) \ ,
\end{equation}
\item[(ii)] in $D$, $D_{\rm poly}$ is replaced by 
its $(3,2)$ Pad\'e approximant; in $D_{\log}$ the $d_4^{\log} u^4$ term is 
factored out and the remainder similarly resummed with a $(0,1)$ Pad\'e:
\begin{equation}
	\label{eq:Dorb_logresum}
	D_{\rm orb} (u_c) = P^3_2 [D_{\rm poly}] + d_4^{\rm log} u_c^4 P^0_1 \left[\dfrac{D_{\rm log}}{d_4^{\rm log} u_c^4}\right] \log(u_c) \ .
\end{equation}
\end{itemize}

\subsection{Waveform and radiation reaction}
\label{sec:waveform}

Let us recall our convention for the multipolar waveform decomposition:\be
h_+-ih_\times = \dfrac{1}{D_L}\sum_{\ell=2}^{\infty}\sum_{m=-\ell}^\ell h_{\lm}{}_{-2}Y_{\lm} \, ,
\ee
where $D_L$ is the luminosity distance to the source and ${}_{-2}Y_{\lm}$ are the $s=-2$ 
spin-weighted spherical harmonics. For each multipolar mode, the waveform is factorized as 
\be
h_\lm = h_\lm^N\hat{h}_\lm \hat{h}_{\ell m}^{\rm NQC}\ ,
\ee
where $h_\lm^N$ is the Newtonian prefactor (given in closed form for the circular case 
in Ref.~\cite{Damour:2008gu} and in Appendix~B of Ref.~\cite{Albanesi:2021rby} for the generic case)
and $\hat{h}_\lm$ is a PN correction. Following~\cite{Damour:2008gu}, this latter is formally
factorized as
\be
\hat{h}_\lm = \hat{S}_{\rm eff}T_\lm e^{i\delta_\lm}(\rho_\lm)^\ell \ ,
\ee
where $\hat{S}_{\rm eff}$ is the effective source, equal to either the effective energy $\hat{S}_{\rm eff} = \hat{H}_{\rm eff}$
if $\ell + m$ is even, or the angular momentum $\hat{S}_{\rm eff} = p_\varphi/(r_{\Omega} v_\varphi)$
(see below for the definitions of $r_{\Omega}$ and $v_\varphi$) if $\ell + m$ is odd.
$T_\lm$ is the tail factor, while $\rho_\lm$ and $\delta_\lm$ are the residual amplitude and phase corrections. 
The latter two are obtained in the circular limit as functions, respectively, of $x = \Omega^{2/3}$, 
where $\Omega = \dot{\varphi}$ is the orbital frequency, and $y = E \Omega$, 
with $E$ denoting the energy along circular orbits. 
Following standard practice in \TEOBResumS{}~\cite{Damour:2007xr,Damour:2008gu,Damour:2012ky,Damour:2014sva}, 
they are evaluated along the dynamics as functions of formally equivalent, but different parameters.
We use $x = (r_{\Omega} \Omega)^2 = v_\varphi^2$, where $v_\varphi$ is a non-Keplerian azimuthal velocity~\cite{Damour:2006tr} and
$r_\Omega$ follows from Eq.~(70) of Ref.~\cite{Damour:2014sva}. It explicitly reads
\be
\label{eq:rOmg}
r_\Omega \equiv \left\{ \dfrac{(r_c^3\psi_c)^{-1/2}+\tilde{G}}{H_{\rm EOB}}\right\}^{-2/3}_{p_{r_*}=0} \ 
\ee
with
\be
\psi_c = \dfrac{2}{A'}\left(u_c'+\dfrac{\tilde{G}'}{u_c A}\sqrt{\dfrac{A}{p_\varphi^2+u_c^2 A}}\right) \ ,
\ee
where the prime indicates the derivative with respect to $r$. 
The definition~\eqref{eq:rOmg} is such that during the adiabatic 
quasi-circular inspiral a Kepler-like law holds, $1=r_\Omega^3 \Omega^2$.
As for $y$, we replace the circular energy $E$ with the actual value of the complete energy 
(with its dependence on $p_{r_*}^2$), $H_{\rm EOB}/M$, when implemented in the full model.
The tail factor explicitly reads
\be
\label{eq:Tlm}
T_\lm=\dfrac{\Gamma\left(\ell+1-2{\rm i}\hat{\hat{k}}\right)}{\Gamma(\ell+1)}e^{\pi \hat{\hat{k}}}e^{2{\rm i}\hat{\hat{k}}\log(2 k r_0)} \ .
\ee
In the above equation we have $\hat{\hat{k}}\equiv m E \Omega$, $k\equiv m\Omega$ and $r_0=2/\sqrt{e}$~\cite{Fujita:2010xj}.
For noncircularized dynamics, we follow the prescription of Ref.~\cite{Chiaramello:2020ehz},
so the Newtonian prefactor is taken in its {\it generic} form, i.e., with explicit dependence on the time-derivatives
of the coordinates as computed along the EOB dynamics.

Finally, the waveform model includes next-to-quasi-circular (NQC) corrections, represented by the factor $\hat{h}_{\ell m}^{\rm NQC}$,
that are NR-informed functions designed to improve the behavior of the analytical waveform during the 
plunge so as to smoothly connect it to the post-merger signal~\cite{Damour:2007xr,Damour:2009kr}. The correcting factor reads:
\begin{equation}
	\hat{h}_{\ell m}^{\rm NQC} = (1 + a_1^{\ell m} n_1^{\ell m} + a_2^{\ell m} n_2^{\ell m}) e^{i (b_1^{\ell m} n_3^{\ell m} + b_2^{\ell m} n_4^{\ell m})}\, ,
	\label{eq:nqcfactor}
\end{equation}
where the $n_i^{\ell m}$ are functions that explicitly depend on the radial momentum and acceleration, 
and the $a_i^{\ell m}$ and $b_i^{\ell m}$ are numerical coefficients. The NQC basis for the $\ell = 2, m = 2$ mode is:
\begin{subequations}
\begin{align}
	n_1^{22} &= \left(\dfrac{p_{r_*}}{r \Omega}\right)^2  \ ,\\
	n_2^{22} &= n_1^{22} p_{r_*}^2 \ ,\\
	n_3^{22} &= \dfrac{p_{r_*}}{r \Omega} \ ,\\
	n_4^{22} &= (r \Omega)p_{r_*} \,.
\end{align}
\end{subequations}
The set $n_i^{\ell m}$ for the other multipoles differs slightly from this, as detailed in previous works~\cite{Nagar:2024dzj,Nagar:2021gss,Nagar:2019wds}. 
We collect here their current form in \TEOBd~for future reference. In most cases, we have $n_k^{\lm} = n_k^{22}$.
The exceptions are the following: 
\begin{itemize}
\item[(i)] $n_2^{21} = n_1^{22} \Omega^{2/3}$.
\item[(ii)] $\ell = 3$ modes: $n_2^{32} = n_1^{22} \Omega^{2/3}$ and $n_4^{32} = n_3^{22} \Omega^{2/3}$.
\item[(iii)]$\ell = 4$ modes: $n_2^{43} = n_2^{42} = n_1^{22} \Omega^{2/3}$ and $n_4^{43} = n_3^{22} \Omega^{2/3}$.
\item[(iv)]$\ell = 5$ modes: $n_2^{55} = n_1^{22} \Omega^{2/3}$, and $n_4^{55} = n_3^{22} \Omega^{2/3}$.
\end{itemize}
% $n_4^{21} = (r\Omega) p_{r_*}$~\cite{Nagar:2024dzj}; $n_2^{32} = n_1^{32} \Omega^{2/3}$; $n_2^{43} = n_1^{43} \Omega^{2/3}$;
% $n_2^{42} = n_1^{42} \Omega^{2/3}$; $n_4^{42} = n_3^{42} (r \Omega)^2$; finally, for the $(\ell, m) = (3,3)$ and $(4,4)$ modes,
% $n_4^{\ell m} = n_3^{\ell m} \Omega^{2/3}$ for negative spins, while $n_4^{\ell m} = n_3^{\ell m} (r \Omega)^2$ if they are positive~\cite{Nagar:2021gss}.
The NQC corrections are informed by NR in that the numerical coefficients $a_i^{\ell m}, b_i^{\ell m}$ are found by imposing,
for each multipole $(\ell, m)$, a $C^{1}$ match between the EOB and NR waveform amplitude and frequency at a specific time
after the merger. More explicitly, for each mode, the numerical coefficients entering Eq.~\eqref{eq:nqcfactor} are found
by solving the system:
\begin{subequations}
\begin{align}
	A_{\ell m}^{\rm NR} (t^{\rm NR}_{A_{\ell m}^{\rm peak}} + 2)            &= A_{\ell m}^{\rm EOB}            (t^{\rm EOB}_{\rm NQC} + \Delta t_{\ell m}^{\rm NR}) \ ,\\
	\dot{A}_{\ell m}^{\rm NR} (t^{\rm NR}_{A_{\ell m}^{\rm peak}} + 2)      &= \dot{A}_{\ell m}^{\rm EOB}      (t^{\rm EOB}_{\rm NQC} + \Delta t_{\ell m}^{\rm NR}) \ ,\\
	\omega_{\ell m}^{\rm NR} (t^{\rm NR}_{A_{\ell m}^{\rm peak}} + 2)       &= \omega_{\ell m}^{\rm EOB}       (t^{\rm EOB}_{\rm NQC} + \Delta t_{\ell m}^{\rm NR}) \ ,\\
	\dot{\omega}_{\ell m}^{\rm NR} (t^{\rm NR}_{A_{\ell m}^{\rm peak}} + 2) &= \dot{\omega}_{\ell m}^{\rm EOB} (t^{\rm EOB}_{\rm NQC} + \Delta t_{\ell m}^{\rm NR})\  ,
\end{align}
\end{subequations}
where $t^{\rm EOB}_{\rm NQC} \equiv t^{\rm EOB}_{A_{22}^{\rm peak}} + 2$, and $\Delta t_{\ell m}^{\rm NR}$ is the delay in the peak amplitude of the $(\ell, m)$ mode with respect to the $(2,2)$ mode,
fitted from NR data~\cite{Nagar:2019wds}. The location of the amplitude peak for the EOB waveform is defined by
$t_{A_{22}^{\rm peak}}^{\rm EOB} = t_{\Omega_{\rm orb}}^{\rm peak} - 2 - \Delta t_{\rm NQC}$, where $\Delta t_{\rm NQC} = 1$~\cite{Nagar:2023zxh} 
and $t_{\Omega_{\rm orb}}^{\rm peak}$ is the time when the {\it pure orbital frequency} $\Omega_{\rm orb}$, i.e., the contribution to the orbital frequency coming from 
$\hat{H}_{\rm eff}^{\rm orb}$ only, reaches its maximum value\footnote{We use this ``partial" orbital frequency in part 
because the complete EOB orbital frequency $\Omega$ does not reach a peak at the end of the dynamics for high spins
(for equal-mass systems, this happens when $\chi_{\rm eff} \gtrsim 0.7$),
but also because it is generally effective in locating the actual peak wave amplitude~\cite{Damour:2014sva, Harms:2014dqa}.}~\cite{Damour:2014sva}:
\begin{equation}
\label{eq:Omg_orb}
	\Omega_{\rm orb} = \dfrac{M p_\varphi A u_c^2}{H_{\rm EOB} \hat{H}_{\rm eff}^{\rm orb}} .
\end{equation}
The inspiral-plunge-merger (insplunge-mrg) waveform described thus far is completed 
by an NR-informed phenomenological ringdown waveform~\cite{Damour:2014yha}, 
so that the complete signal can be represented as:
\begin{equation}
	h_{\ell m}= \theta (t^{\rm match}_{\ell m} - t) h_{\ell m}^{\rm insplunge-mrg} + \theta (t - t^{\rm match}_{\ell m}) h_{\ell m}^{\rm rng},
\end{equation}
where $\theta(t)$ is the Heaviside step function. The matching time when the ringdown signal is attached to the inspiral-merger
waveform is $t^{\rm match}_{\ell m} = t^{\rm EOB}_{A_{\ell m}^{\rm peak}} + 2 \equiv t^{\rm EOB}_{A_{22}^{\rm peak}} + \Delta t_{\ell m}^{\rm NR} + 2$.
See Refs.~\cite{Nagar:2020pcj,Nagar:2019wds} for details on the post-merger model
$h_{\ell m}^{\rm rng}$ and in particular for the explicit expression of the NR-informed functions involved.
The treatment of the $(2,1)$ mode is an exception to both the NQC corrections and the post-merger waveform 
described above~\cite{Nagar:2023zxh}. Due to the need to match the $(2,1)$ EOB and NR waveforms {\it before} their peak 
amplitude to avoid the problematic behavior of some of the NQC basis functions in the strong field, the NQC parameters in 
this case are determined earlier, namely at the time of the peak of the $(2,2)$ mode amplitude, 
$t_{A_{22}^{\rm peak}}^{\rm EOB}$; in addition, the ringdown portion of the waveform is attached at this same time.
Since the ringdown model otherwise implemented in \TEOBResumS{} from Ref.~\cite{Nagar:2020pcj} 
does not extend to times before each mode's peak amplitude, we employ the \SEOBNRv{5}  ringdown 
fits introduced in~\cite{Pompili:2023tna} in the case of the $(2,1)$ 
mode\footnote{Note that in Ref.~\cite{Nagar:2023zxh} it was mentioned that also the ringdown for the
$\ell=m=3$ and $\ell=m=4$ modes can be optionally constructed following Ref.~\cite{Pompili:2023tna} within
the \TEOBResumS{} model. This feature is inherited also by the model considered here.}.

The structure of the multipolar waveform is inherited by the angular component of the 
radiation reaction force $\hat{\cal F}_{\varphi}$. Following the prescription of 
Ref.~\cite{Nagar:2021xnh} (see Sec.~IIC there), it reads:
\begin{equation}
	\label{eq:radreac}
	\hat{\mathcal{F}}_{\varphi} = -\dfrac{32}{5} \nu r_\Omega^4 \Omega^5 \hat{f} \left(x\right) + \hat{\mathcal{F}}_{\varphi}^{\rm H}\ ,
\end{equation}
where $\hat{\mathcal{F}}_\varphi^{\rm H}$ is the angular momentum flux absorbed by
the two black holes~\cite{Damour:2014sva}. The factor $\hat{f} \left(x\right)$ is 
a sum of modes $\hat{f}_{\ell m}$, each normalized by the leading-order, 
$\ell = 2, m = 2$ Newtonian flux $\hat{F}_{22}^{\rm Newt} = -\frac{32}{5} \nu \Omega^{10/3}$, 
and factorized and resummed as detailed above for the waveform multipoles:
\begin{subequations}
\begin{align}
        \label{eq:sumflm}
	\hat{f} (x) &= \sum_{\ell = 2}^{8} \sum_{m = -\ell}^{\ell} \hat{f}_{\ell m} \hat{f}_{\ell m}^{\rm noncircular}  \ ,\\
	\hat{f}_{\ell m} &= \dfrac{\hat{F}^{\rm Newt}_{\ell m}}{\hat{F}^{\rm Newt}_{22}} |\hat{h}_{\ell m}|^2 \ .
	\label{eq:reducedflux}
\end{align}
\end{subequations}
When considering generic orbital dynamics, each of these modes can be complemented 
by the noncircular correcting factor $\hat{f}_{\ell m}^{\rm noncircular}$, of which \TEOBResumS~currently
uses only the leading-order $(\ell, m) = (2,2)$ term, expressed in terms of explicit time-derivatives of the
dynamical variables. More details on its expression and implementation are given in 
Appendix~\ref{app:derivs}, which integrates and completes the information shared in 
previous works~\cite{Chiaramello:2020ehz,Nagar:2021xnh}.

The radial component of the radiation reaction is factorized as 
$\hat{\mathcal{F}}_{r_*} = -5/3 p_{r_*}/p_\varphi \hat{\cal F}_\varphi \hat{f}_{p_{r_*}}$,
with the polynomial function $\hat{f}_{p_{r_*}}$ Pad\'e-resummed, as specified in and around
Eqs.~(6) and (7) of Ref.~\cite{Nagar:2023zxh}.

\subsubsection{New resummation of the waveform amplitude corrections}

\begin{figure}[t]
	\center	
	\includegraphics[width=0.45\textwidth]{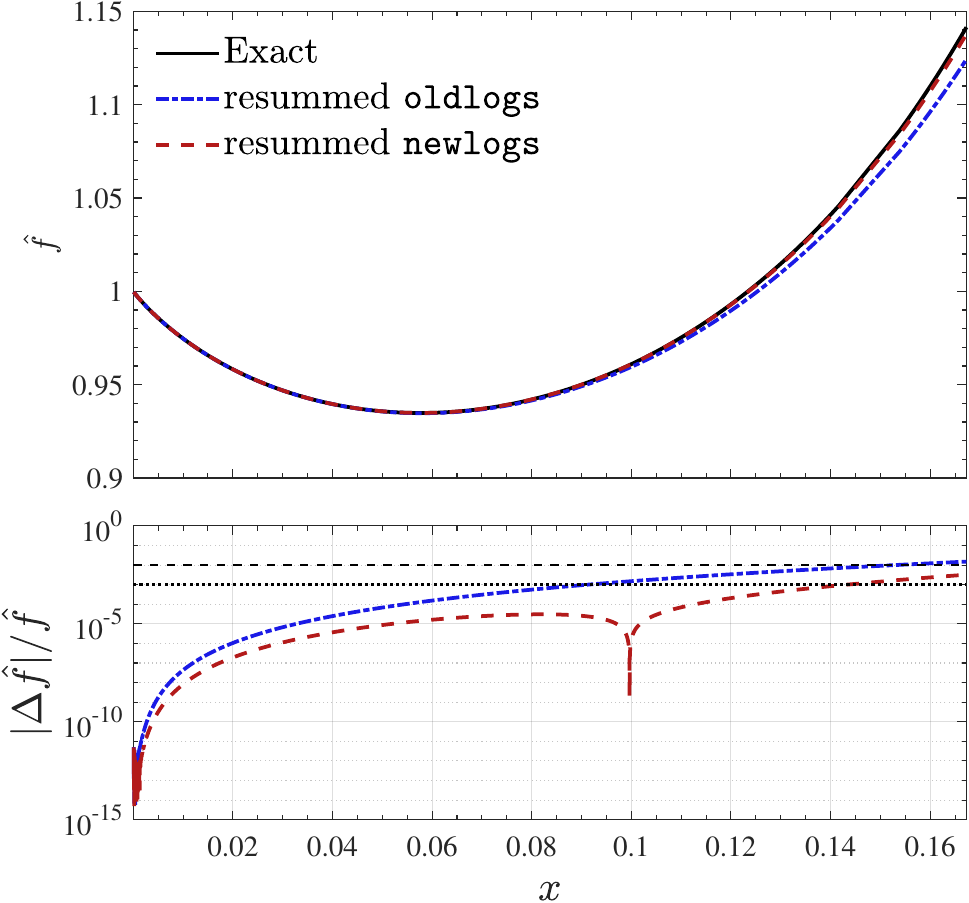}
	\caption{\label{fig:hatf_comparison}Newton-reduced energy flux function of a test mass on circular orbits around a 
	Schwarzschild black hole. The top panel displays the function $\hat{f}(x)$ defined in
	Eqs.~\eqref{eq:sumflm}-\eqref{eq:reducedflux}. The black curve, considered exact, 
	is obtained solving the Teukolsky equation numerically~\cite{Hughes:2005qb}.
	The dashed and dash-dotted curves represent analytical functions obtained factorizing 
	and resumming post-Newtonian results as described in the text. The {\tt newlogs} is computed by setting $\nu = 0$
	in Eq.~\eqref{eq:sumflm} and with the resummation choices outlined in Sec.~IIB and the Pad\'e 
	approximants of Table~\ref{tab:rholmresum}; the {\tt oldlogs} is the same but with the $\rho_{\ell m}$ 
	functions considered in Paper~I, with the $\log(x)$ replaced by constants while resumming. 
	All multipoles up to $\ell=8$ included are considered. 
	In the bottom panel are the relative differences, 
	$\Delta \hat{f}=(\hat{f}_{\rm Exact}-\hat{f}_{\rm X})/\hat{f}_{\rm Exact}$ with X either {\tt newlogs}
	or {\tt oldlogs}. The new treatment of the logarithms delivers uniform improvement, with 
	the difference at $x_{\rm LSO} = 1/6$ smaller by approximately one order of magnitude.}
\end{figure}

The changes with respected to Paper~I concern the $\rho_{\lm}$ functions, that are here 
resummed differently and include more PN information for the $\ell=7$ and $\ell=8$ modes.
In general, each $\rho_{\ell m}$ comprises an orbital part $\rho_{\ell m}^{\rm orb}$ and
a spin contribution $\rho_{\ell m}^{\rm S}$.
Beginning with the former, $\rho_{22}^{\rm orb}$ is here taken at full 4PN, as computed
in Paper I, while the other functions, up to $\ell=8$, are taken at $3^{+3}$~PN order, that 
is, the 3PN $\nu$-dependent information is hybridized with test-mass terms so as to reach 
(formal) 6PN accuracy. This is in particular an update over Paper~I for what concerns the $\ell = 7, 8$
multipoles, which were there kept in Taylor expanded form and only up to 3PN\footnote{The rationale behind this choice
is to have the flux analytically consistent with that used in the EOB waveform model 
developed for extreme mass ratio inspirals (EMRIs)~\cite{Albanesi:2021rby}, although we are here 
implementing a different treatment of the logarithmic terms. An updated version of said model
will be presented in a forthcoming work~\cite{Panzeri:2024prep}. 
See also Refs.~\cite{Albertini:2022dmc,Albertini:2023aol,Albertini:2024rrs}.}.
The 4PN-accurate $\rho_{22}^{\rm orb}$ function is resummed according to the procedure
of Appendix~A of Paper~I, starting from Eq.~(A7) therein, separating it in a rational
part $p_{22}^{0} (x)$, resummed with a $(2,2)$ Pad\'e approximant, and a remainder that, 
after factoring out a $\log(x)$, is a polynomial $p_{22}^{\log} (x)$ that is resummed using a $(0,1)$ Pad\'e approximant:
\begin{equation}
	\label{eq:rho22resum}
	\rho_{22}^{\rm orb} (x) = P_2^2\left[p_{22}^{0} (x)\right] + P_1^0 \left[p_{22}^{\log} (x)\right] \log (x)  \ .
\end{equation}
All other orbital modes up to $\ell=8$ can be formally written as:
\be
\label{eq:rholm}
\rho^{\rm orb}_\elm(x) = p_{\ell m}^{0}(x) + p_{\ell m}^{\log, 1}(x)\log(x) + p_{\ell m}^{\log,2}\log^2(x) \ ,
\ee
where $p_{\ell m}^{0,\log 1,\log 2}(x)$ are polynomials in $x$ of the form:
\begin{subequations}
\begin{align}
	p_{\ell m}^{0}(x) &= 1 + x + x^2 + x^3 + x^4 + x^5 + x^6 \ , \\
	p_{\ell m}^{\log, 1}(x) &= x^3 + x^4+x^5+x^6 \ , \\
	p_{\ell m}^{\log,2}(x) &=x^6 \ .
\end{align}
\end{subequations}
The choice of Pad\'e approximant for each $\rho_\lm^{\rm orb}$ is driven by comparisons with 
the functions computed numerically by solving the Teukolsky equation for
a test-mass orbiting a Kerr black hole~\cite{Hughes:2005qb}, following the 
standard approach used in the past to validate resummation 
choices (see e.g. Refs.~\cite{Damour:2008gu,Messina:2018ghh} 
and in particular Appendix~A of Paper~I for the $\ell=m=2$ mode). 
Our final choices are summarized in Table~\ref{tab:rholmresum}.
Although a comprehensive analysis is postponed to a forthcoming  
study~\cite{Panzeri:2024prep}, for completeness we report here some results in the nonspinning
case of a particle orbiting a Schwarzschild BH. 
Although each Pad\'e approximant was chosen by inspecting the 
behavior of each $\rho_\lm$ function separately, for brevity we discuss
here the quality of the nonspinning reduced flux function $\hat{f}(x)$ in the circular 
test-mass limit (i.e., Eq.~\eqref{eq:sumflm} above with $\hat{f}_{\lm}^{\rm noncircular}=1$ and $\nu=0$). 
This is done in \figref{fig:hatf_comparison}, where the function obtained with the 
new resummation (dubbed {\tt newlogs}) is contrasted with both the exact curve 
obtained from numerical solutions of the Teukolsky equation and the 
analytical EOB one (in the limit $\nu=0$, dubbed {\tt oldlogs}) of Paper~I 
(and largely based on the choices of Ref.~\cite{Nagar:2020pcj,Messina:2018ghh}).
As shown by the relative differences plotted in the bottom panel, the {\tt newlogs} option
leads to better agreement across the board; in particular, at the LSO, accuracy to numerical data
increases by an order of magnitude.

Concerning the spin terms $\rho_{\ell m}^{\rm S}$, as in Paper~I, for all but the $(2,2)$ mode 
we again follow the prescriptions described in Sec.~II~A of Ref.~\cite{Nagar:2020pcj}, differentiating the treatment of even-$m$
and odd-$m$ modes and employing the resummations summarized in Table~I therein.
In past models, and in particular in Paper~I and Ref.~\cite{Nagar:2021xnh}, $\rho_{22}^S$ was taken at full NNLO 
complemented by suitably $\nu$-dressed test-mass terms~\cite{Messina:2018ghh}, 
see Eq.~(19) of Paper~I. 
Despite the availability of the NNLO (and beyond) knowledge~\cite{Henry:2022dzx, Bohe:2016gbl}, the model 
we discuss here only retains the NLO contributions (spin-orbit, spin-spin, and spin-cube) in $\rho_{22}^{\rm S}$,
\begin{equation}
	\label{eq:rho22S}
    \rho_{22}^{\rm S}= c_{\rm SO}^{\rm LO} x^{3/2} +c_{\rm SS}^{\rm LO} x^2 +c_{\rm SO}^{\rm NLO} x^{5/2} + c_{\rm SS}^{\rm NLO}x^3 + c_{\rm S^3}^{\rm LO} x^{7/2}.
\end{equation}
This choice is motivated by the fact that, as already mentioned in Ref.~\cite{Nagar:2016ayt}, 
the beyond-NLO terms give rather large contributions towards merger that eventually reduce the flexibility 
of the model and the accuracy of the NR-completion procedure. Sec.~\ref{sec:results}
below will show that a highly accurate EOB/NR phasing agreement is possible even while neglecting these terms.
For consistency, we also drop from $\rho_{22}^{\rm S}$ the aforementioned test-mass terms from Paper~I
(Eqs.~(26)-(27) there).

% The $p_0(x)$ term is typically resummed with a $(4,2)$ Pad\'e approximant,
% except for the $(\ell,m)$ modes $(3,1)$, $(5,1)$, $(6,2)$, $(6,1)$, $(7,1)$, $(8,2)$ e $(8,1)$
% that are taken in Taylor-expanded form;
% $p^1_{\log}(x)$, the coefficient of the logarithm, is usually taken in Taylor-expanded form. 
% Exceptions to this rule are the $(5,2)$, $(6,3)$, $(6,5)$, $(7,2)$, $(7,4)$, $(7,6)$, $(8,3)$, $(8,5)$ modes, for which we 
% use a $P^1_2$ approximant, and the $(2,1)$ and $(5,4)$ modes, for which we use a $P^2_1$ approximant.
% These choices are driven by comparisons with test-mass data (that also involve the black hole spin), 
% and will be detailed in a forthcoming study~\cite{Panzeri:2024prep}. 
%======================
% Tables for the approximants
%======================
\begin{table}[t]
	\caption{\label{tab:rholmresum}
	Choice of resummation for the $\rho^{\rm orb}_{\lm} (x)$ functions, for all $\ell \leq 8$ and $1 \leq m \leq \ell$, save the $(\ell, m) = (2,2)$ mode. 
	Here, $p_{\lm}^{0}$ and $p_{\lm}^{\log, 1}$ are, respectively, the rational part and the polynomial coefficient of $\log(x)$ in Eq.~\eqref{eq:rholm}.
	$P^n_k$ stands for a $(n,k)$ Pad\'e approximant; an $(n,0)$ Pad\'e is the same as a Taylor series to $n$-th order ($T_n$ below).}
	  \begin{center}
		% \begin{ruledtabular}
   \begin{tabular}{c l | c l} 
	\hline \hline \\[-0.99em]
   \multicolumn{2}{c|}{\boldmath $p^0_{\lm}$ \unboldmath} & \multicolumn{2}{|c}{\boldmath $p^{\log, 1}_{\lm}$ \unboldmath} \\[0.1em]
   \hline
   \multirow{6}{1.75cm}{\centering \boldmath $P^6_0 \equiv T_6$ \unboldmath} & $(3,1)$                     & \multirow{2}{1.75cm}{\centering \boldmath $P^2_1$ \unboldmath} & $(2,1)$ \\
                                                                             & $(5,1)$                     &                                                                & $(5,4)$ \\
																	         \cline{3-4}
																	         & $(6,2), (6,1) \hphantom{1}$ & \multirow{4}{1.75cm}{\centering \boldmath $P^1_2$ \unboldmath} & $(5,2)$ \\
																	         & $(7,1)$                     &                                                                & $(6,5), (6,3)$ \\
																	         & $(8,2), (8,1)$              &                                                                & $(7,6), (7,4), (7,2) \hphantom{1}$ \\
																	         &                             &                                                                & $(8,5), (8,3)$ \\
	\hline																  			
    \boldmath $P^4_2$ \unboldmath                                         & All others & \boldmath $P^3_0 \equiv T_3$ \unboldmath & All others \\[0.1em]
	\hline \hline
      \end{tabular}
% \end{ruledtabular}
\end{center}
\end{table}

%=================================
% Test-mass limit: validating the new-logs flux
%=================================

\subsection{Resummation of the NNLO spin-orbit waveform amplitude correction}
\label{sec:NLOvsNNLO}
\begin{figure}[t]
	\center	
	\includegraphics[width=0.4\textwidth]{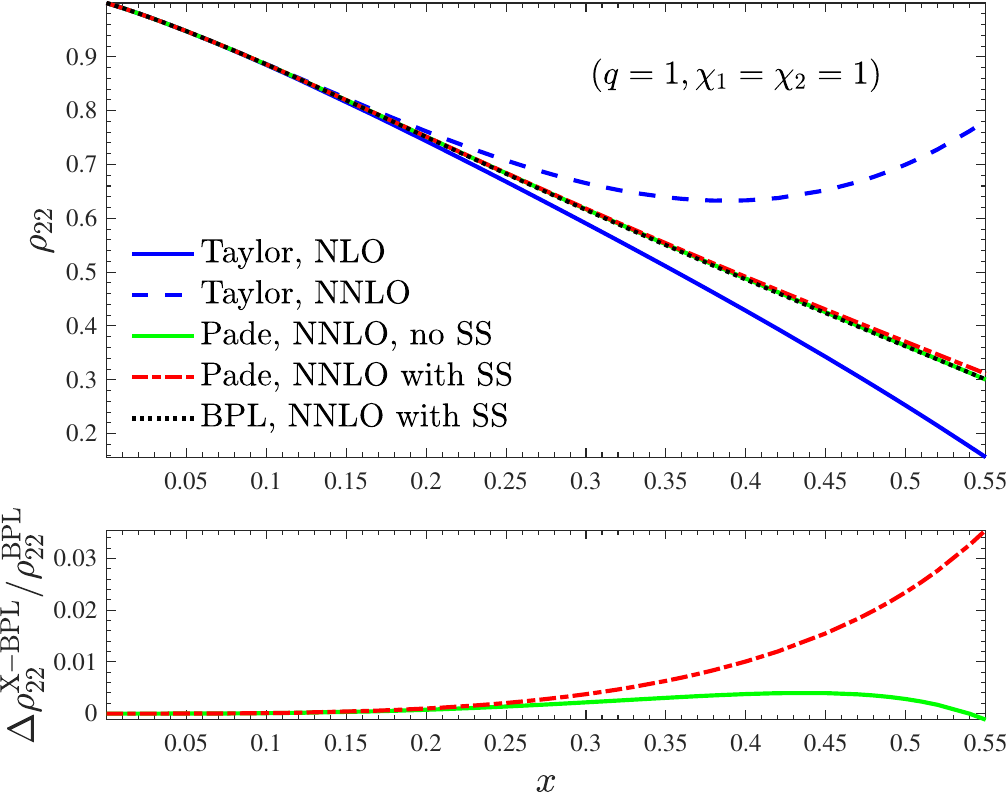 }
        \caption{\label{fig:rho22_resum}
		Comparing the effect of resummed and nonresummed
        versions of the spin contribution $\rho_{22}^{\rm S}$, Eq.~\eqref{eq:rho22S}, to $\rho_{22}$.
        The plot highlights the mutual consistency between three different resummed 
        expressions for $\rho_{22}^{\rm S}$, including the Borel-Pad\'e-Laplace (BPL).
        Bottom panel: relative differences between the two Pad\'e resummed expressions and 
        the BPL. See text for details.}
\end{figure}

Before discussing the performance of the updated model presented in this work, let us explore 
in some detail the strong field behavior of the function $\rho_{22}^S$ at NNLO accuracy; we neglect in this section
the spin-cube term $c_{\rm S^3}^{\rm LO}$. 
The complete $\rho_{22}$ reads
\begin{align}
\rho_{22}=\rho_{22}^{\rm orb}+\rho_{22}^{\rm S} \ ,
\end{align}
where
\begin{align}
\label{eq:rho22s_nnlo}
\rho_{22}^{\rm S}=& c_{\rm SO}^{\rm LO} x^{3/2} +c_{\rm SS}^{\rm LO} x^2 +c_{\rm SO}^{\rm NLO} x^{5/2} \nonumber \\
                 & + c_{\rm SS}^{\rm NLO}x^3 + c_{\rm SO}^{\rm NNLO} x^{7/2} \ ,
\end{align}
and the explicit expression of the various coefficients is reported in Eqs.~(20)-(23) of Paper~I.
Fig.~\ref{fig:rho22_resum} compares various representations of $\rho_{22}$ in the extremal
case with $q=1$ and $\chi_1=\chi_2=1$.
The figure illustrates that the effect of the NNLO contribution in the (Taylor-expanded) 
Eq.~\eqref{eq:rho22S} is rather large, so that the function effectively changes
shape moving from NLO to NNLO truncation. Such a large difference between NLO and NNLO
indicates that one should attempt a resummation. This was the spirit of Ref.~\cite{Nagar:2016ayt}, that proposed to 
factorize the spin-independent part and resum with certain Pad\'e approximants both the orbital
and spin parts. Here we explore a different, more ``mathematically natural'' procedure. 
We consider the $\rho_{22}$ global function as the sum of two polynomials, one in
integer powers of $x$ and another with semi-integer powers of $x$, and we resum the two parts
separately. Formally, we write the functions as
\be
\label{eq:rho22_tot}
\rho_{22}=\rho_{22}^{\rm orb} + c_{\rm LO}^{\rm SS} x^2 + c_{\rm NLO}^{\rm SS} x^3 +c_{\rm SO}^{\rm LO}x^{3/2}(1+c_{\rm nlo} x + c_{\rm nnlo} x^2)
\ee
where $c_{\rm nlo}\equiv c_{\rm SO}^{\rm NLO}/c_{\rm SO}^{\rm LO}$ and $c_{\rm nnlo}\equiv c_{\rm SO}^{\rm NNLO}/c_{\rm SO}^{\rm LO}$.
The function $(1+c_{\rm nlo} x + c_{\rm nnlo} x^2)$ is then resummed as a $(1,1)$ Pad\'e approximant. 
For the even-in-spin part (the $c^{\rm SS}$ terms above, plus the orbital part) we can adopt two different procedures, 
compared with the Taylor-expanded results in Fig.~\ref{fig:rho22_resum}. We can: 
(1) resum just $\rho_{22}^{\rm orb}$ as discussed above (Eq.~\eqref{eq:rho22resum}), with the spin-spin terms
added in Taylor-expanded form (``Pad\'e, NNLO, no SS'' in Fig.~\ref{fig:rho22_resum}); or 
(2) incorporate the spin-spin terms in the rational part of $\rho_{22}^{\rm orb}$ and resum them together 
via a $(2,2)$ Pad\'e approximant (``Pad\'e, NNLO, with SS'' in Fig.~\ref{fig:rho22_resum}).
These two approaches are broadly consistent, though fractional differences of a few percent can be appreciated at large frequencies.
In a forthcoming study dealing with the fluxes in the test-mass limit~\cite{Panzeri:2024prep}, we will show that treating 
separately integer and semi-integer powers of $x$ yields a better agreement with the numerical data and 
thus this choice should be a priori preferred also in the comparable mass case.

Fig.~\ref{fig:rho22_resum} shows that, even when resummed, the NNLO expressions remain larger than the
the NLO \textit{nonresummed} version, and thus their use prevents us from attaining a comparable level
of NR-faithfulness to the EOB model we discuss in this work.
Evidently this is true keeping the same treatment of the spin-orbit interaction in the conservative part 
of the dynamics. We do not exclude that a different approach to the spin-orbit sector could allow us to 
more effectively use the resummed expression. Additional investigations aimed at exploring
this possibility are postponed to future work.

%\subsubsection{Borel--Pad\'e--Laplace resummation}
%\rg{New sub-sub section. Imho, this could/should be upgraded as a ``proper'' subsection, 
%with an example application.}
%So far, we have based our resummation procedures on straight Pad\'e approximants of the $\rho_{22}$ functions 
%in the $x$-space. 
%===================
% Test borel summed thing
%===================

\subsubsection{Borel-Pad\'e-Laplace resummation}
To gain more insight on the robustness of our Pad\'e approximants in $x$-space in view of future applications, it is
also instructive to explore the outcome of applying Borel summation to the spin-orbit terms.

Assuming $x$ is a complex variable, Borel summation assigns to a formal power series $\tilde{f}(x)\in x^\alpha \mathbb{C}[\![x]\!]$ a 
complex analytic function $F(x)$ defined in a sector of opening angle less than $\pi$. The function $F$ is called the Borel sum of $\tilde{f}$ 
and it has the property of being uniform Gevrey asymptotic~\footnote{Let $\tilde{f}(x)=\sum_{n=0}^{\infty}a_n x^{n}\in\mathbb{C}[\![x]\!]$. 
An analytic function $F\colon\mathcal{D}\to\mathbb{C}$ is uniformly $1$-Gevrey asymptotic to $\tilde{f}\in \mathbb{C}[\![x]\!]$ if 
there exist some constants $A,C>0$ such that
\be
\Big|F(x)-\sum_{n=0}^{N-1}a_n x^{n}\Big|\le C A^N N! |x|^{N}\,,
\ee
for $N>0$, and $x\in\mathcal{D}$; see Ref.~\cite[Definition~5.21]{Sauzin}.} to $\tilde{f}$; see~\cite[Theorem~5.20]{Sauzin}. This property makes Borel summation special: 
indeed, if an analytic function $\hat{F}$ is Gevrey asymptotic to $\tilde{f}$ in a large enough sector, then $\hat{F}$ agrees with the Borel sum of 
$\tilde{f}$. In other words, the Borel sum is the unique analytic function with uniform Gevrey asymptotics $\tilde{f}$ in a large enough 
sector~\cite{nevanlinna,sokal1980improvement}\footnote{For a modern proof we refer to~\cite[Theorem~B.15]{nikolaev2023exact}. 
For a generalization to power series with fractional power exponents we refer to~\cite{delabaere--rosoamanana}.}. 

The paradigm of Borel summation consists of three steps:
\begin{itemize}
\item[(i)] The first step is given by the Borel transform $\mathcal{B}\colon x^\alpha\mathbb{C}[\![x]\!]\to s^{\alpha}\mathbb{C}[\![s]\!]$. 
By definition, for every $n=1,2,\ldots$ and $\alpha\in\mathbb{Q}\setminus\mathbb{Z}^\times$,
\footnote{We denote by $\mathbb{Q}$ the set of rationals, and by $\mathbb{Z}^\times$ the set of non zero integers.} 
the Borel transform maps monomials $x^{n+\alpha+1}$ into monomials 
$\frac{s^{n+\alpha}}{\Gamma(n+\alpha+1)}$, and it is extended by countable linearity to formal power series,
\begin{align*}
\mathcal{B}\Big( x^\alpha\sum_{n=0}^\infty a_n x^n\Big)=s^\alpha\sum_{n=1}^\infty a_n\frac{s^{n-1}}{\Gamma(n+\alpha)}\,.
\end{align*}
The Borel transform of a formal series $\tilde{f}(z)$ will be denoted by $\hat{f}(s)$. 
\item[(ii)]The second step consists of summing the series $\hat{f}(s)$ and possibly extending it analytically outside 
the disk of convergence. Indeed, when the series $\tilde{f}(x)$ is divergent, its Borel sum $\hat{f}(s)$ has singularities 
in the $s$-space, and it is important to know their structures in view of the third step.    
\item[(iii)]The third step is given by the Laplace transform $\mathcal{L}$, which acts on locally integrable exponentially 
decaying functions on the positive reals and gives real-analytic functions,\footnote{In fact, it gives complex-analytic 
functions in the half-plane $\Re(x)>c$, where $c$ is a constants that depends on the exponential decay of $\hat{f}(s)$.}
\be\label{Laplace}
\big[\mathcal{L}\hat{f}\big](x):=\int_0^{+\infty}e^{-\frac{s}{x}}\hat{f}(s) \, ds\,.
\ee
\end{itemize} 
In numerical computations, step~$(\rm ii)$ usually consists of taking a Pad\'e approximant in $s$-space. 
If fractional powers of $s$ are present, it is preferable to use Pad\'e on the integral part after the factorization of the 
lowest fractional exponent (see Eq.~\eqref{eq:Pade} below for an example).
We refer to this resummation paradigm as Borel--Pad\'e--Laplace (BPL).

The choice of Pad\'e approximant in $s$-space should be made so that no poles lie on the positive reals, 
as it will otherwise not be possible to perform the Laplace transform. 
Still, even if every choice of Pad\'e approximant did have poles on the positive real axis, it would yet be possible to take the so called median resummation 
(see for instance~Eq.~(214) in Ref.~\cite{Fantini:2024snx}).

Borel summation is extensively used to resum divergent power series~\cite{Sauzin,costin}, namely it can be applied 
to formal series $\tilde{f}$ with zero radius of convergence. However, if $\tilde{f}$ is convergent, then its Borel 
transform $\hat{f}$ is an entire function of bounded exponential type~\cite[Lemma~5.7]{Sauzin}. Therefore, the Laplace 
transform of $\hat{f}$ is well defined, and its Taylor series expansion is given by $\tilde{f}$ itself. 

We now return to our original application. To cross check the Pad\'e resummation of the semi-integer powers part of $\rho_{22}^{\rm S}$ in Eq.~\eqref{eq:rho22S}, 
we implement a BPL resummation. As an illustrative example, let us consider a function of the form:
\be
f = f_{3/2} x^{3/2}+f_{5/2} x^{5/2} + f_{7/2} x^{7/2} .
\ee
Then, its Borel transform is 
\be
\big[\mathcal{B}f\big](s)=f_{3/2} \frac{s^{1/2}}{\Gamma\big(\tfrac{3}{2}\big)}+f_{5/2} \frac{s^{3/2}}{\Gamma\big(\tfrac{5}{2}\big)} + f_{7/2} \frac{s^{5/2}}{\Gamma\big(\tfrac{7}{2}\big)}\,.
\ee
Since there are fractional exponents, we use a (1,1) Pad\'e on the polynomial 
\be\label{eq:Pade}
s^{-\frac{1}{2}}\big[\mathcal{B}f\big](s)= \frac{f_{3/2}}{\Gamma\big(\tfrac{3}{2}\big)}+ \frac{f_{5/2}}{\Gamma\big(\tfrac{5}{2}\big)} s +  \frac{f_{7/2}}{\Gamma\big(\tfrac{7}{2}\big)} s^2\,,
\ee
and we obtain a rational function 
\be
\hat{f}(s)=s^{\frac{1}{2}} \frac{p_0+p_1s}{1+q_{1}s}\,,
\ee
for some coeffcients $p_0,p_1,q_1$.
The last step consists of taking the Laplace transform of $\hat{f}$, which gives a real-analytic function in the orginal $x$-space. 

The BPL-resummed spin-orbit part is then added to the orbital part resummed in $x$-space, that also
incorporates the spin-spin terms $c_{\rm SS}^{\rm LO}x^2$ and $c_{\rm SS}^{\rm NLO}x^3$ in the $(2,2)$
Pad\'e approximant (what we called option (2) above). The result is displayed as a black solid line in the top panel of Fig.~\ref{fig:rho22_resum}.
The bottom panel shows the fractional difference between each Pad\'e resummed expression and the BPL
one, i.e $\Delta\rho_{22}^{\rm X-BPL}\equiv \rho_{22}^{\rm X}-\rho_{22}^{\rm BPL}$ where $X$ can be
either one of the models (1) and (2) mentioned.
The consistency between the three resummed expressions is evident, although the BPL 
one is found to be closer to the Pad\'e-resummed one with the spin-spin terms simply added (model (1)) instead of the 
one that includes them within the Pad\'e (model (2)), as would be a priori expected.
In any case, this comparison indicates that the BPL procedure is a useful tool to cross-validate 
Pad\'e resummations  in $x$-space in the absence of other validation tools (e.g., exact data).

\subsection{Synthesis}
We summarize here the changes with respect to Paper~I introduced in this work for the conservative dynamics, 
waveform and radiation reaction. They will be tested in the following sections through NR comparisons.
\begin{enumerate}
\item[(i)] A new resummed form is introduced for the $A$ and $D$ functions entering the Hamiltonian,
            starting from the same 5PN-expanded versions considered in Paper~I (see also Ref.~\cite{Nagar:2021xnh}). 
            In each function, the rational and logarithmic terms in the PN expansion are 
            resummed separately according to Eqs.~\eqref{eq:Aorb_logresum} and~\eqref{eq:Dorb_logresum}.
\item[(ii)] The orbital part of the residual amplitude correction $\rho_{22}$ is taken at 4PN accuracy; it   
            is resummed following the same logic as the potentials (Eq.~\eqref{eq:rho22resum}). 
			The spin-dependent terms $\rho_{22}^{\rm S}$ up to NLO are added to $\rho_{22}^{\rm orb}$ in Taylor-expanded form.
			We exclude the NNLO (3.5PN) contributions to $\rho_{22}^{\rm S}$, as we find that they have a large
			impact that limits the flexibility of the model and prevents accurate phasing agreement with NR,
			as well as some hybrid, test-mass, $\geq \rm N^3LO$ SO terms that were used in Paper~I.
\item[(iii)] The remaining $\rho_{\lm}$ functions, up to $\ell = 8$, are taken at $3^{\rm +3}$ PN accuracy in their orbital part, 
             hybridizing complete 3PN information with 6PN test-mass results. The choice of resummation depends on the multipole (see
			 Table~\ref{tab:rholmresum}), but the same idea of treating rational and logarithmic terms separately is applied.
			 The spin contributions $\rho_{\lm}^{\rm S}$ are implemented as in Paper~I, i.e., following the prescriptions
			 detailed in Sec.~II.A of Ref.~\cite{Nagar:2020pcj}.
\end{enumerate}

\section{Binaries on bound orbits}
\label{sec:results}     
\subsection{Informing the model using NR simulations}
\label{sec:nrinfo}

%=================================
% Analytical summary of the spinning models
%=================================
\begin{table*}[t]
	\caption{\label{tab:c3_coeff}Coefficients for the fit of $c_3$ given by Eq.~\eqref{eq:c3fit} on the first-guess values for 
	 $c_3$ reported in Tables~\ref{tab:c3_eqmass}-\ref{tab:c3_uneqmass} for the SXS datasets listed in the same tables.}
	  \begin{center}
		\begin{ruledtabular}
   \begin{tabular}{c |c c c c c c | c c c c c c} 
   Model   & \multicolumn{11}{c}{\hspace{-28mm}$c_3^{=}\equiv p_0\left(1 + n_1\tilde{a}_0 + n_2\tilde{a}_0^2 + n_3\tilde{a}_0^3 + n_4\tilde{a}_0^4\right)/\left(1 + d_1\tilde{a}_0\right)$}\\
			   &  \multicolumn{11}{c}{$c_3^{\neq}\equiv \left(p_1\tilde{a}_0 + p_2\tilde{a}_0^2  + p_3\tilde{a}_0^3\right)\sqrt{1-4\nu}+ p_4\tilde{a}_0\nu \sqrt{1-4\nu} + \left(p_5\tilde{a}_{12}+ p_6\tilde{a}_{12}^2\right)\nu^2$}\\
			   \hline
	{\tt TEOBResumS*}      & $p_0$ & $n_1$ & $n_2$   &  $n_3$   &     $n_4$  & $d_1$  & $p_1$ & $p_2$ & $p_3$ & $p_4$ & $p_5$ & $p_6$\\
   \hline
   %{\tt Dali}$_{\tt newlogs}$  & 41.683 & $-1.656$ & $0.967$  & $-0.171$ & $-0.0456$ & $-0.688$   &  12.2697  & $-2.5116$ & $3.4525$  & $-56.4450$  & $67.7644$ & $-93.5733$   \\
   {\tt Dali}$_{\tt newlogs}$  & 43.588 & $-1.623$ & $0.913$  & $-0.103$ & $-0.0769$ & $-0.653$   &  12.0202 & $-2.7103$ & $-0.35956$  & $-36.6435$  & $34.8067$ & $-85.9733$   \\
   \end{tabular}
   \end{ruledtabular}
   \end{center}
   \end{table*}

%==========================
% Figure for a6c
%==========================
\begin{figure}[t]
	\center	
	\includegraphics[width=0.45\textwidth]{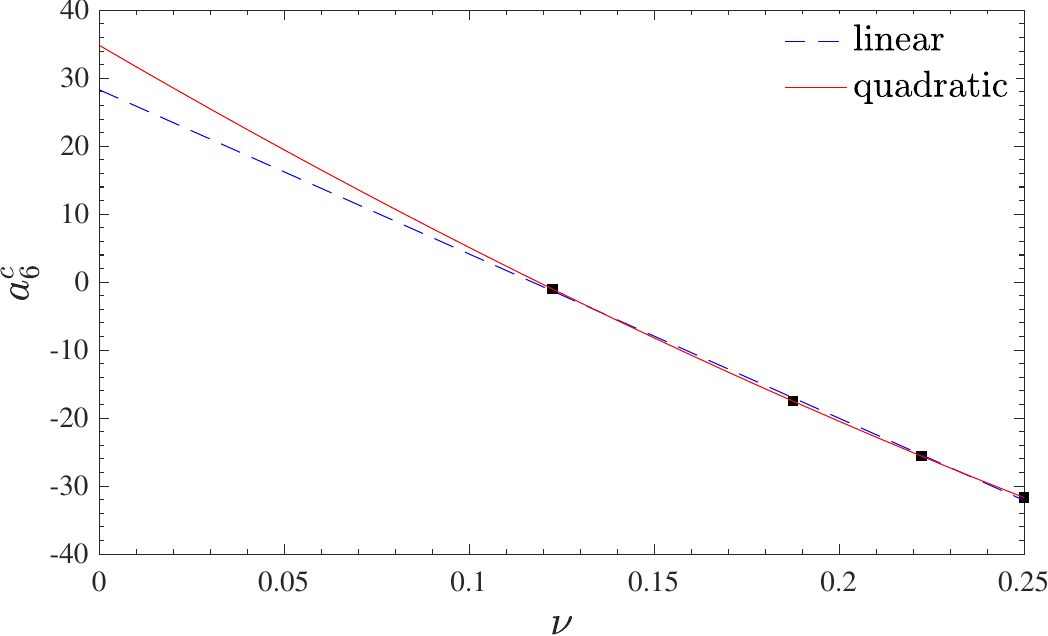} 
		\caption{\label{fig:a6c}NR-informed values of $a_6^c$  (points), as taken from the 6th column of Table~\ref{tab:a6cs0}, 
		and their fits. The quadratic function, Eq.~\eqref{eq:a6c}, yields an EOB dynamics that is NR-faithful also for $q>6$ ($\nu \lesssim 0.122$).}
\end{figure}

%==============================
% Figure for calibration: phase difference
%==============================
\begin{figure*}[t]
	\center	
	\includegraphics[width=0.32\textwidth]{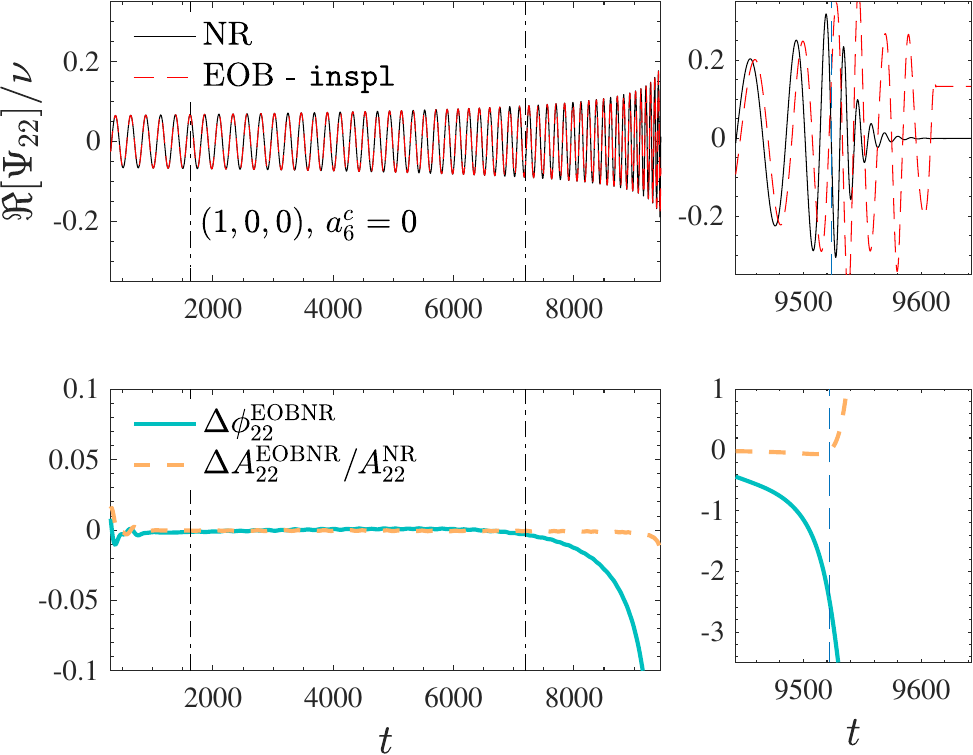}
	\includegraphics[width=0.32\textwidth]{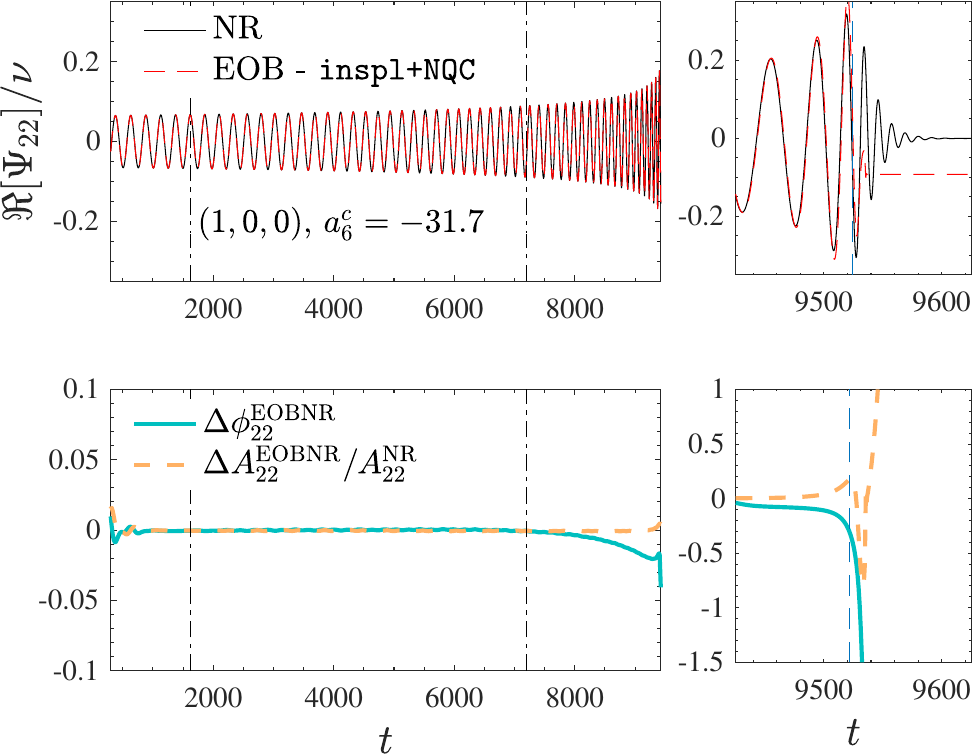}
	\includegraphics[width=0.32\textwidth]{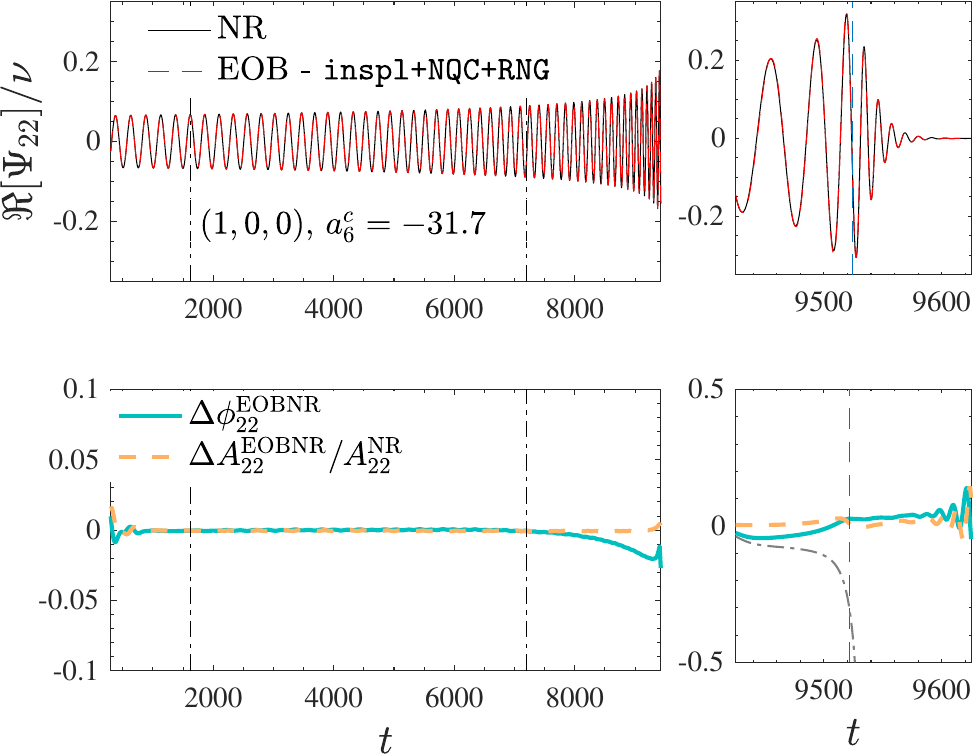}
	\caption{\label{fig:a6c_determination_steps}
	Illustrating the impact of the various NR-informed elements of the waveform for the $q=1$ case.
	The top plots display the real part of the waveform function $\Psi_{22} \equiv A_{22} \e^{-\i \phi_{22}}$, 
	while the EOB/NR amplitude and phase differences are in the bottom row.
	The EOB and NR waveforms are aligned in the frequency interval $(\omega_L,\omega_R)=(0.026,0.04)$, whose temporal bounds are marked by
	the vertical lines in the left part of each panel. Left panel: the EOB waveform, without NQC corrections (merger) and ringdown, and with $a_6^c=0$.
	Middle panel: the same waveform with $a_6^c=-31.7$, considered the best first-guess value of the parameter. Right panel: 
	performance of the complete waveform with $a_6^c=-31.7$, NQC corrections applied and ringdown attached. Note that the value
	of $a_6^c$ is obtained inspecting the full NQC-ringdown-completed waveform, not just the inspiral part.}
\end{figure*}

%=====================================
% Figure for calibration: Amplitude and Frequency
%=====================================
\begin{figure}[t]
	\center	
	\includegraphics[width=0.4\textwidth]{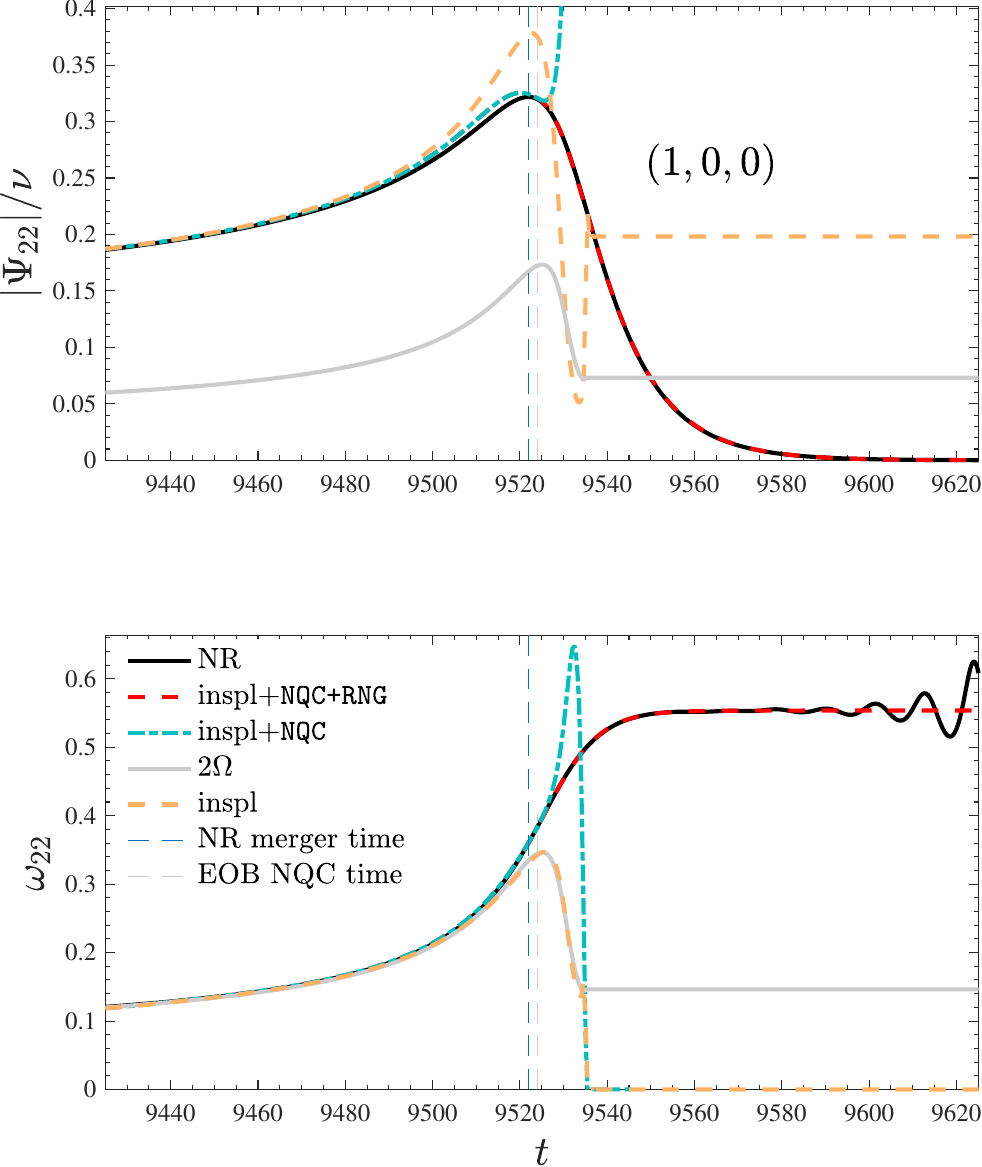}	
	\caption{\label{fig:a6c_Amp_Freq}
	Complement to Fig~\ref{fig:a6c_determination_steps} to highlight 
	the effect of the NQC corrections (light-blue dashed lines) as well as the completion of the EOB analytic waveform 
	(dubbed as {\tt inspl}, orange lines) with the ringdown ({\tt RNG}) attachment (red lines) on the $(2,2)$ mode. Waveform amplitude (top)
	and frequency (bottom). The vertical dashed blue line marks the time of merger in the NR data, while the grey one indicates
	$t_{\rm NQC}$, the time where the NQC corrections to the waveform are determined and the {\tt inspl+NQC} 
	waveform is attached to the NR-informed description of the ringdown.
	The gray line represents the orbital frequency $\Omega$ in the top panel and $2\Omega$ in the bottom panel, 
	to highlight the consistency with both the NR and the {\tt inspl} (non-NQC corrected) curve.}
\end{figure}

Let us turn now to tuning our model to NR through the determination of the effective 5PN parameter $a_6^c(\nu)$ 
and of the effective N$^3$LO spin-orbit parameter $c_3(\nu,\chi_{1,2})$ introduced in Sec.~\ref{sec:dynamics}. 
Here we review our method for doing so in detail, while also showcasing the impact of the other sources 
of NR information in the model, namely the NQC corrections and the post-merger waveform.
The procedure we follow for the calibration of the two parameters is the same as previous works
(see Paper~I and references therein, notably~\cite{Damour:2014sva,Nagar:2018zoe,Nagar:2020pcj,Nagar:2021xnh})
and consists of finding values that reduce (and possibly minimize) the accumulated phase difference 
at merger (whenever possible) when the EOB and NR waveforms are aligned in the time-domain 
during the early inspiral.
The alignment frequency region is chosen such that over it the phase difference,
which features small-amplitude oscillations because of residual eccentricity,
approximately averages zero.
In practice one seeks values of the parameters such that the phase
difference is comparable with (or below) the error on the NR phase at merger. This
error can be estimated (conservatively) by taking the difference between the highest and second 
highest resolutions available in the SXS catalog (see e.g. Refs.~\cite{Nagar:2019wds,Nagar:2020pcj}).
Once the values of the best-fitting parameters have been determined for a sample of a few tens
of NR datasets, global fits are performed, so that $a_6^c$ is function of $\nu$, while $c_3$ depends on
both $\nu$ and the BH spins. The performance of these NR-informed models is finally evaluated
by computing the EOB/NR unfaithfulness (or mismatch) for all available NR simulations (see Sec.~\ref{sec:testing} below).
As pointed out in Ref~\cite{Nagar:2023zxh} and stressed again in Paper~I, the lowest values 
of the EOB/NR unfaithfulness are reached when the EOB/NR phase difference is monotonically
decreasing and negative around merger.

For the determination of the 5PN coefficient $a_6^c$, we use 4 NR datasets with mass ratios
$q \in \left\{1, 2, 3, 6\right\}$, detailed in the sixth column of Table~\ref{tab:a6cs0} that 
we report in Appendix~\ref{app:newa6}. We also collect there for each the corresponding first-guess 
value of $a_6^c$, chosen as explained above and specifically allowing phase differences accumulated 
up to the merger of $\sim 0.1$~rad.
These points are accurately fitted (see Fig.~\ref{fig:a6c}) by the following quadratic function of $\nu$:
\be
\label{eq:a6c}
a_6^c(\nu) = 208.19\nu^2-318.26\nu+34.85.
\ee
Although at first sight the points seem to be approximately aligned
(and can be well fitted by $a_6^{c-{\rm linear}}(\nu)=8.3053-241.662\nu$), we find that 
a linear fitting function leads to extrapolated values of $a_6^c$ that yield large 
phase differences at merger for higher mass ratios (e.g., $\sim 1.9$~rad for $q=9.5$).

%==========================
% spin-dependent parameter space
%==========================
\begin{figure}[t]
	\center	
	\includegraphics[width=0.4\textwidth]{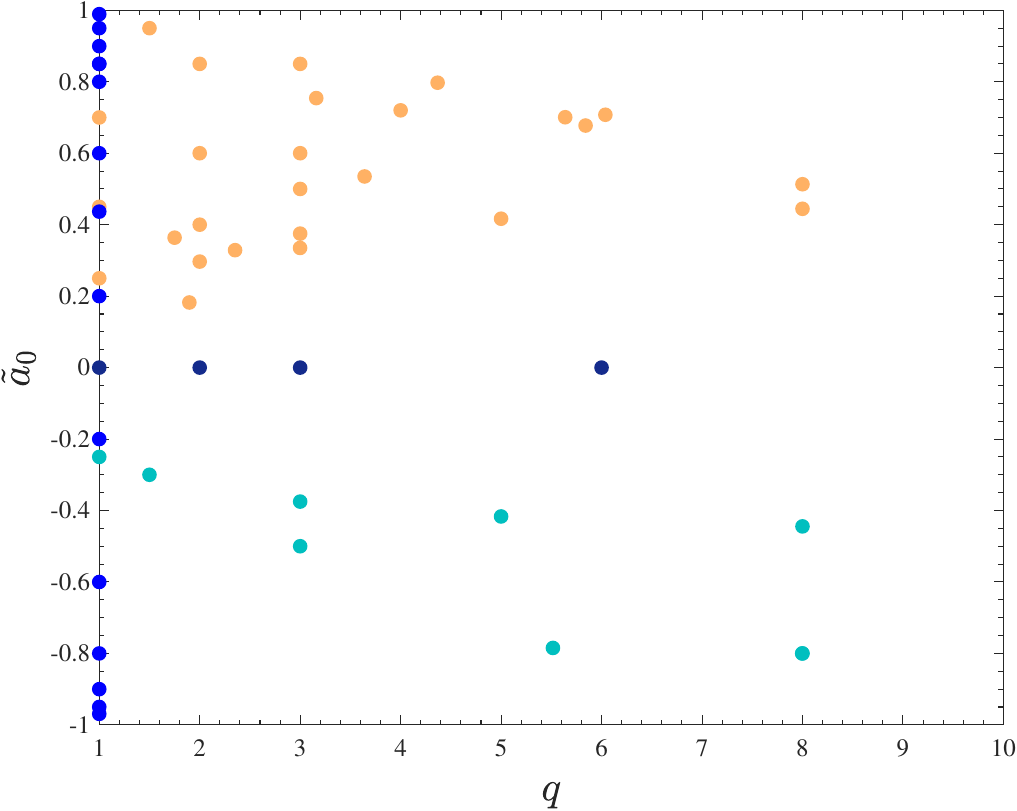}\\
	\vspace{2mm}	
	\includegraphics[width=0.4\textwidth]{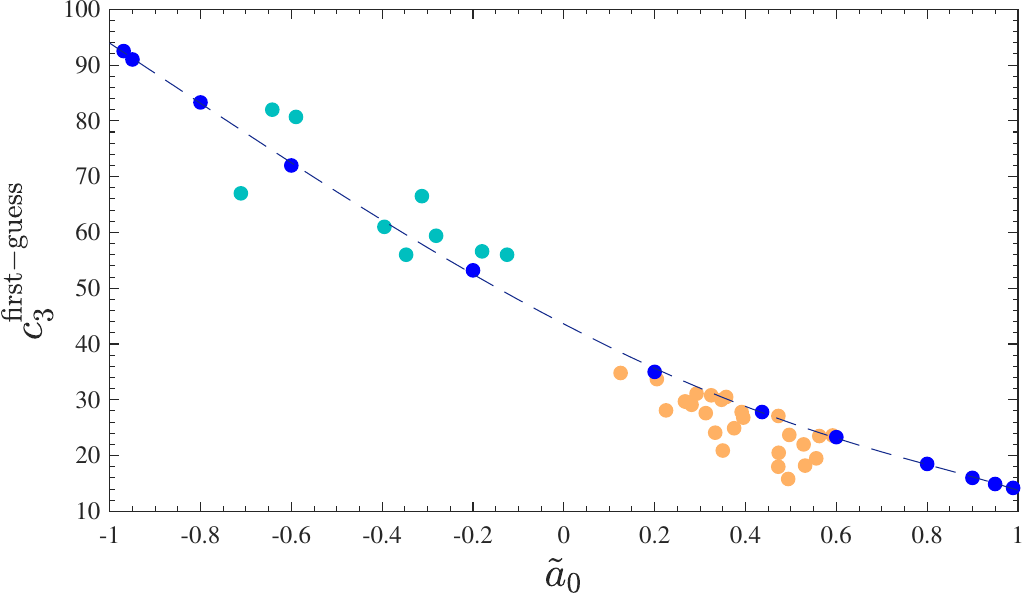}	
	\caption{\label{fig:c3}Top panel: parameters of the NR simulations used to inform 
	$a_6^c$ (dark blue), see Eq.~\eqref{eq:a6c} and $c_3$ (blue, equal-mass and spin; 
	light blue, $\tilde{a}_0 <0$; yellow, $\tilde{a}_0 >0$), see Table~\ref{tab:c3_coeff}. 
	We use a total of only 50 NR datasets, 4 of which are nonspinning. Bottom panel: 
	the selected values of $c_3$, dubbed $c_3^{\rm first-guess}$, are then fitted 
	according to Table~\ref{tab:c3_coeff}. Note that some of the dataset with positive 
	$\tilde{a}_0$ are redundant, as they have similar effective spin and very close values of $c_3^{\rm first-guess}$. 
	The dashed line is the resulting fitting function $c_3^{=}(\tilde{a}_0)$ of Table~\ref{tab:c3_coeff} 
	that fits the values corresponding to the equal-mass,  equal-spin portion of the parameter space.}
\end{figure}

As for the N$^3$LO effective spin-orbit parameter $c_3$~\cite{Damour:2014sva}, we consider the 
same 46 spin-aligned SXS configurations used in Paper~I and in~\cite{Nagar:2023zxh}, 
which we report for completeness in Tables~\ref{tab:c3_eqmass}-\ref{tab:c3_uneqmass} 
together with the first-guess values of $c_3$. This information is also visualized 
in Fig.~\ref{fig:c3}: the top-panel shows the binary configurations used to NR-inform the
model (including the four nonspinning ones mentioned above), the bottom panel the
corresponding first-guess values of $c_3$, before the global fit. The
functional dependence of the equal-mass, equal-spin $c_3^{\rm first-guess}$ values 
(lighter blue in the figure) on $\tilde{a}_0$ is not far from a linear behavior. 
It is also interesting that the $c_3^{\rm first-guess}$ values for the other selected points
are relatively close to the equal-mass, equal-spin ones. The functional form used to fit $c_3$ 
is also the same considered in Ref.~\cite{Nagar:2023zxh}, and explicitly reads:
\begin{align}
  \label{eq:c3fit}
c_3(\nu,\tilde{a}_0,\tilde{a}_{12})= c^{=}_{3}+c_3^{\neq} \ ,
\end{align}
where
\begin{subequations}
	\label{eq:c3fit_pieces}
\begin{align}
c_3^{=}&\equiv p_0\dfrac{1 + n_1\tilde{a}_0 + n_2\tilde{a}_0^2 + n_3\tilde{a}_0^3 + n_4\tilde{a}_0^4}{1 + d_1\tilde{a}_0}\\
c_3^{\neq}&\equiv \left(p_1\tilde{a}_0 + p_2\tilde{a}_0^2  + p_3\tilde{a}_0^3\right)\sqrt{1-4\nu} \nonumber\\
                 &+ p_4\tilde{a}_0\nu \sqrt{1-4\nu} + \left(p_5\tilde{a}_{12}+ p_6\tilde{a}_{12}^2\right)\nu^2. \label{eq:c3fit_pieces_uneq}
\end{align}
\end{subequations}
The coefficients of the fit are listed in Table~\ref{tab:c3_coeff}, while the function $c_3^{=}(\tilde{a}_0)$ 
is also displayed in the bottom panel of Fig.~\ref{fig:c3}. We use a hierarchical fitting procedure:
the $c_3^{=}(\tilde{a}_0)$ function is fitted first using the equal-mass, equal-spin configurations only; 
then, the values of the resulting function are subtracted from the $c_3^{\rm first-guess}$ of 
the unequal-spin and unequal-mass configurations, and this difference is fitted according 
to the model of Eq.~\eqref{eq:c3fit_pieces_uneq}.
% is calculated over the other configurations of
% the parameter space, it is subtracted to the corresponding $c_3^{\rm first-guess}$ values for the 
% unequal-spin and unequal-mass configurations and the difference is eventually fitted.

As an illustrative example of the role of each NR-informed building block in our waveform model, 
Fig.~\ref{fig:a6c_determination_steps} focuses on a $q=1$ nonspinning binary and displays three 
EOB/NR phasing comparisons obtained using different amounts of NR information. 
In the left panel, we see the performance of the purely analytical model, that is, with $a_6^c=0$, 
no NQC correcting factor $\hat{h}_{22}^{\rm NQC}$, and without the attachment of the ringdown portion.
The plot highlights the excellent phasing agreement along the full inspiral obtained even with these omissions, 
with a phase difference displaying small ($\sim 10^{-3}$) oscillations about zero, unresolved on the plot scale.
We can see from the misaligned amplitude peaks that the merger predicted by the EOB model is delayed with respect to the
NR one (indicated by a vertical dashed blue line). Physically, this means that the modeled 
gravitational attraction between the two bodies is too weak.
On the dynamical level, we infer that the attractive character of the EOB potential $A$ should be increased.
This is found to be possibile by progressively reducing the value of $a_6^c$. The phasing plot in the middle 
panel is obtained with $a_6^c=-31.7$. It is remarkable that with this modification the two waveforms are 
visually (almost) indistinguishable up to the merger, with a phase difference of the order of $\sim 0.2$~rad there.
The rightmost panel of Fig.~\ref{fig:a6c_determination_steps} shows the performance of the final waveform,
with the NQC corrections applied and ringdown part attached. The bottom right subpanel highlights the effect of these 
around merger, their inclusion decreasing $\Delta\phi_{22}^{\rm EOBNR} |_{\rm mrg}$ in absolute value from $\sim 0.25$~rad 
(dot-dashed gray line) to $\sim 0.0264$.
To better appreciate the effect of the various NR-informed elements around merger, consider also the time
evolution of the amplitude and frequency shown in Fig.~\ref{fig:a6c_Amp_Freq}.
Notice there the important effect of the NQC corrections, which both lower the height of the amplitude peak 
and increase the frequency to the correct value around the inflection point of $\omega_{22}^{\rm NR}$,
leading to an excellent match between the EOB and NR waveform.

\subsection{Testing: quasi-circular and eccentric binaries}
\label{sec:testing}

%==========================
% Figure unfaithfulness nonspinning
%==========================
\begin{figure}[t]
	\center	
	\includegraphics[width=0.235\textwidth]{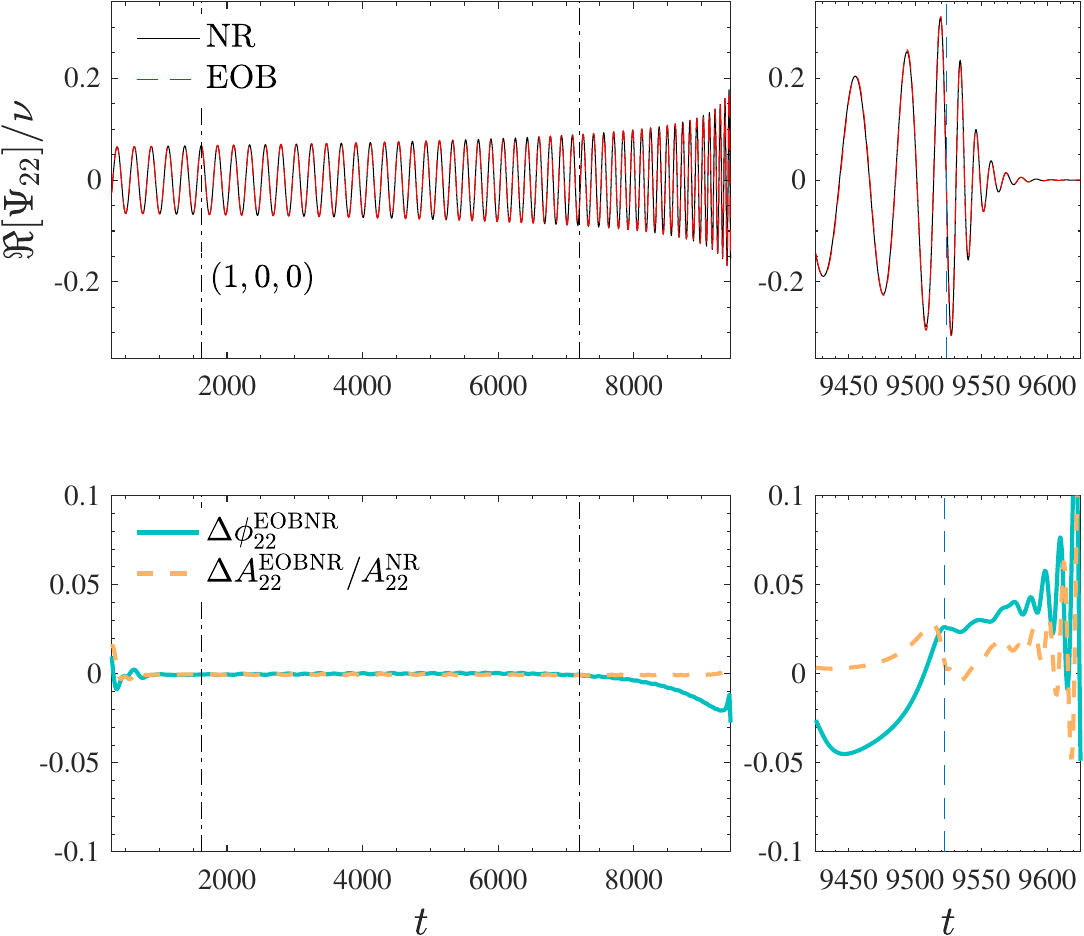}
	\includegraphics[width=0.235\textwidth]{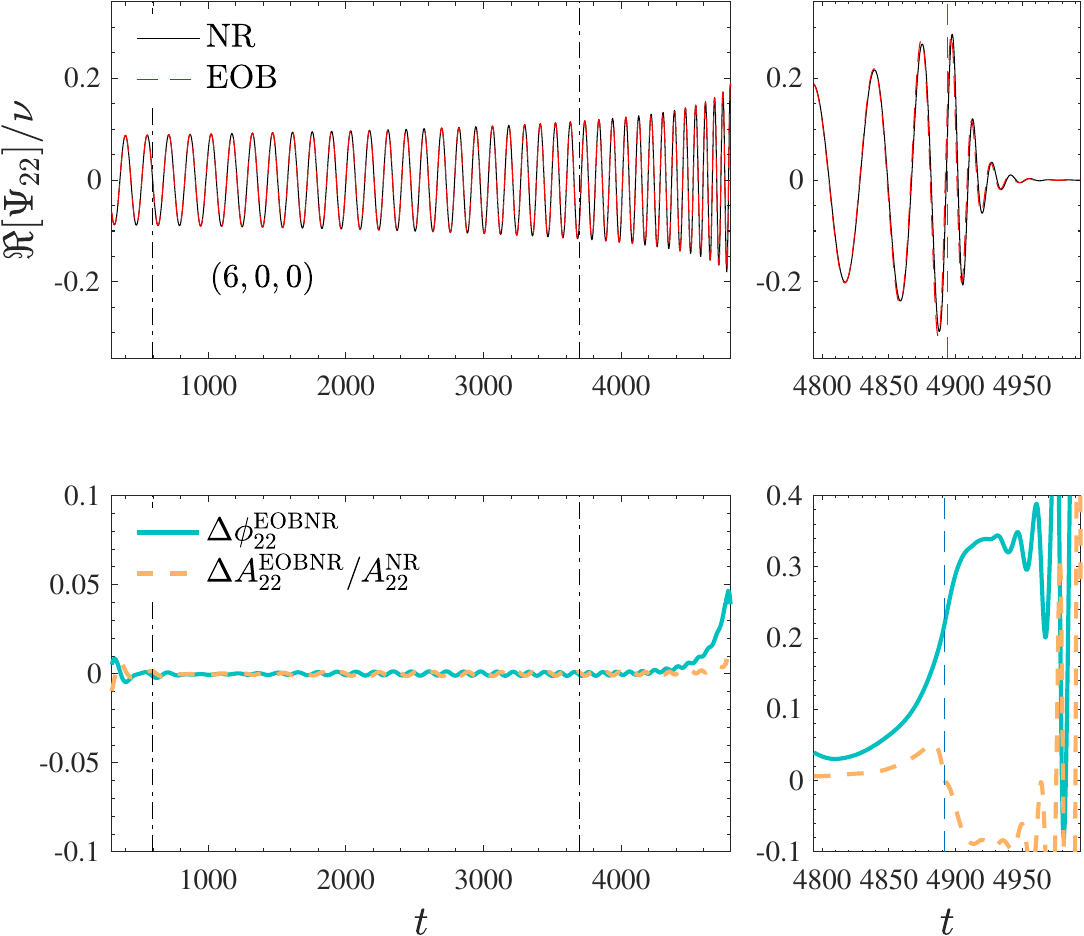}\\
	\vspace{2mm}
	\includegraphics[width=0.235\textwidth]{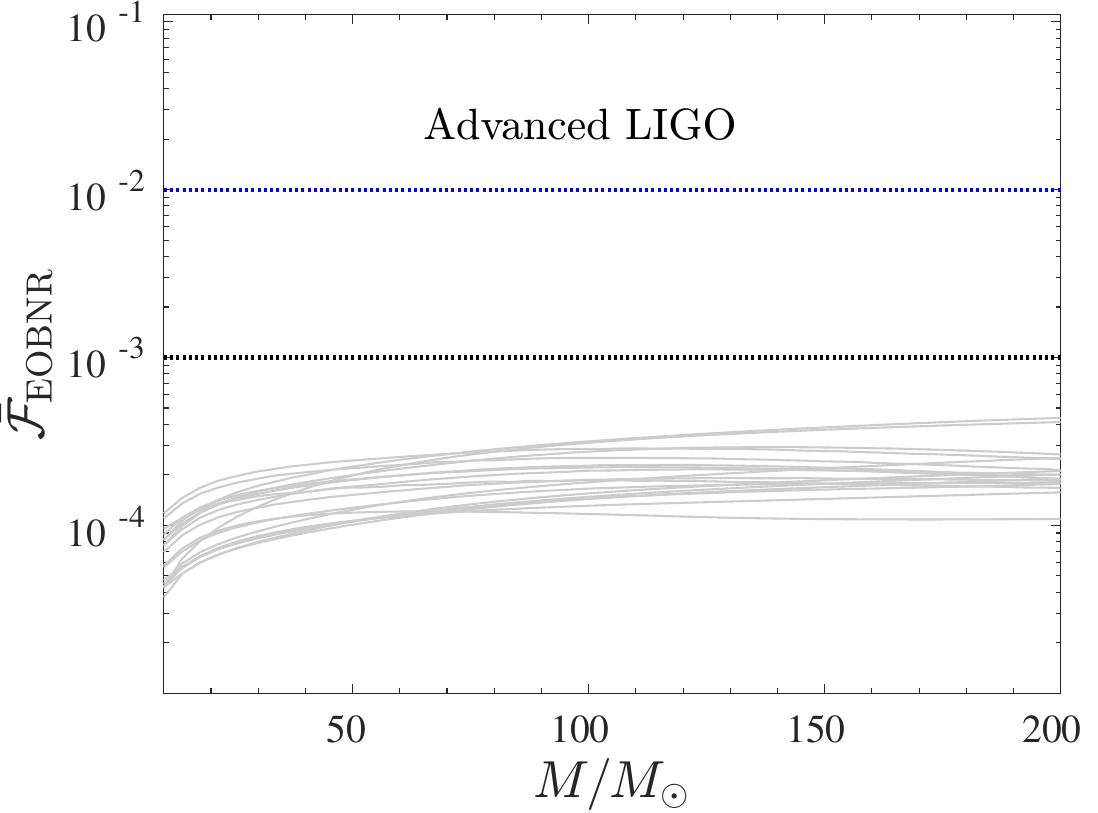}
	\includegraphics[width=0.235\textwidth]{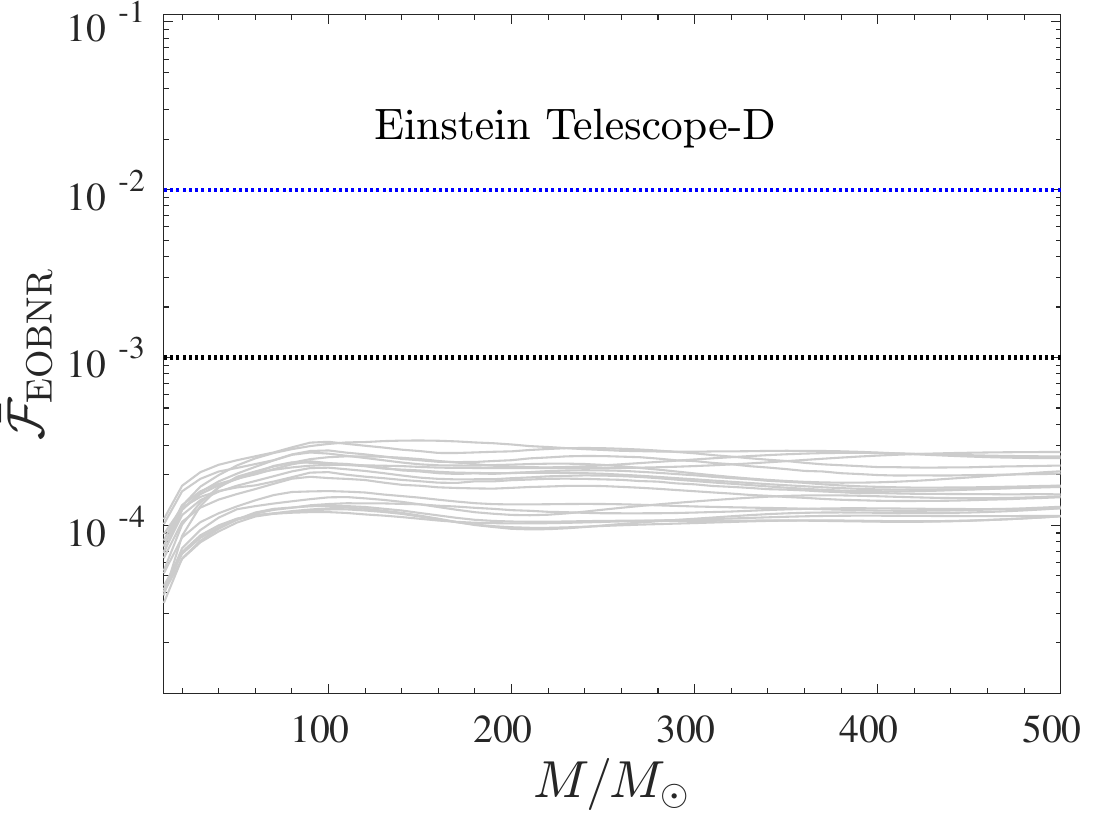}
	\caption{\label{fig:barFs0}
	Nonspinning case. Top panels: two illustrative time-domain EOB/NR phasing comparisons for the $\ell=m=2$ waveform mode.
	The dash dotted vertical lines in the left part of each panel indicate the alignment interval, while the dashed line in the right panel show the merger
	time, i.e. the peak of the NR waveform.
	Bottom panels: EOB/NR unfaithfulness for the sample of SXS nonspinning datasets considered in this paper 
	(including the $q=15$ case, see Fig.~\ref{fig:q15} and Table~\ref{tab:nonspinning_datasets}) with both the Advanced-LIGO and 
	ET-D~\cite{Hild:2008ng,Hild:2009ns, Hild:2010id} sensitivity designs. The small phase differences accumulated up to 
	merger yield $\bar{\cal F}^{\rm max}_{\rm EOBNR}\sim 10^{-4}$. The largest value 
	of $\bar{\cal F}^{\rm max}_{\rm EOBNR}$ for the Advanced-LIGO sensitivity is $5.1\times 10^{-4}$, obtained for $q=1$.}
\end{figure}

\begin{figure}[t]
	\center	
	\includegraphics[width=0.44\textwidth]{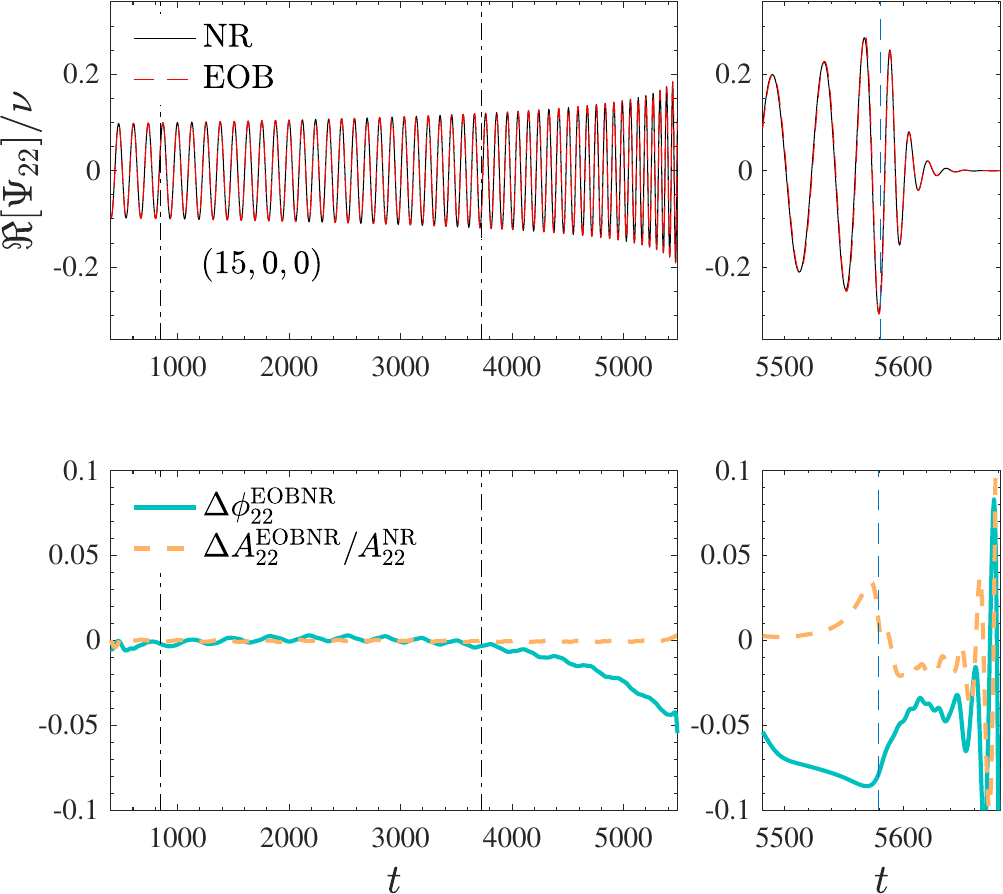}
	\caption{\label{fig:q15}
	EOB/NR time domain phasing for $q=15$, nonspinning. 
	It is remarkable that the model performs so well for such a large mass ratio, even if the 
	function $a_6^c(\nu)$ of Eq.~\eqref{eq:a6c}  was determined using only 
	NR data with mass ratios $q=(1,2,3,6)$. This time-domain phasing corresponds to 
	$\bar{\cal F}^{\rm max}_{\rm EOBNR}=4.13\times 10^{-4}$ with the Advanced-LIGO sensitivity.}
\end{figure}

%====================
% Figure with spin phasings
%====================
\begin{figure*}[t]
	\center	
	\includegraphics[width=0.31\textwidth]{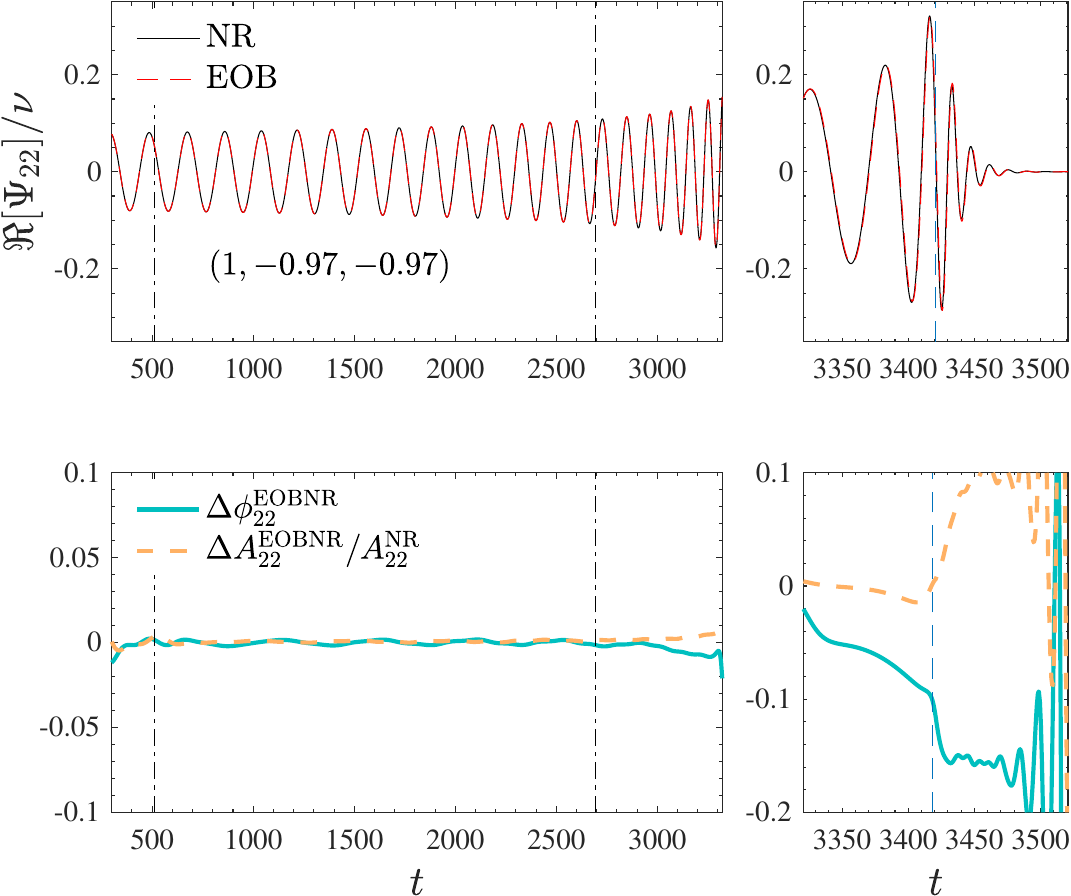}
	\includegraphics[width=0.31\textwidth]{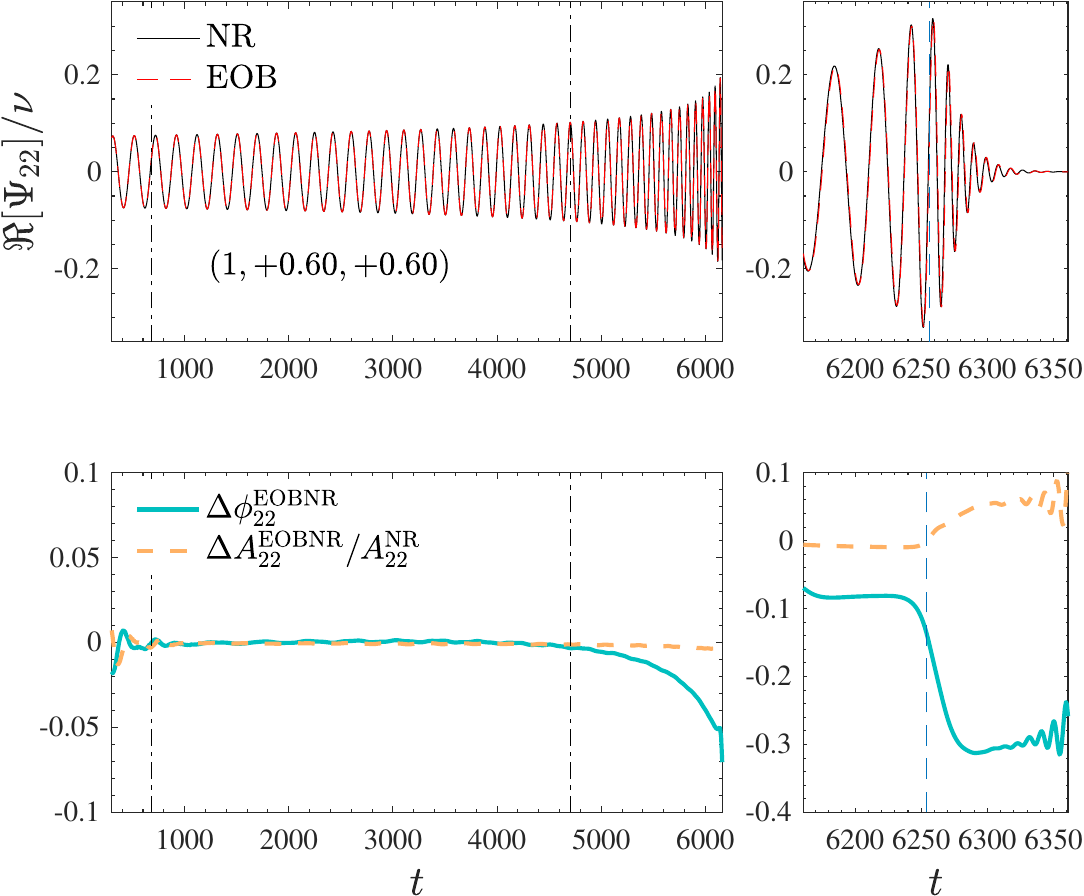}
	\includegraphics[width=0.31\textwidth]{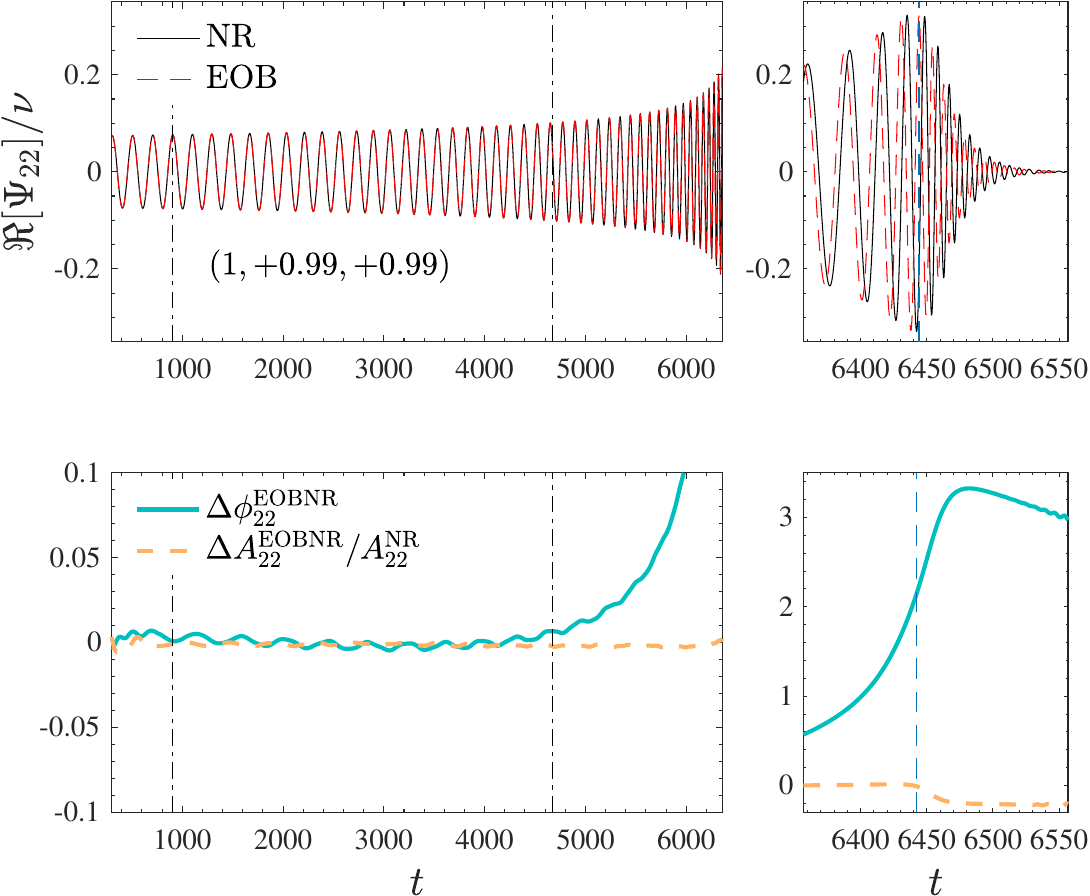}\\
	\vspace{5mm}	
	\includegraphics[width=0.31\textwidth]{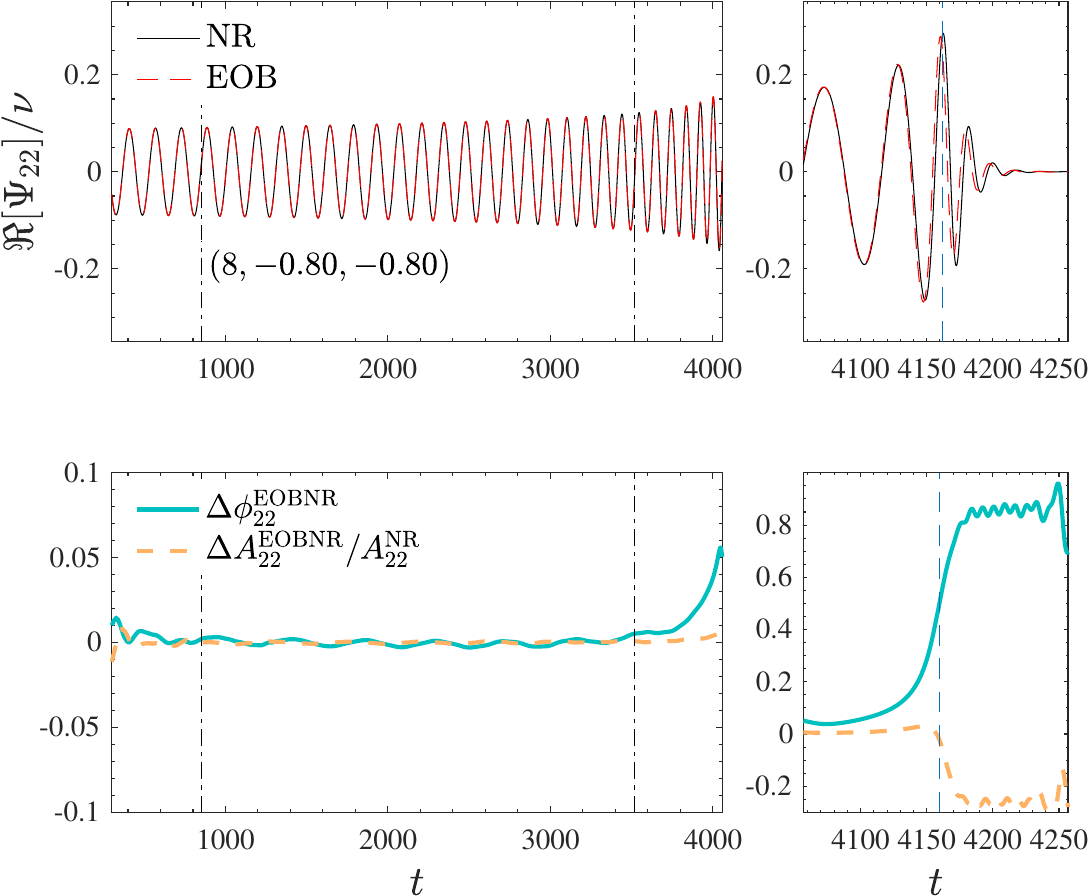}
	\includegraphics[width=0.31\textwidth]{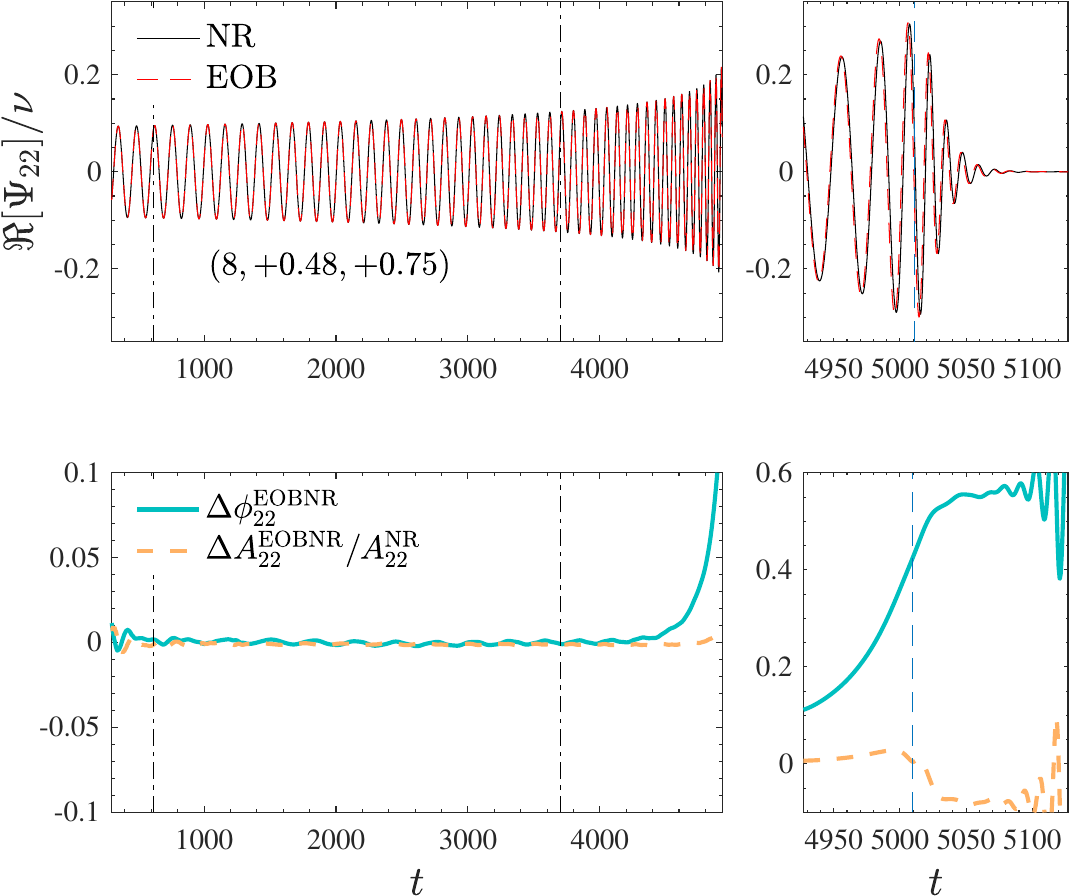}
	\includegraphics[width=0.31\textwidth]{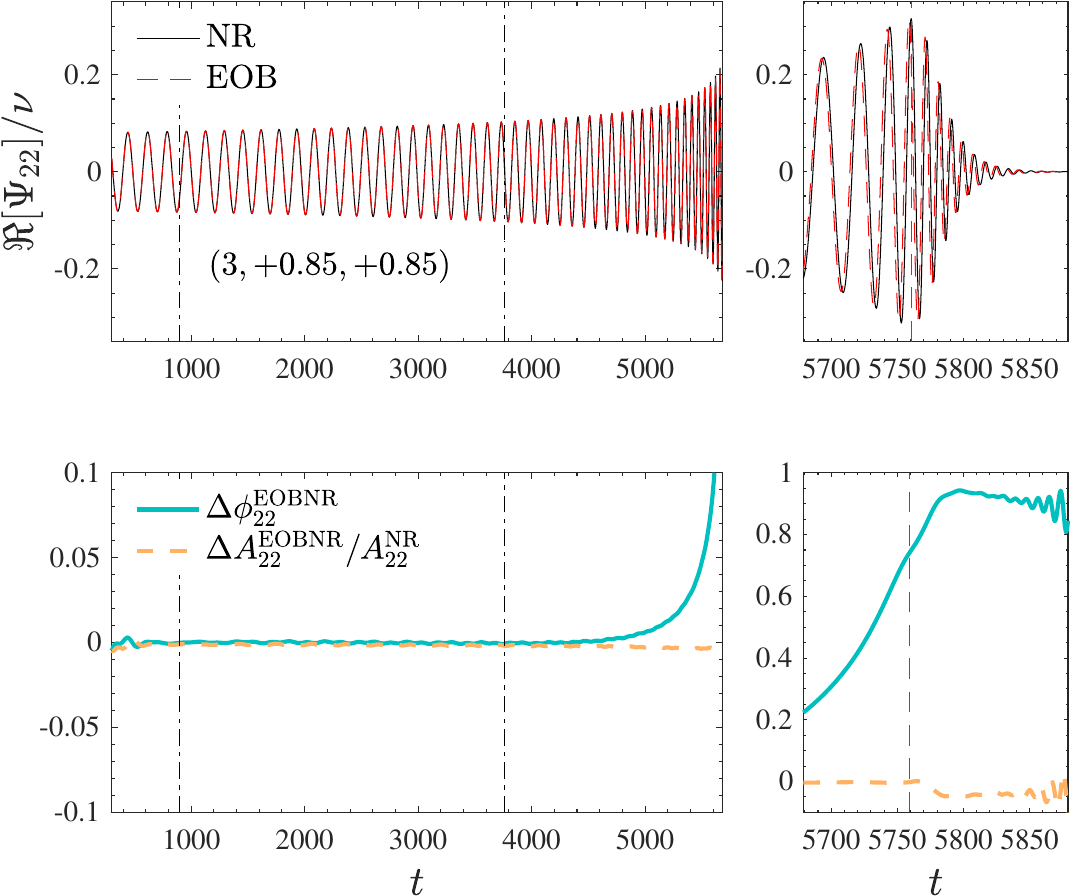}
	\caption{\label{fig:phasings}
	EOB/NR time-domain phasings for a few illustrative spin-dependent
	configurations all over the parameter space. The largest EOB/NR dephasings are found around the
	equal-mass, large positive spin corner. Phase differences at merger of $\sim2$~rad for $(1,+0.99,+0.99)$ yield 
	maximum unfaithfulness values $\sim 10^{-2}$; by contrast, dephasings of $\sim 0.1$~rad 
	yield $\bar{\F}^{\rm max}_{\rm EOBNR}\sim 10^{-4}$, see Fig.~\ref{fig:barF_full}. 
	The vertical dash-dotted lines indicate the temporal bounds coming from the alignment
	frequency interval in the inspiral, similarly to Fig.~\ref{fig:a6c_determination_steps}.}
\end{figure*}

%====================
% Figure with spin phasings
%====================
\begin{figure*}[t]
	\center	
	\includegraphics[width=0.325\textwidth]{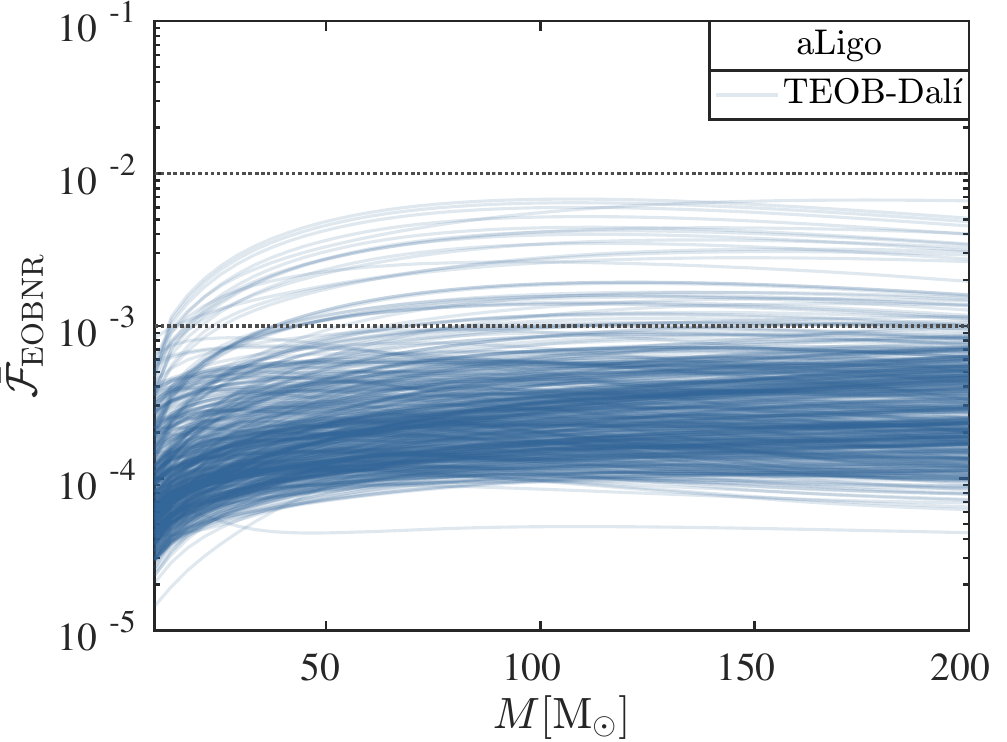}
	\includegraphics[width=0.31\textwidth]{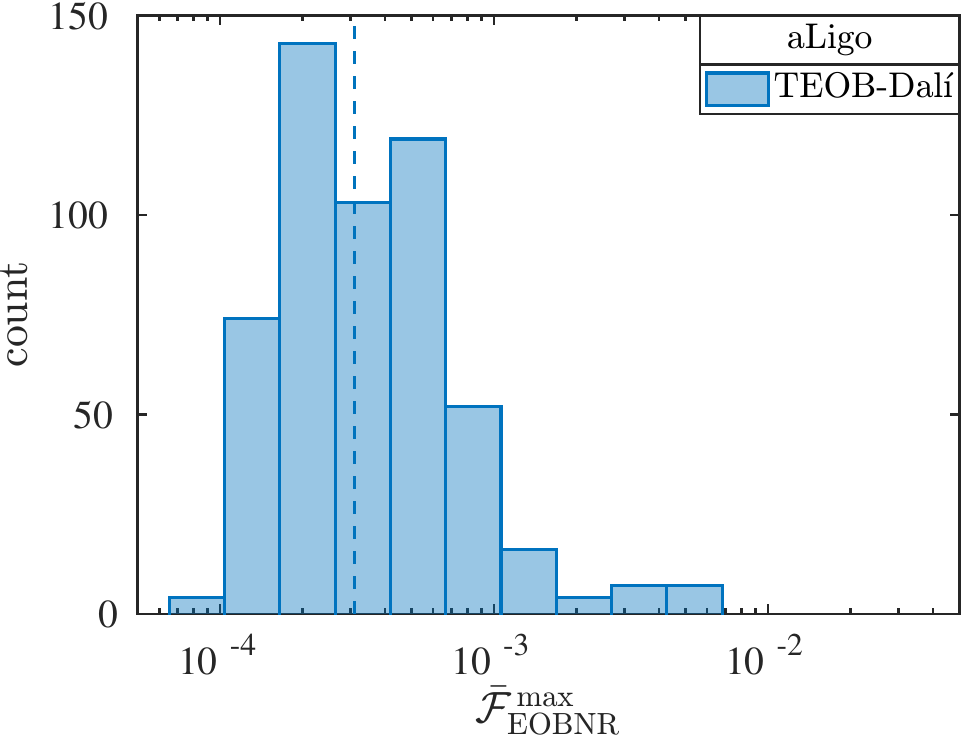}
	\includegraphics[width=0.345\textwidth]{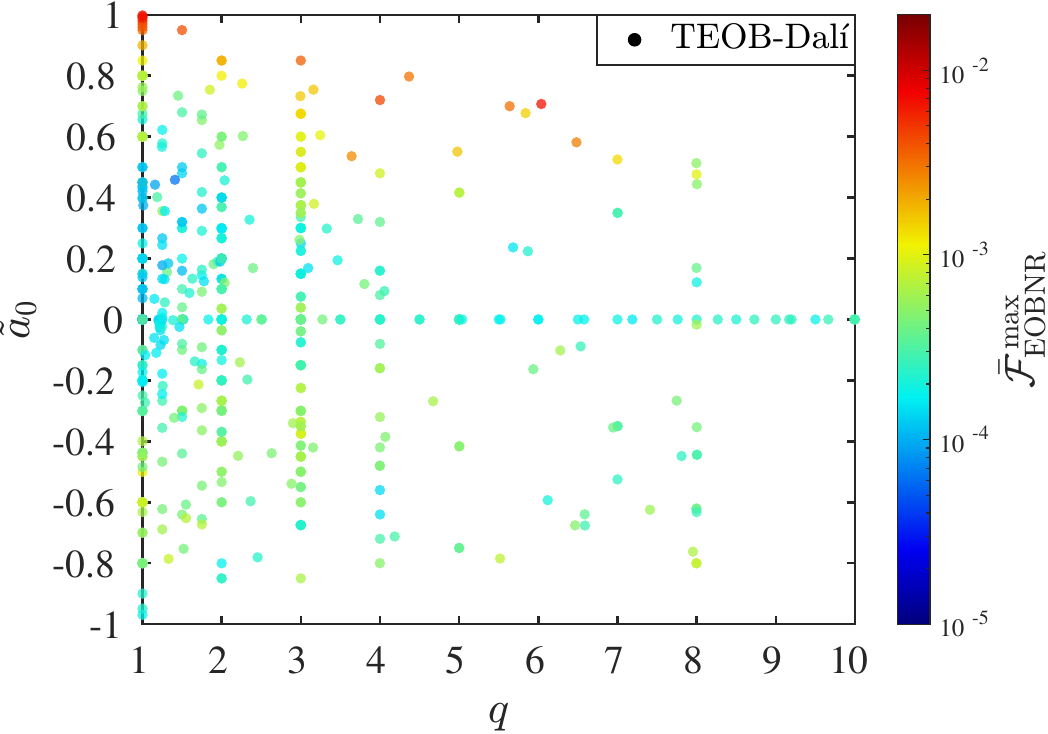}
	\caption{\label{fig:barF_full}
	Left panel: EOB/NR unfaithfulness for the $\ell=m=2$ mode versus the binary 
	total mass $M$ all over the 530 spin-aligned simulations of the SXS catalog. 
	Middle panel: distribution of the maximum unfaithfulness $\bar{\cal F}_{\rm EOBNR}^{\rm max}$,
	whose median  is $3.09\times 10^{-4}$ (marked by the vertical dashed line). 
	Right panel: $\bar{\cal F}_{\rm EOBNR}^{\rm max}$ versus $(\tilde{a}_0,q)$.
	The EOB/NR disagreement is at most $6.8\times 10^{-3}$, for $q=1$ and $\chi_1=\chi_2=0.998$.
	Only 5 configurations have $\bar{\cal F}_{\rm EOBNR}^{\rm max}>5\times 10^{-3}$. Four are equal 
	mass and equal spin ones with $\chi_1=\chi_2\gtrsim 0.98$, while the other has $q=6.038$, 
	$\chi_1=0.8$ and $\chi_2=0.14752$, SXS:BBH:1437. Note the improvement with respect to 
	Fig.~11 of Ref.~\cite{Nagar:2024dzj} despite the absence of NR-tuning of radiation reaction.}
\end{figure*}

%----------------------------------------------------------------
%Problematic waveforms are:
%i=58 SXS_BBH_0157 (1.000000,+0.949586,+0.949586) maxFbar=0.620049[%]
%i=59 SXS_BBH_0158 (1.000000,+0.969504,+0.969504) maxFbar=0.531487[%]
%i=66 SXS_BBH_0172 (1.000000,+0.979413,+0.979413) maxFbar=0.629762[%]
%i=70 SXS_BBH_0177 (1.000000,+0.989253,+0.989253) maxFbar=0.731766[%]
%i=71 SXS_BBH_0178 (1.000000,+0.994246,+0.994247) maxFbar=1.082829[%]
%i=299 SXS_BBH_1124 (1.000001,+0.998001,+0.998001) maxFbar=0.836142[%]
%----------------------------------------------------------------

We begin the comprehensive testing of the model with the quasi-circular configurations.
As usual, EOB and NR waveforms are compared both
computing the EOB/NR phase difference in the time domain and the EOB/NR unfaithfulness for
a given detector noise. This quantity, addressed as $\bar{\F}$, is defined as follows. 
Given two waveforms $(h_1,h_2)$,  $\bar{\F}$ is a function of the total 
mass $M$ of the binary:
\be
\label{eq:barF}
\bar{\cal F}(M) \equiv 1-{\cal F}=1 -\max_{t_0,\phi_0}\dfrac{\langle h_1,h_2\rangle}{||h_1||||h_2||},
\ee
where $(t_0,\phi_0)$ are the initial time and phase. We used $||h||\equiv \sqrt{\langle h,h\rangle}$,
and the inner product between two waveforms is defined as 
$\langle h_1,h_2\rangle\equiv 4\Re \int_{f_{\rm min}^{\rm NR}(M)}^\infty \tilde{h}_1(f)\tilde{h}_2^*(f)/S_n(f)\, df$,
where $\tilde{h}(f)$ denotes the Fourier transform of $h(t)$, $S_n(f)$ is the detector power spectral density (PSD),
and $f_{\rm min}^{\rm NR}(M)=M\hat{f}^{\rm NR}_{\rm min}$ is the initial frequency of the
NR waveform at highest resolution. In practice, to avoid contamination from the NR junk radiation,
$f_{\rm min}^{\rm NR}$ is taken to be 1.35 times larger than the actual initial frequency 
of $\tilde{h}_{\rm NR}(f)$ (i.e., 1.35 times the frequency where $|\tilde{h}_{\rm NR}|$ peaks; 
this is the same choice adopted in Ref.~\cite{Pompili:2023tna}). 
Note that, before taking the Fourier transforms, the waveforms are also tapered in the time-domain 
to reduce high-frequency oscillations. For $S_n$, in our comparisons we use either the 
zero-detuned, high-power noise spectral density of Advanced LIGO~\cite{aLIGODesign_PSD} 
or the predicted sensitivity of Einstein Telescope~\cite{Hild:2009ns, Hild:2010id}.
The unfaithfulness computation is carried out using the \texttt{pycbc} package~\cite{Biwer:2018osg}.

Following the same logic of Paper~I, it is instructive to focus first on nonspinning binaries.
The top panel of Fig.~\ref{fig:barFs0} reports time-domain phasing comparisons in terms of the
waveform modes $\Psi_{\ell m} = h_{\ell m}/\sqrt{(\ell+2)(\ell+1)\ell(\ell-1)}$
(normalized following Ref.~\cite{Nagar:2005ea}) for two selected datasets 
with $q=1$ and $q=6$. We focus only on the $\ell=m=2$ mode, that is decomposed 
in phase and amplitude as $\Psi_{22}=A_{22} \e^{-\i\phi_{22}}$.
Considering first the equal-mass case, we highlight that the phase difference accumulated
up to merger is just $\Delta\phi^{\rm EOBNR}_{22}\equiv \phi_{22}^{\rm EOB}-\phi_{22}^{\rm NR}
\sim -0.05$~rad, yielding $\bar{\cal F}_{\rm EOBNR}\sim 10^{-4}$,
with $\bar{\F}_{\rm max}^{\rm EOBNR} \sim 1.21 \times 10^{-4}$. For $q=6$, the cumulative dephasing 
is $\sim 0.2$~rad, yielding $\bar{\F}^{\rm max}_{\rm EOBNR}=1.79\times 10^{-4}$.
The time evolution of $\Delta\phi^{\rm EOBNR}_{22}$ is qualitatively comparable with, 
though quantitatively better than, the corresponding one shown in Fig.~10 of Paper~I. In that case
we were however relying on an ad-hoc NR-calibration of the radiation reaction through the 
$\rho_{22}$ function. This calibration is not needed now, and the improved performance is fully 
due to the new treatment of the logarithmic terms. Consistently, in the bottom panel of the 
figure we report the computation of $\bar{\cal F}_{\rm EOBNR}$, as a function of the total mass $M$, 
with either the Advanced LIGO noise (left panel) or the ET predicted sensitivity curve (right panel). 
The figure is similarly consistent with the bottom panels of Fig.~10 of Paper~I. 
Given how small the accumulated phase difference is along the full evolution, it seems 
hard to further improve the model in this regime without a more stringent estimate of the NR error, 
which the model might be grazing now. For $q=6$ we accepted a phase difference larger 
than the equal-mass case, though further tuning of $a_6^c$ could have decreased it further, 
with the understanding that for such large mass ratio 
the NR uncertainty might be slightly larger than for the $q=1$ case.
For SXS datasets the NR uncertainty is usually estimated comparing two different resolutions 
as given in the public catalog, and this typically yields values of NR-NR unfaithfulness of the 
order of $10^{-4}-10^{-5}$ (see for example Sec.~4 in Ref.~\cite{Boyle:2019kee}, or the left panel 
of Fig.~2 in Ref.~\cite{Nagar:2020pcj}), i.e., still mostly below the current EOB/NR unfaithfulness 
level, but comparable for some configurations.
Nonetheless, it is worth stressing that $\Delta\phi^{\rm EOBNR}_{22}$ remains of the same order
{\it also} for larger mass ratios not considered in the determination of $a_6^c$, and most notably it is even smaller
for $q=15$, as illustrated in Fig.~\ref{fig:q15}. This feature illustrates that the
model presented here is accurate also {\it outside} the region of the parameter space used 
to determine $a_6^c$, and thus its predictive power (and accuracy) generally looks stronger 
than its previous avatars. 
The waveform was tested also against NR simulations with very large mass ratios,
up to $q=128$~\cite{Lousto:2020tnb}. The consistency through merger and 
ringdown is analogous to the one obtained in Ref.~\cite{Nagar:2022icd} for the 
quasi-circular {\tt TEOBResumS-GIOTTO} model.

Moving on now to the more general case of spinning BHs, a few illustrative phasing comparisons 
are shown in Fig.~\ref{fig:phasings}. Fig.~\ref{fig:barF_full} gives instead a global overview of the 
performance of the model, tested over 530 spin-aligned NR simulations of the SXS catalog~\cite{Boyle:2019kee},
displaying $\bar{\F}_{\rm EOBNR}$ as a function of the total mass $M$ (right), the distribution of the maximum
unfaithfulness values (middle), and $\bar{\F}_{\rm EOBNR}^{\rm max}$ across the $(q, \tilde{a}_0)$ parameter 
space (left)\footnote{From these plots we excluded the datasets SXS:BBH:1414, SXS:BBH:1416 and SXS:BBH:1417,
characterized by very long inspirals (over 100 orbits for the former two).  Their discussion is postponed ot Sec.~\ref{eob:comparisons} below.}.
We find that $\bar{\F}_{\rm EOBNR}^{\rm max} \lesssim 10^{-3}$ for most of the dataset, 
with the median of the mismatch distribution being ${\rm Me}[\bar{\cal F}_{\rm EOBNR}^{\rm max}]=3.09\times 10^{-4}$.
This is  smaller than the value ${\rm Me}[\bar{\cal F}_{\rm EOBNR}^{\rm max}]=1.06\times 10^{-3}$ obtained in Paper~I 
while NR-tuning {\it also} the radiation reaction, and comparable to the values obtained with the 
standard quasi-circular EOB models  {\tt TEOBResumS-GIOTTO}~\cite{Nagar:2023zxh} 
or {\tt SEOBNRv5HM}~\cite{Pompili:2023tna}. These two do remain slightly better, especially for high spins, 
since for {\tt TEOBResumS-GIOTTO}~\cite{Nagar:2023zxh} we have $\bar{\F}_{\rm EOBNR}^{\rm max}\lesssim 10^{-3}$,
and  for {\tt SEOBNRv5HM}~\cite{Pompili:2023tna} $\bar{\F}_{\rm EOBNR}^{\rm max}\lesssim 3\times 10^{-3}$ (see Sec.~\ref{eob:comparisons} below).
The model always yields $\bar{\F}_{\rm EOBNR}^{\rm max}< 5\times 10^{-3}$, except for
four equal-mass datasets, SXS:BBH:0172, SXS:BBH:0177, SXS:BBH:0178 and SXS:BBH:1124,
all with with very high spins, $\chi_1 = \chi_2 =(0.979413,\  0.989253, \ 0.994246, \ 0.998001)$
respectively, and the dataset SXS:BBH:1437, that has parameters $(6.038,+0.80,+0.1475)$.

In general, for large, positive spins the tuning of $c_3$ is ineffective in further reducing the phase
difference during the late inspiral, that for $(1,+0.99,+0.99)$ is $\sim 2$~rad. As shown above (Fig.~\ref{fig:rho22_resum}), 
the use of the NNLO spin-orbit correction in $\rho_{22}^S$, even in resummed form, increases the 
waveform amplitude (see Fig.~\ref{fig:rho22_resum}). This, in turn, would further accelerate the inspiral, 
yielding an even larger phase difference at merger for large, positive, spins. This is the reason why
the NNLO term in $\rho_{22}^{\rm S}$ is not included in our model at this time.
This lack of flexibility in the waveform and radiation reaction model in dealing with the issues encountered
for high spins suggests that solutions should be sought elsewhere; for instance, the spin-orbit sector of
the conservative dynamics should perhaps be reconsidered. 
Focusing on the rightmost panel of Fig.~\ref{fig:barF_full}, it is also worth noting that
most of the smallest values of $\bar{\cal F}_{\rm EOBNR}^{\rm max}$ are obtained for 
moderately high, positive spins and mass ratios in the range $1\leq q \leq 4$. This is the consequence
of the fact that most of the datasets used to NR-inform $c_3$ have $q\lesssim 4$ and $\tilde{a}_0>0$ 
(see Table~\ref{tab:c3_uneqmass}), while just 9 have $\tilde{a}_0<0$. 
The solid performance of the model even in the latter case points to the robustness and predictive power
of the analytical spin sector of \TEOBd{}. Still, this fact suggests that improvement in the
negative-spin region of the parameter space might be achieved by just anchoring the determination of 
$c_3$ to a few more datasets with $1 \leq q \leq 4$ and $\tilde{a}_0<0$. Similar conclusions are applicable
also to the region of positive spins with $q\gtrsim 5$.
Indeed, the rather large value of $\bar{\cal F}_{\rm EOBNR}^{\rm max}$ for  $(6.038,+0.80,+0.1475)$ comes from
the fact that the first guess value of $c_3$ chosen in this case, $c_3^{\rm newlogs}=23.6$ in Table~\ref{tab:c3_uneqmass},
is returned as $c_3=23.06$ by the fit: this difference is by itself sufficient to increase the
EOB/NR phasing disagreement around merger from $\bar{\cal F}_{\rm EOBNR}^{\rm max} \sim 5 \times 10^{-3}$
to $\sim 6.7 \times 10^{-3}$.
Since our main purpose here is to explore the effect of the new analytic treatment of the 
logarithmic terms throughout the model without changing the NR information, we postpone a more
in depth study of the optimal calibration dataset to future work.

%i=66 SXS_BBH_0172 (1.000000,+0.979413,+0.979413) maxFbar=0.521804[%]
%i=70 SXS_BBH_0177 (1.000000,+0.989253,+0.989253) maxFbar=0.600282[%]
%i=71 SXS_BBH_0178 (1.000000,+0.994246,+0.994247) maxFbar=0.644006[%]
%i=299 SXS_BBH_1124 (1.000001,+0.998001,+0.998001) maxFbar=0.960743[%]

%===========
% Eccentric data
%===========
\begin{figure}[t]
	\center	
	\includegraphics[width=0.45\textwidth]{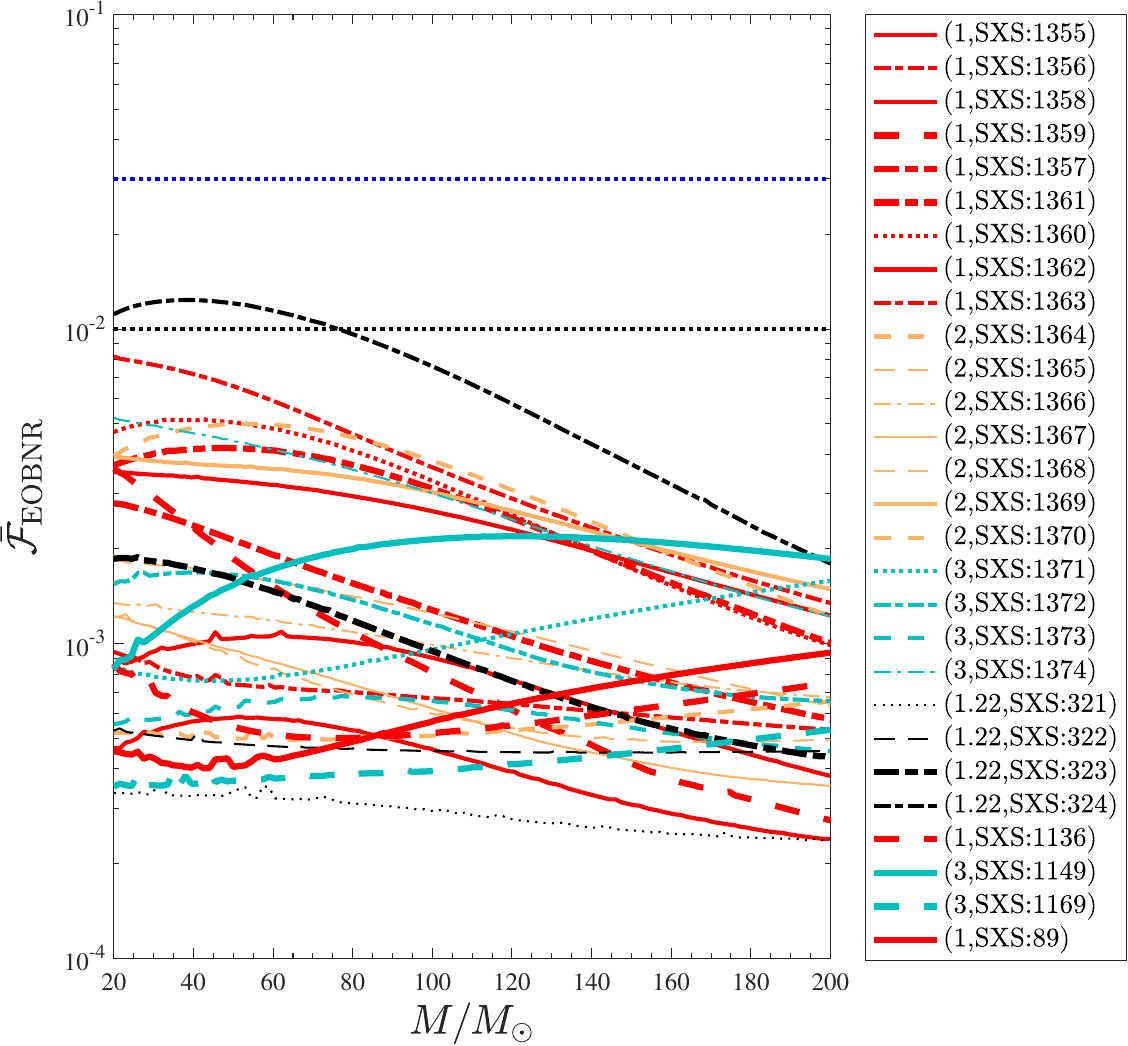 }
        \caption{\label{fig:barF_ecc} EOB/NR unfaithfulness for the $\ell=m=2$ mode 
        calculated over the 28 publicly available SXS eccentric simulations. }
\end{figure}

Finally, the EOB/NR unfaithfulness comparisons for eccentric binaries are
shown in Fig.~\ref{fig:barF_ecc}. As usual, the model is compared with the 28 publicly available SXS
datasets~\cite{Hinder:2017sxy}, whose binary properties (mass ratio and spins) can be found in
Table~IV of Paper~I. The performance is much improved with respect to 
Paper~I (see Fig.~8 or Fig.~12 there), except for the case SXS:BBH:0324, 
corresponding to $(q, \chi_1, \chi_2) = (1.22, +0.33, -0.44)$, that is characterized by
the largest initial NR eccentricity at apastron ($e_{\omega_a}\sim 0.20)$ 
and a rather noisy (and possibily slightly inaccurate) inspiral. 
We do note that the initial data (eccentricity and apastron frequency) were slightly retuned with respect to Paper~I
to compensate for the various analytical modifications that were implemented in the model concerning the inspiral.
Although the initial data choice could be probably refined further 
(see e.g. Ref.~\cite{Bonino:2024xrv} and Ref.~\cite{Gamboa:2024hli}),
it is worth to remark that for SXS:BBH:1149, corresponding to $(q, \chi_1, \chi_2) = (3, +0.70, +0.60)$, we find
$\bar{\F}^{\rm max}_{\rm EOBNR} \sim 2 \times 10^{-3}$, which improves on any version of the model considered
in Paper~I. In this respect, let us also recall that the unfaithfulness computation is performed using for initial
frequency values $Mf_{\rm min}$ listed in Table~IV of Paper~I. These value were chosen sufficiently low to
include the full eccentric inspiral of each NR dataset. However, because of this, the small-mass part of the 
curves in Fig.~\ref{fig:barF_ecc} is probably partly contaminated by spurious effects coming from junk
radiation, numerical noise and boundary effects.

\subsection{Comparison with other EOB models}
\label{eob:comparisons}

%===============
% Comparing models
%===============
\begin{table}[t]
	\caption{\label{tab:model_performance}
	Median (Me) and maximum (Max) values of $\bar{\cal F}^{\rm max}_{\rm EOBNR}$,
	the unfaithfulness between EOB and SXS $\ell=m=2$ waveform data for four state-of-the-art EOB models in the spin-aligned, 
	quasi-circular case. Note than only \TEOBd{}$_{\tt newlogs}$ can also deal with eccentric binaries and scattering configurations. 
	Note also that the values for the {\tt SEOB} family that are extracted from the literature are computed on 441 SXS NR datasets 
	(partly private) that do not fully overlap with the 530 public ones used for the {\tt TEOBResumS} family. In this respect, the last
	row of the table reports the {\tt SEOBNRv5HM} median on the public simulations.}
	  \begin{center}
		\begin{ruledtabular}
   \begin{tabular}{l l l c} 
   Model  & ${\rm Me}[\bar{\cal F}^{\rm max}_{\rm EOBNR}]$ & ${\rm Max}[\bar{\cal F}^{\rm max}_{\rm EOBNR}]$ & Reference   \\
   \hline
   {\tt SEOBNR-PM}            & $6.1\times 10^{-4}$   & $\sim 2\times 10^{-1}$ & \cite{Buonanno:2024byg} \\
   \TEOBd{}$_{\tt newlogs}$   & $3.09\times 10^{-4}$  & $6.80\times 10^{-3}$   & this paper              \\
   {\tt TEOBResumS-GIOTTO}    & $4.0\times 10^{-4}$   & $\sim 10^{-3}$         & \cite{Nagar:2023zxh}    \\
   {\tt SEOBNRv5HM}           & $1.99\times 10^{-4}$  & $\sim 2\times10^{-3}$  & \cite{Pompili:2023tna}  \\
   {\tt SEOBNRv5HM}           & $1.47\times 10^{-4}$  & $2.98 \times10^{-3}$   & this paper
   \end{tabular}
\end{ruledtabular}
\end{center}
\end{table}

%===============
% Plot barF for outlier
%===============
\begin{figure}[t]
	\centering
	\includegraphics[width=0.43\textwidth]{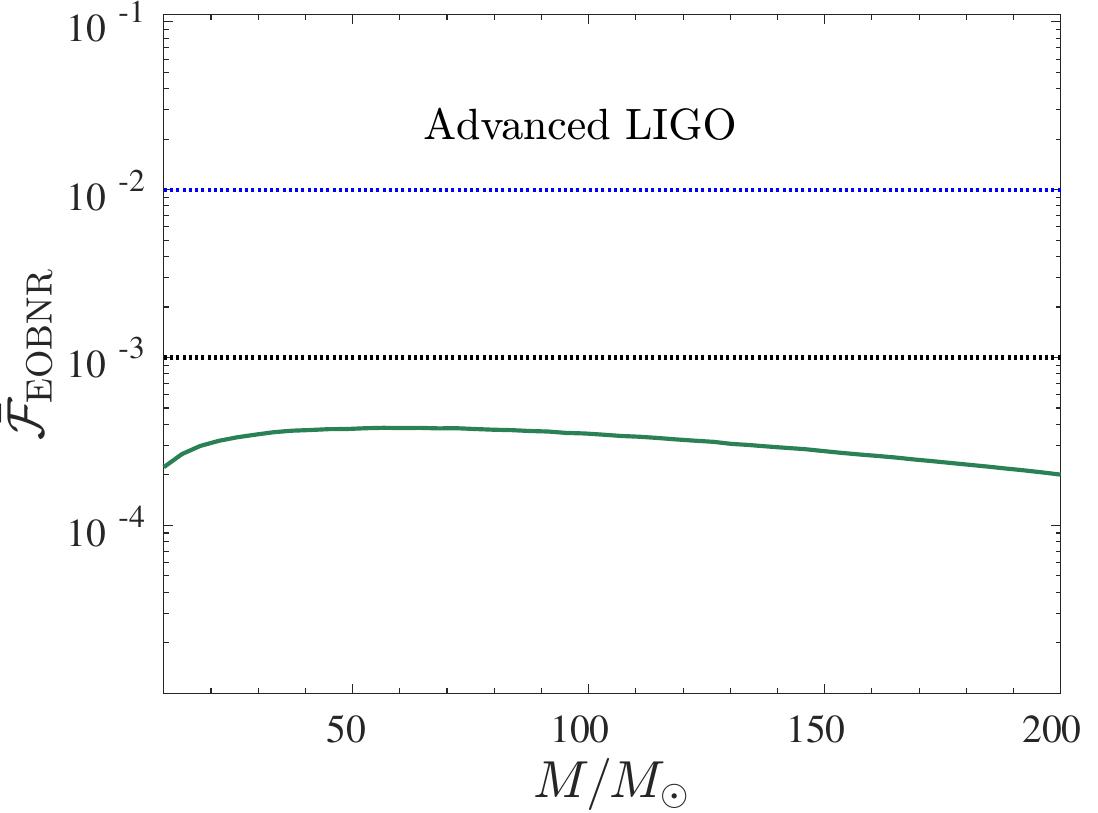} \\
	\vspace{3mm}
        \includegraphics[width=0.43\textwidth]{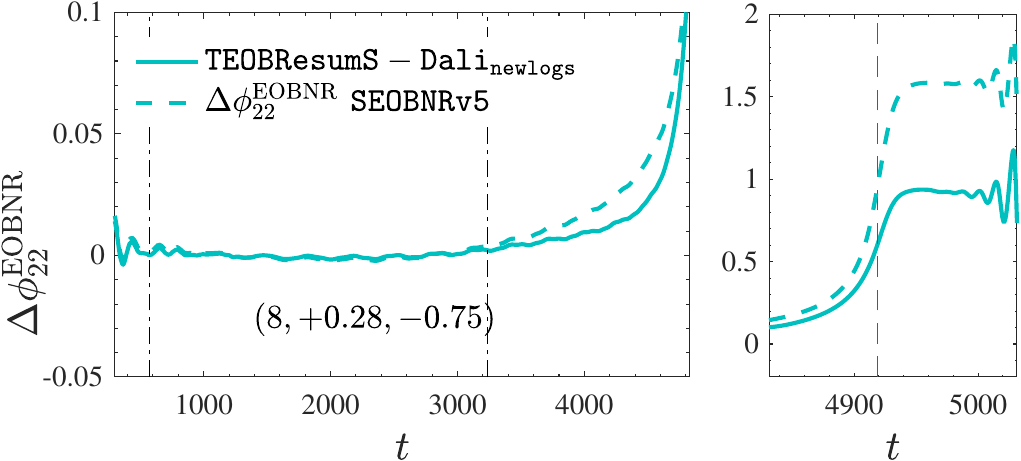} 
	\caption{\label{fig:outlier}
	Performance  of our new version of \TEOBd{} for SXS:BBH:1430, with $(q, \chi_1, \chi_2) = (8,+0.284,-0.751)$,
	corresponding to an effective spin $\tilde{a}_0=0.165$ and $\bar{\cal F}^{\rm max}_{\rm EOBNR}=3.81\times 10^{-4}$. 
	For this configuration, the {\tt SEOBNRv5HM} model delivers its worst EOB/NR agreement, with 
	$\bar{\cal F}^{\rm max}_{\rm EOBNR}\sim 2\times 10^{-3}$ (see Fig.~5 of Ref.~\cite{Pompili:2023tna}).
	This is understood by comparing the two EOB/NR time-domain phasings for the two models (bottom panel). 
	The alignment window is the same for both waveforms (vertical dash-dotted lines).}
\end{figure}

%================
% Plot barF for seobv5
%================
\begin{figure}[t]
	\centering
	\includegraphics[width=0.5\textwidth]{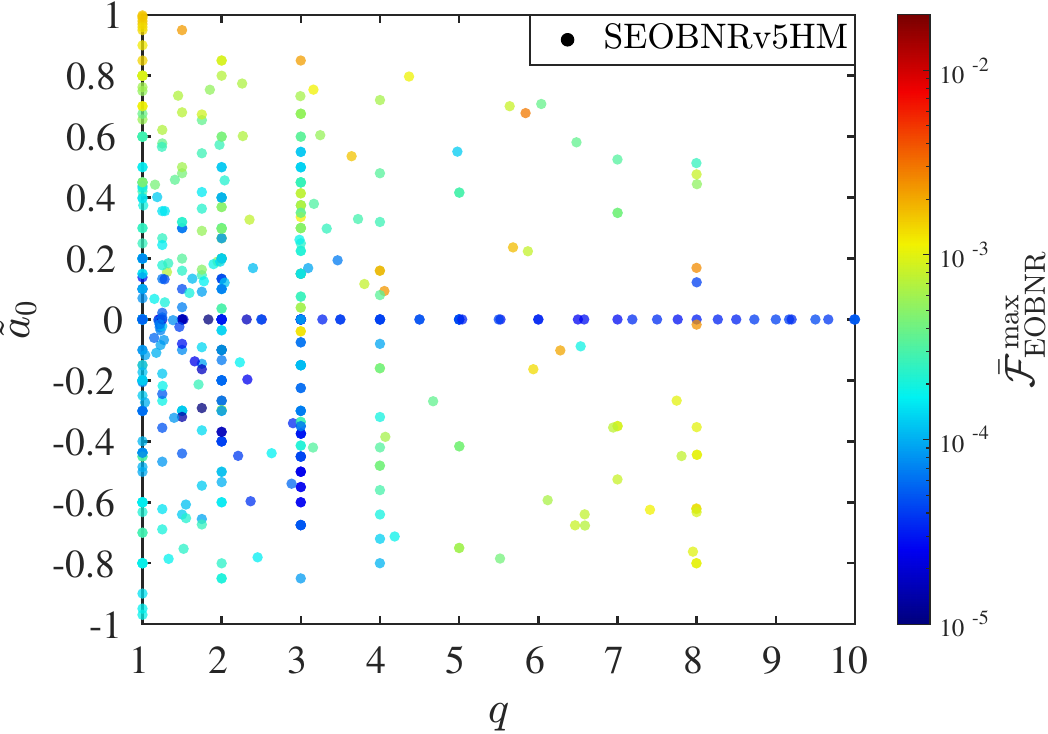}		
	\caption{\label{fig:barF_seobv5}Performance of ${\tt SEOBNRv5HM}$ in terms of $\bar{\cal F}_{\rm EOBNR}^{\rm max}$
	over the 530 public SXS datasets used to validate $\TEOBd{}$. The color scale is the same as the one used in the rightmost 
	panel of Fig.~\ref{fig:barF_full} for \TEOBd{}. For nonspinning binaries, the performance of ${\tt SEOBNRv5HM}$ 
	is better than that of \TEOBd{} by almost an order of magnitude. However, it is interesting
	that for ${\tt SEOBNRv5HM}$ small changes in $\tilde{a}_0$ may yield large jumps in $\bar{\cal F}_{\rm EOBNR}^{\rm max}$,
	in some cases as large as two orders of magnitude. This is in contrast with \TEOBd{}, which displays a more homogeneous
	behavior. Also, the performance of \TEOBd{} is generally superior for large mass ratio with spins, despite the
	smaller number of datasets used to inform the model.}
\end{figure}

%================
% Plot barF for seobv5
%================
\begin{figure}[t]
	\centering
	\includegraphics[width=0.45\textwidth]{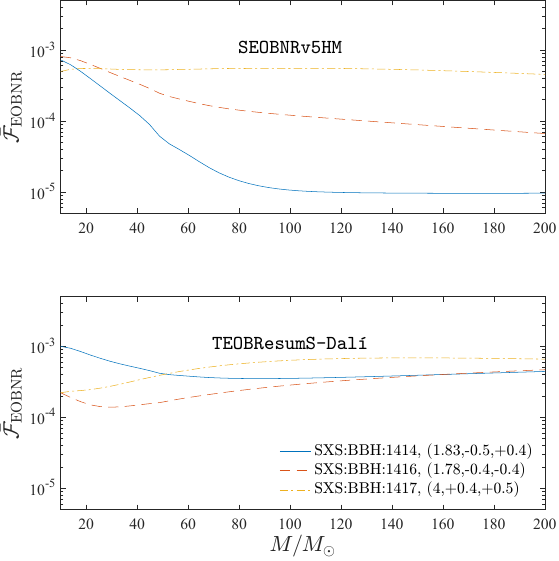}		
	\caption{\label{fig:long_match}
	Unfaithfulness comparison with three long-inspiral (more than 100 orbits) datasets.
	Although the NR-inspiral is likely to be affected by boundary effects and various systematics  
	(see e.g. Ref.~\cite{Szilagyi:2015rwa}), this analysis suggests a slightly stronger consistency 
	between \TEOBd{} and NR for low masses. By contrast, {\tt SEOBNRv5} performs better for high masses.
	} 
\end{figure}

\begin{table*}[t]
	\caption{\label{tab:model_performance_spin_ns}
	Median (Me) and maximum (Max) values of $\bar{\cal F}^{\rm max}_{\rm EOBNR}$,
	the unfaithfulness between EOB and SXS $\ell=m=2$ waveform data for four state-of-the-art EOB models in the spin-aligned, 
	quasi-circular case.  The first row of values refers to the collection of nonspinning simulations in Table~\ref{tab:nonspinning_datasets}.
	The second row of values to the restricted sample of the spinning configurations used to determine $c_3$ and 
	listed in Tables~\ref{tab:c3_eqmass} and~\ref{tab:c3_uneqmass}.
	The comparison shows that while {\tt SEOBNRv5HM} is better able to model nonspinning systems, the three models deliver
	substantially comparable results for spinning binaries.}
	  \begin{center}
		\begin{ruledtabular}
   \begin{tabular}{c c c c c} 
                                                     &                                                  & \TEOBd{}$_{\tt newlogs}$ & {\tt TEOBResumS-GIOTTO} & {\tt SEOBNRv5HM}        \\
   \hline
    \multirow{2}{1.5cm}{\centering nonspinning}      & ${\rm Me}[\bar{\cal F}^{\rm max}_{\rm EOBNR}]$   & $2.13 \times 10^{-4}$    & $1.89 \times 10^{-4}$   & $4.20 \times 10^{-5}$   \\
	                                                 & ${\rm Max}[\bar{\cal F}^{\rm max}_{\rm EOBNR}]$  & $4.37 \times 10^{-4}$    & $8.78 \times 10^{-4}$   & $5.38 \times 10^{-5}$   \\
	\hline       
	\multirow{2}{1.5cm}{\centering spinning}         & ${\rm Me}[\bar{\cal F}^{\rm max}_{\rm EOBNR}]$   & $6.23 \times 10^{-4}$    & $3.96 \times 10^{-4}$   & $4.44 \times 10^{-4}$   \\
	                                                 & ${\rm Max}[\bar{\cal F}^{\rm max}_{\rm EOBNR}]$  & $6.70 \times 10^{-3}$    & $1.12 \times 10^{-3}$   & $2.98 \times 10^{-3}$   \\											   
  %  \hline       
%	\multirow{2}{1.5cm}{\centering \textbf{Overall}} & ${\rm med}$ & $3.2\times 10^{-4}$      & $4.0 \times 10^{-4}$    & $1.99 \times 10^{-4}$   & $6.1 \times 10^{-4}$    \\
%											         & ${\rm max}$ & $\sim 10^{-2}$           & $\sim 10^{-3}$          & $\sim 2 \times 10^{-3}$ & $\sim 2 \times 10^{-1}$ \\										   
   \end{tabular}
\end{ruledtabular}
\end{center}
\end{table*}

To put our results in perspective, it is instructive to quantitatively compare and contrast our model's
construction and performance with those of other EOB-based waveform models; 
namely, the quasicircular {\tt TEOBResumS-GIOTTO}~\cite{Nagar:2023zxh}, {\tt SEOBNR-PM}~\cite{Buonanno:2024byg, Buonanno:2024vkx} 
and {\tt SEOBNRv5HM}~\cite{Pompili:2023tna}.

First of all, the median maximum unfaithfulness we obtain for quasi-circular configurations, 
${\rm Me}[\bar{\cal F}^{\rm max}_{\rm EOBNR}]=3.09\times 10^{-4}$, is approximately half of the value 
$\sim 6.1\times 10^{-4}$ obtained by the recent {\tt SEOBNR-PM} model
based on the incorporation of Post Minkowskian information within the EOB 
formalism\cite{Damour:2016gwp,Damour:2017zjx,Damour:2019lcq,Vines:2017hyw,Vines:2018gqi,Damour:2020tta,Rettegno:2023ghr,Damour:2022ybd}.
Although the analytical structure of the two models is different, it is worth remarking
that they are calibrated to NR to similar extent, in the sense that only two
functions are NR-informed for both, one in the nonspinning sector and one in the spinning sector.
In particular, for $\TEOBd{}$ the NR-calibration delivers $a_6^c$, Eq.~\eqref{eq:a6c} (nonspinning)
and $c_3$, Eq.~\eqref{eq:c3fit} (spinning). A total of 15 NR-informed fitting coefficients,
3 in $a_6^c$ plus 12 in $c_3$, are determined using only 50 NR datasets, 4 nonspinning and 46 spinning
(see Tables~\ref{tab:c3_eqmass}-\ref{tab:c3_uneqmass}). For the {\tt SEOBNR-PM} model, 441 NR simulations are 
used instead to calibrate the $\Delta t_{\rm NR}$ function\footnote{$\Delta t_{\rm NR}$ parametrizes the distance between the 
peak of the $(2,2)$ mode amplitude and the time $t_{\rm ISCO}$ when $r=r_{\rm ISCO}$, where $r_{\rm ISCO}$
is the radius of the Innermost Stable Circular Orbit (ISCO) of a Kerr spacetime
with the same mass and spin as the remnant black hole at the end of the coalescence; see~\cite{Pompili:2023tna}.} 
(see Eq.~(C1) of Ref.~\cite{Buonanno:2024byg}) by means of 21 NR-fitted parameters, 5 for the nonspinning part and 
16 for the spinning part, treated separately through hierarchical fits. 
Differently from $\TEOBd{}$, this parameter impacts only the waveform but not the dynamics.
Given that $\TEOBd{}$ model remains based on resummed PN expansions, though with a more 
careful analytical treatment, our results seem to suggest that in building EOB-based waveform models 
the choice of resummation schemes might be somewhat more important than the amount of 
analytical information incorporated.

Similar accounting applies also for the quasi-circular {\tt TEOBResumS-GIOTTO} model, notably the version 
{\tt v4.3.2} of Ref.~\cite{Nagar:2023zxh}. This model is based on the same mild NR-calibration of the two 
functions $(a_6^c,c_3)$ discussed above, although 54 NR datasets were used for the task in this case.
On the other hand, 46 fitting parameters (10 nonspinning and 36 spinning) are present in the {\tt SEOBNRv5HM} model
(see Eqs.~(78)-(81) of Ref.~\cite{Pompili:2023tna}), where four different functions are NR-informed 
(two for the nonspinning part and two for the spinning part) using 441~NR datasets. The model eventually
delivers a median unfaithfulness $\sim 1.99\times 10^{-4}$, that is smaller by a factor $\sim 2/3$ than 
the one of $\TEOBd{}$ and by a factor $2$ than {\tt TEOBResumS-GIOTTO}.
It is notable that {\tt TEOBResumS}{} (in either its {\tt Dal\'i} or {\tt GIOTTO} avatars),
despite being calibrated to only $\sim 50$ NR datasets, seems capable of attaining comparable performance
to the PN-based {\tt SEOBNRv5HM}, with the median maximum unfaithfulness at most twice larger.

For further insight in the strength of each of these three models across the parameter space,
we proceeded to compute the EOB/NR unfaithfulness for {\tt SEOBNRv5HM} and ${\tt TEOBResums-GIOTTO}$ 
over the same 530 public SXS datasets used for \TEOBd{} (see Ref.~\cite{Nagar:2020pcj,Riemenschneider:2021ppj} 
for the full list).
The resulting values of $\bar{\cal F}_{\rm EOBNR}^{\rm max}$ for {\tt SEOBNRv5HM} are shown in 
Fig.~\ref{fig:barF_seobv5}.
Table~\ref{tab:model_performance_spin_ns} instead compares the models' performance by
listing the median and the maximum of $\bar{\cal F}^{\rm max}_{\rm EOBNR}$ on restricted samples 
of data: (i) the sample of nonspinning datasets considered in Fig.~\ref{fig:barFs0}, and 
(ii) the set of spinning data used in the calibration of $c_3$; see App.~\ref{app:tables}
for tables listing the properties of these simulations. These results reveal that the overall
model hierarchy is explained by the rather large edge {\tt SEOBNRv5HM} has over the others in its
modelization of nonspinning systems (it achieves a median unfaithfulness over the restricted sample
considered smaller by a factor of $\sim 5$ than \TEOBd{}). Differences in performance are much less stark
when looking at the spinning case, with {\tt TEOBResumS-GIOTTO} and {\tt SEOBNRv5HM} equivalent and only
slightly better overall than \TEOBd{}.

An important difference in the NR calibration procedure between both \TEOBd{} and {\tt TEOBResums-GIOTTO}
and the {\tt SEOB} family is the fact that the former two only calibrate the binary dynamics, limiting
NR information in the waveform to NQC corrections and the post-merger model. Rather than using
a function such as {\tt SEOBNRv5HM}'s $\Delta t_{\rm ISCO}$, the merger point is instead located by tying it
to the dynamical evolution, namely to the peak of the pure orbital frequency $\Omega_{\rm orb}$ by means
of the parameter $\Delta t_{\rm NQC}$.
A preliminary analysis, based also on time-domain comparisons, showed that the modelization of the
merger-ringdown phase in the waveform plays a large role in explaining the better results of {\tt SEOBNRv5HM}.
It would be an interesting exercise to investigate the outcome of calibrating \TEOBd{} to NR
following a procedure similar to the one used for {\tt SEOBNRv5HM}, that is, by focusing on
the mismatch, rather than the phase difference, as well as including calibration functions
acting on the plunge-merger waveform.
A thorough analysis of this will be reported in future work (see however Ref.~\cite{Albanesi:2023bgi} for 
exploratory investigations in the large mass ratio limit).

Although the analysis of the EOB/NR unfaithfulness throughout the parameter space covered by numerical
simulations gives a good overall measure of the quality of these models, a more precise quantification of
their differences for particular cases is also informative.
Let us focus on the SXS:BBH:1430 dataset, a configuration with $(q, \chi_1, \chi_2) = 
(8, +0.284, -0.751)$, giving $\tilde{a}_0 = +0.165$.
This is the configuration where {\tt SEOBNRv5HM} was found to have its worst EOB/NR agreement, with 
$\bar{\cal F}^{\rm max}_{\rm EOBNR}\sim 2\times 10^{-3}$, as illustrated
in the top-right panel of Fig.~5 of Ref.~\cite{Pompili:2023tna}; we similarly find $\bar{\cal F}^{\rm max}_{\rm EOBNR}=2.61\times 10^{-3}$.
The top-panel of Fig.~\ref{fig:outlier} shows the unfaithfulness for the model presented in this work, 
that is always well below $10^{-3}$, reaching at most $4\times 10^{-4}$ for the highest masses considered. 
% The shape of the curve as the total mass increases is also different with respect to the {\tt SEOBNRv5HM} one. 
This is better understood looking at the corresponding time-domain phasings, shown in the bottom panel of the 
figure: there we see that the time evolution of the phase difference $\Delta\phi_{22}^{\rm EOBNR}$ is similar for
the two models, but at merger its value for {\tt SEOBNRv5HM} is almost twice as much as \TEOBd{}'s.
Our model's performance is in line with the results presented earlier in this section,
where we saw that \TEOBd{} is strongest around the zero-spin line and for moderate, positive spins, 
progressively worsening as the effective spin grows to very high values. 
In a sense, the behavior for this small value of $\tilde{a}_0$ is numerically consistent with the 
nonspinning configurations, as it is evident from the $\bar{\cal F}_{\rm EOBNR}^{\rm max}$ 
plot in the rightmost panel of Fig.~\ref{fig:barF_full}.
By contrast, it is surprising to see that the behavior for {\tt SEOBNRv5HM} is different: even
though the model is extremely NR-faithful for nonspinning systems (independently of the mass
ratio), outcomes appear to degrade rather significantly even for rather small values of $\tilde{a}_0$.
%
% To gain a wider perspective, we proceeded to compute the EOB/NR unfaithfulness for {\tt SEOBNRv5HM} 
% all over the 530 public SXS  datasets used for \TEOBd{} (see Ref.~\cite{Nagar:2020pcj,Riemenschneider:2021ppj} 
% for the full list) and plot $\bar{\cal F}_{\rm EOBNR}^{\rm max}$ against $(q,\tilde{a}_0)$, see Fig.~\ref{fig:barF_seobv5}.
% This is now done employing the minimization procedure implemented within
% the {\tt pyCBC} package for convenience in using the ${\tt pySEOB}$ framework.
Fig.~\ref{fig:barF_seobv5} highlights that the behavior found for SXS:BBH:1430 is 
not an isolated case: there are other configurations with rather small effective spin, 
$|\tilde{a}_0|\sim 0.1$, where $\bar{\cal F}_{\rm EOBNR}^{\rm max}$ increases by even up to two 
orders of magnitude with respect to the corresponding nonspinning case.
It is also surprising to find, for configurations with almost identical values of $(q,\tilde{a}_0)$,
different results for the unfaithfulness between NR and {\tt SEOBNRv5HM}, whereas
a more consistent behavior is found for $\TEOBd{}$.
An evident example of this case are four datasets with $q\sim 4$ and 
$0.1\lesssim \tilde{a}_0\lesssim 0.2$. 
In addition, the comparison between the parameter space plots in Figs.~\ref{fig:barF_full}
and~\ref{fig:barF_seobv5} also shows that $\TEOBd{}$ is globally more accurate for large mass 
ratios, despite the large number of simulations used to NR-calibrate {\tt SEOBNRv5HM}.

As a last remark, in Fig.~\ref{fig:long_match} we show $\bar{\cal F}_{\rm EOBNR}$ for three
very long datasets from the SXS catalog, with inspirals of up to $\sim 140$ orbits.
Keeping in mind that the NR inspirals could be affected by some systematics for simulations of this length~\cite{Szilagyi:2015rwa}, 
the plot suggests that, for some configurations, the \TEOBd{} model appears to deliver better agreement with 
NR than {\tt SEOBNRv5HM} during the inspiral, as evidenced by the lower unfaithfulness values found for small 
total mass $M$ in the cases of SXS:BBH:1416 and 1417. Conversely, the high quality of the merger-ringdown 
representation in the latter model is again clear here. Note that, as for the other datasets, the waveform 
is cut at a frequency that is 1.35 times larger than the initial frequency of the simulation. We note that without this
cut, one typically finds that $\bar{F}_{\rm EOBNR}\sim 10^{-2}$ for low masses.

These quantitative discrepancies between {\tt SEOBNRv5HM} and \TEOBd{} deserve
more detailed investigations regarding the analytic structure of the two models and the role 
played in each one by the NR-informed functions (either in the dynamics or in the waveform),
which we hope to carry out in future work.
Here we only note that the results presented in this section suggest that \TEOBd{} seems to
deliver more robustness or consistency across the parameter space than {\tt SEOBNRv5HM},
even while the latter overall, and especially in the case of nonspining systems, remains
the more accurate model in terms of EOB/NR unfaithfulness.

\section{Scattering configurations}
\label{sec:scattering}

%=====================
% Scattering angle
%=====================
\begin{table}[t]
	\caption{\label{tab:chi_scattering}
		Comparison between EOB and NR scattering angle for nonspinning binaries from
		the sequence of Ref.~\cite{Damour:2014afa} (fixed energy, varying angular momentum, first 10 rows)
		and from Ref.~\cite{Hopper:2022rwo} (fixed angular momentum and varying energy, last 5 rows).}
	  \begin{center}
		\begin{ruledtabular}
   \begin{tabular}{c c c c |c c c} 
   $\#$  & $E_{\rm in}^{\rm NR}/M$ & $J_{\rm in}^{\rm NR}/M^2$ & $\chi^{\rm NR}$ & 
   $\chi^{\rm EOB}_{\rm newlogs}$ & $\Delta\chi_{\rm newlogs}^{\rm EOBNR}[\%]$ \\
   \hline
   1     &  1.0225555  &  1.099652   &  305.8(2.6)  &  $plunge$  &  $\dots$  \\
   2     &  1.0225722  &  1.122598   &  253.0(1.4)  &  310.720   &  22.81    \\
   3     &  1.0225791  &  1.145523   &  222.9(1.7)  &  252.77    &  13.39    \\
   4     &  1.0225870  &  1.214273   &  172.0(1.4)  &  181.356   &  5.44     \\
   5     &  1.0225884  &  1.260098   &  152.0(1.3)  &  157.389   &  3.54     \\
   6     &  1.0225907  &  1.374658   &  120.7(1.5)  &  122.330   &  1.35     \\
   7     &  1.0225924  &  1.489217   &  101.6(1.7)  &  102.122   &  0.514    \\
   8     &  1.0225931  &  1.603774   &  88.3(1.8)   &  88.470    &  0.193    \\
   9     &  1.0225938  &  1.718331   &  78.4(1.8)   &  78.426    &  0.033    \\
   10    &  1.0225932  &  1.832883   &  70.7(1.9)   &  70.631    &  0.097    \\
   \hline
   \hline
   11    &  1.035031   &  1.1515366  &  307.13(88)  &  $plunge$  & $\dots$ \\
   12    &  1.024959   &  1.151845   &  225.54(87)  &  263.901   & 17.01   \\
   13    &  1.0198847  &  1.151895   &  207.03(99)  &  224.790   & 8.58    \\
   14    &  1.0147923  &  1.151918   &  195.9(1.3)  &  204.306   & 4.27    \\
   15    &  1.0045678  &  1.1520071  &  201.9(4.8)  &  202.774   & 0.43 
   \end{tabular}
\end{ruledtabular}
\end{center}
\end{table}

%
%====================
% Scattering angle with spin
%====================
\begin{table}[t]
	\caption{\label{tab:chi_scattering_spin}
	Comparison between EOB and the (average) NR scattering angle 
	for some of the equal-mass, spin-aligned, configurations of Ref.~\cite{Rettegno:2023ghr}. 
	All datasets share the same initial angular momentum $J_{\rm in}^{\rm NR}/M^2=1.14560$. 
	Datasets are ordered according to the values of the effective spin $\tilde{a}_0$. The remarkable
	EOB/NR agreement obtained for small values of the scattering angle degrades progressively
	as the threshold of direct capture is approached.}
	  \begin{center}
		\begin{ruledtabular}
   \begin{tabular}{c c c c  c | c c} 
   $\chi_1$  & $\chi_2$ & $\tilde{a}_0$ & $E_{\rm in}^{\rm NR}/M$  & $\chi^{\rm NR}$ &  $\chi^{\rm EOB}_{\rm newlogs}$ & $\Delta\chi_{\rm newlogs}^{\rm EOBNR}[\%]$   \\
   \hline
$-0.30$  &  $-0.30$  &  $-0.30$  &  1.02269  &  $plunge$ &  $plunge$  &  $\dots$  \\
$-0.25$  &  $-0.25$  &  $-0.25$  &  1.02268  &  367.55   &  $plunge$  &  $\dots$  \\
$-0.23$  &  $-0.23$  &  $-0.23$  &  1.02267  &  334.35   &  $plunge$  &  $\dots$  \\
$-0.20$  &  $-0.20$  &  $-0.20$  &  1.02266  &  303.88   &  $plunge$  &  $\dots$  \\
$-0.15$  &  $-0.15$  &  $-0.15$  &  1.02265  &  272.60   &  $plunge$  &  $\dots$  \\
$-0.10$  &  $-0.10$  &  $-0.10$  &  1.02265  &  251.03   &  314.99    &  25.48    \\
$-0.05$  &  $-0.05$  &  $-0.05$  &  1.02264  &  234.57   &  277.44    &  18.28    \\
\hline
\hline
$\;0.0$  &  $\;0.0$  & $\;0.0$   &  1.02264  &  221.82   &  253.17    &  14.13    \\
$+0.05$  &  $-0.05$  & $\;0.0$   &  1.02264  &  221.87   &  253.15    &  14.10    \\
$+0.15$  &  $-0.15$  & $\;0.0$   &  1.022650 &  221.89   &  253.20    &  14.11 \\
$+0.20$  &  $-0.20$  & $\;0.0$   &  1.02266  &  221.82   &  253.25    &  14.17    \\
$+0.40$  &  $-0.40$  & $\;0.0$   &  1.02274  &  221.85   &  253.71    &  14.36    \\
$+0.60$  &  $-0.60$  & $\;0.0$   &  1.02288  &  222.08   &  254.53    &  14.61    \\
$+0.80$  &  $-0.80$  & $\;0.0$   &  1.02309  &  221.68   &  255.83    &  15.40    \\
\hline
\hline
$+0.10$  &  $+0.10$  &  $+0.10$  &  1.02265  &  202.61   &  221.53    &  9.34    \\
$+0.80$  &  $-0.50$  &  $+0.15$  &  1.02294  &  198.99   &  210.76    &  5.91    \\
$+0.70$  &  $-0.30$  &  $+0.20$  &  1.02284  &  190.41   &  201.03    &  5.58    \\
$+0.20$  &  $+0.20$  &  $+0.20$  &  1.02266  &  187.84   &  200.74    &  6.87    \\
$+0.30$  &  $+0.30$  &  $+0.30$  &  1.02269  &  176.59   &  185.60    &  5.11    \\
$+0.60$  &  $0.0$    &  $+0.30$  &  1.02276  &  177.63   &  185.70    &  4.54    \\
$+0.40$  &  $+0.40$  &  $+0.40$  &  1.02274  &  167.54   &  173.87    &  3.78    \\
$+0.80$  &  $+0.20$  &  $+0.50$  &  1.02288  &  162.07   &  164.44    &  1.46    \\
$+0.60$  &  $+0.60$  &  $+0.60$  &  1.02288  &  154.14   &  156.53    &  1.55    \\
$+0.80$  &  $+0.50$  &  $+0.65$  &  1.02295  &  152.30   &  153.09    &  0.52    \\
$+0.80$  &  $+0.80$  &  $+0.80$  &  1.02309  &  145.36   &  144.09    &  0.87
   \end{tabular}
\end{ruledtabular}
\end{center}
\end{table}

Let us finally analyze scattering configurations, focusing on the same equal-mass data used in Paper~I. The NR data 
are taken from Refs.~\cite{Damour:2014afa} (fixed energy, varying angular momentum, nonspinning), Ref.~\cite{Hopper:2022rwo} 
(fixed angular momentum, varying energy, nonspinning) and Ref.~\cite{Rettegno:2023ghr} (fixed energy and angular momentum
while varying the initial spins). The NR and EOB data are compared in Tables~\ref{tab:chi_scattering} and ~\ref{tab:chi_scattering_spin}. 
The most interesting finding is that we obtained an excellent (and unprecedented) EOB/NR agreement for the smallest values of the scattering angle,
that is well below the $1\%$ level for several configurations, notably datasets $\#9$, $\#10$ and $\#15$ in Table~\ref{tab:chi_scattering}
or the last rows of Table~\ref{tab:chi_scattering_spin}. Comparing with the corresponding data in Tables~V and VI of Paper~I,
we see the improved EOB/NR consistency obtained now thanks to the updated analytic treatment of logarithmic terms in
resummations. The case with $\chi_1=\chi_2=+0.80$ is particularly remarkable. It yields $\Delta\chi^{\rm EOBNR}_{\rm newlogs}\sim 0.87\%$,
to be contrasted with the $\sim 5\%$ relative difference obtained in Paper~I, despite using there an NR-calibrated flux.

A few considerations are in order. First, this finding is a remarkable consistency test (and the first of its kind) 
between the EOB dynamics for bound and unbound configurations.
Second, it exemplifies the fact that suboptimal (if not really incorrect) analytic choices {\it cannot} be compensated by suitable calibrations to 
NR results. Last, the excellent EOB/NR agreement for (relatively) small values of the scattering angle is contrasted
with its progressive worsening as the scattering angle increases, with incorrect analytical predictions as
the threshold of direct capture is approached. This is consistent with a recent EOB/NR analysis~\cite{Albanesi:2024xus}
that used NR simulations to probe the threshold between scattering and dynamical capture.
Our findings indicate that the strong-field description of the EOB potentials in scattering configurations should be
improved. In other words, in terms of the effective potential $A$, the description of the circular {\it unstable} orbits
needs to be modified. At the same time, the EOB dynamics also depends on the potentials $D$ and $Q$, and small
changes in their strong-field behavior may give rise to visible differences. Let us recall in this respect that,
though $D$ is resummed, $Q$ is not and we are considering only its local-in-time part, in the sense of Ref.~\cite{Bini:2019nra}. 
A systematic analysis of the strong-field behavior of $(D,Q)$ (and possibly their resummations) might improve 
the EOB/NR agreement for the scattering angle. The role of $D$ (notably working up to 6PN) was partly investigated 
in Refs.~\cite{Nagar:2020xsk} and~\cite{Nagar:2023zxh}. These studies should be updated with the current choice of radiation reaction, but this
is postponed to future work.

Before we move on to our councluding remarks, let us draw the reader's attention to the numerical and EOB 
scattering angles the middle block of Table~\ref{tab:chi_scattering_spin}, corresponding to equal-mass binaries 
with opposite spins ($\chi_2 = -\chi_1$).
The NR angles all lie within a tiny interval, $0.4^\circ$ wide, with small variations due partly to the slightly
different incoming energies. By contrast, the $\chi^{\rm EOB}$ (approximately) monotonically grow as the magnitude
of the spins increases, displaying much larger variance ($\sim 2.5^\circ$) than $\chi^{\rm NR}$. While this is mostly due to the
small differences in energy being amplified by the more strongly attractive nature of the EOB dynamics, it
is interesting because this kind of comparison can point to very specific places in the analytical model
for improvement. Namely, in this case, assuming equal initial energy, angular momentum and radius, due to
the spin combinations $\tilde{a}_0, \hat{S}, \hat{S}_*$ all vanishing, any differences in the evolution of 
these systems can only arise from a single NLO term in the centrifugal radius,
plus a few of the spin contributions in the residual waveform amplitude corrections 
that depend on $\tilde{a}_{12}$ and enter the radiation reaction (see in particular Table~I of Ref.~\cite{Nagar:2020pcj}).

\section{Conclusions}
\label{sec:end}
Systematic biases related to inaccurate models threaten to limit our ability 
to extract astrophysical information from GW signals~\cite{Purrer:2019jcp, Kapil:2024zdn, Dhani:2024jja}.
Building on the insight gained in previous work~\cite{Nagar:2024dzj} we have presented here a new and improved avatar of
the \TEOBd{} model for generic, spin-aligned binaries that is based on analytically more consistent 
Pad\'e resummation schemes. More precisely, the logarithms that appear in both the conservative and nonconservative 
sectors of the model are factorized and their polynomial coefficients are Pad\'e 
resummed whenever needed. For what concerns the radiation reaction, this procedure is applied on 
the spin-independent part of all multipoles up to $\ell=8$. 
Except for the $\ell=m=2$ mode, that is considered at 4PN accuracy 
following~\cite{Nagar:2024dzj} (and is obtained from the factorization of the results of 
Refs.~\cite{Blanchet:2023bwj,Blanchet:2023sbv} following~\cite{Damour:2008gu}), 
all other modes are hybridized with test-mass information up to 6PN order. 
The small, though important, change in the analytic treatment of the logarithmic terms is such that 
our usual approach to NR-inform the model allows a reduction of the EOB/NR differences down to a remarkable level,
with a median EOB/NR unfaithfulness ${\rm Me}[{\cal F}_{\rm EOBNR}]=3.09\times 10^{-4}$ all over 
the SXS public catalog of 530 quasi-circular, spin-aligned datasets.
This is comparable to other quasi-circular models like {\tt TEOBResumS-GIOTTO}~\cite{Nagar:2023zxh}
and {\tt SEOBNRv5HM}~\cite{Pompili:2023tna} and better by almost one order of magnitude than the
eccentric {\tt SEOBNRv4E} model~\cite{Ramos-Buades:2021adz} (compare the middle panel
of our Fig.~\ref{fig:barF_full} with Fig.~3 of Ref.~\cite{Ramos-Buades:2021adz}), as discussed
extensively in Sec.~\ref{eob:comparisons} above.
When considering eccentric inspiral configurations, the EOB/NR agreement is generally improved with
respect to Paper~I (especially for large, positive,  spins) with $\bar{\F}_{\rm EOBNR}$ mostly below $\sim5\times 10^{-3}$,
with two outliers; one of them, SXS:BBH:324, is however characterized by a rather noisy NR inspiral.
The largest value of $\bar{\F}_{\rm EOBNR}^{\rm max}$ that we find for quasi-circular configurations is
$\sim 10^{-2}$, corresponding to an accumulated phase difference at merger over the last $\sim 10$ orbits of $\sim 2$~rad,
for an equal-mass system with spins $\chi_1=\chi_2 \simeq+0.99$. This is more generally the region of the parameter space where 
our model is least faithful to NR. We conducted preliminary investigations of possible resummation schemes for the inclusion of
the NNLO spin-orbit contribution in the residual wave amplitude $\rho_{22}^{\rm S}$, also testing Borel-Pad\'e-Laplace summation,
finding mostly consistent results. However, these seem to always lead to even worse performance in the most challenging regime
of high spins, prompting us to hypothesize that the problem resides elsewhere, such as in a suboptimal modelization of 
spin effects in the Hamiltonian and their interplay with the radiation reaction. 
This conclusion calls for a systematic investigation of the properties of the spin-orbit part 
of the Hamiltonian, going beyond the standard treatment used here (that is still based on Ref.~\cite{Damour:2014sva}) 
and building upon some recent, though partly inconclusive, investigations regarding 
${\rm N^3LO}$ spin-orbit effects~\cite{Antonelli:2020aeb,Antonelli:2020ybz,Nagar:2021xnh} or gauge choices~\cite{Placidi:2024yld}.

For scattering configurations, we find an unprecedented EOB/NR agreement ($< 1\%$) for relatively small 
values of the scattering angle, though differences progressively increase as the threshold of immediate 
capture is approached\footnote{Note that no quantitative testing on open orbits is currently available 
for the {\tt SEOBNRv4E} model~\cite{Ramos-Buades:2021adz}, but only qualitative results.}. 
This is the case for both nonspinning (see Table~\ref{tab:chi_scattering}) and 
spinning (see Table~\ref{tab:chi_scattering_spin}) configurations, thus suggesting that improvements 
to the conservative dynamics in this part of the binary parameter space are needed. Our findings are 
also consistent with other recent investigations involving NR simulations at the threshold 
of immediate capture~\cite{Albanesi:2024xus}.

In conclusion, we believe that the analysis presented here gives us confidence in the reliability 
of the dissipative sector of \TEOBd{} and its resummation as in the new resummed expressions. 
Further improvement in the model is likely to come from a new, more systematic analysis of the conservative sector, possibly including higher-order 
PN effects in the $(D,Q)$ functions, in the spin-orbit contributions, and attempting new resummation strategies. 
For example, preliminary analyses concerning the impact of the (resummed) $D$ function at 6PN can be 
found in Ref.~\cite{Nagar:2020xsk} (see Fig.~10 and Table~VI therein), while analogous investigations 
about the mutual compensation of the impact of $(D,Q)$ at different PN orders are discussed in 
Ref.~\cite{Nagar:2023zxh} (see Fig.~6 there). In particular, in its current form, the \TEOBd~Hamiltonian
defines conservative dynamics lacking a Last Stable Orbit (LSO) for high, positive spins 
($\chi_{\rm eff} > 0.7$ for the $q=1$ case); related to this is the fact that the orbital frequency 
$\Omega $ does not reach a peak value at the end of the inspiral in this regime, 
but rather increases indefinitely. The many choices made in the construction of $\hat{H}_{\rm eff}^{\rm SO}$ 
merit a careful review; we should like to explore whether other options might alter these suboptimal 
properties of the conservative dynamics, while at the same time delivering better phasing agreement 
with NR immediately before and around merger\footnote{We mention in passing that the dynamical effect of the angular momentum 
flux absorbed by the two BHs~\cite{Taylor:2008xy,Chatziioannou:2012gq} 
might also merit renewed analysis. This effect is incorporated at leading order in the model (see
Ref.~\cite{Damour:2014sva}), but a careful assessment of its properties as well as the impact of the
most recent analytical results~\cite{Saketh:2022xjb} is still lacking.}.
We believe that such a systematic analysis will be key to further increase the EOB/NR faithfulness of
\TEOBd{} all over the parameter space, notably for large, positive, spins and towards the threshold of 
immediate capture for unbound orbits.

\acknowledgments
D.C. acknowledges support from the Italian Ministry of University and Research (MUR) via the PRIN 2022ZHYFA2, {\it GRavitational wavEform models for coalescing compAct binaries
with eccenTricity} (GREAT).
S.A. acknowledges support from the Deutsche Forschungsgemeinschaft (DFG) project ``GROOVHY'' 
(BE 6301/5-1 Projektnummer: 523180871).
S.~B. acknowledges support by the EU Horizon under ERC Consolidator Grant, no. InspiReM-101043372.
V.~F. is supported by the ERC-SyG project ``Recursive and Exact New Quantum Theory'' (ReNewQuantum), 
which received funding from the European Research Council (ERC) within the 
European Union's Horizon 2020 research and innovation program under Grant No. 810573.
R.~G. acknowledges support from NSF Grant PHY-2020275
(Network for Neutrinos, Nuclear Astrophysics, and Symmetries (N3AS)).
P. ~R. and S.~B. thank the hospitality and the stimulating environment 
of the IHES, where part of this work was carried out.
The present research was also partly supported by the ``\textit{2021 Balzan Prize for 
Gravitation: Physical and Astrophysical Aspects}'', awarded to Thibault Damour.
We also thank S.~Hughes for providing us with the numerical fluxes 
in the test-particle limit.

\appendix
\section{Radiation reaction force: structure, implementation and testing}
\label{app:derivs}

\begin{figure*}[t]
	\centering
	\includegraphics[width=0.49\textwidth]{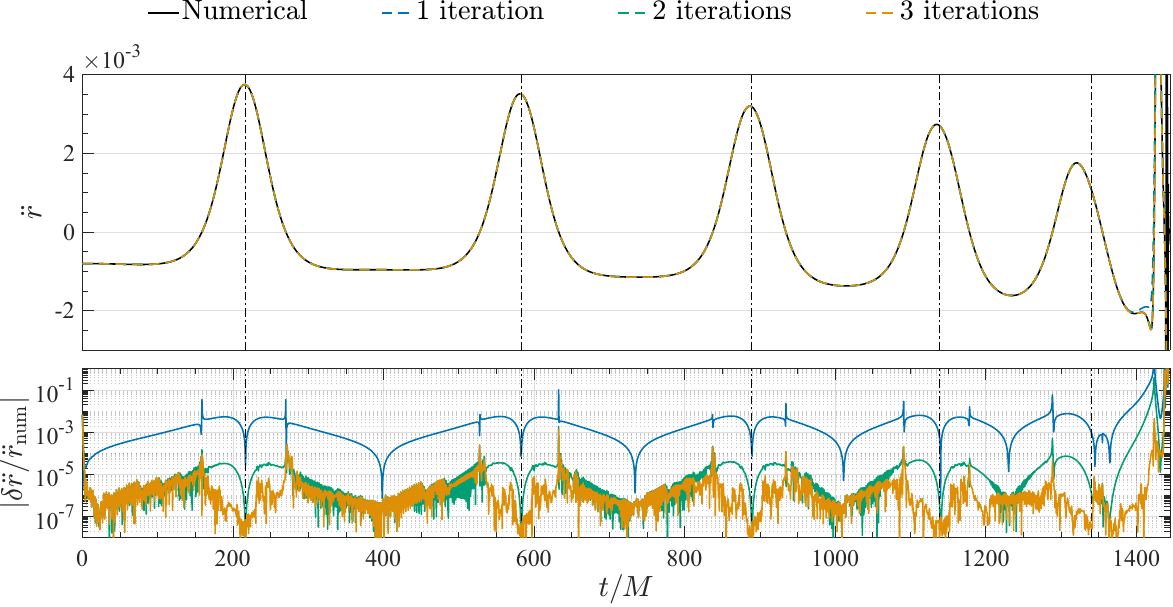}
	\includegraphics[width=0.49\textwidth]{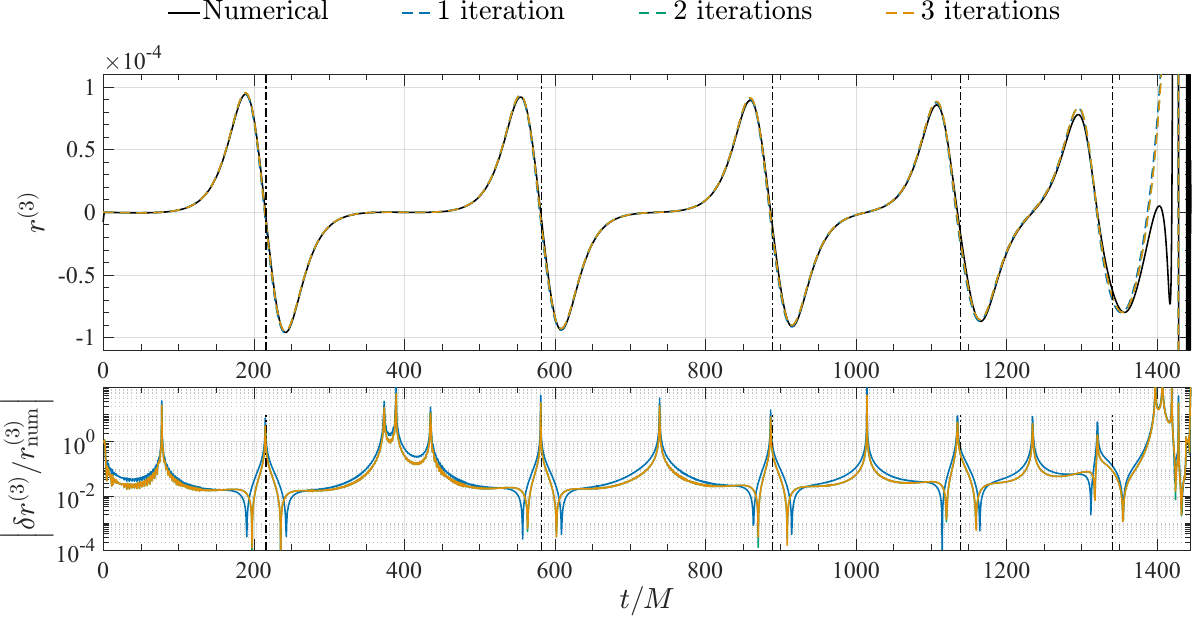}
	\caption{\label{fig:iters}
	Comparisons between $\ddot{r}$ and $r^{(3)} = \dddot{r}$ as calculated by numerically differentiating 
	the dynamics with a $4^{\rm th}$ order scheme and using the iterative procedure, with up to 3 iterations, for an equal-mass system with $\chi_1 = \chi_2 = 0.6$ 
	and initial eccentricity $0.5$. Direct comparisons in the top panels, and relative 
	differences between analytical and numerical values in the bottom panels. Vertical lines mark periastron passages.
	Convergence is evident in the case of $\ddot{r}$; 
	as mentioned in the text, neglected terms in its analytical expression cause errors larger
	than between successive iterations in the case of $\dddot{r}$. Note though that the relative error in $\dddot{r}$ is 
	highest when the derivative is near zero, which should dampen the effect of this level of inaccuracy.}
\end{figure*}

\begin{figure*}[t]
	\centering
	\includegraphics[width=0.49\textwidth]{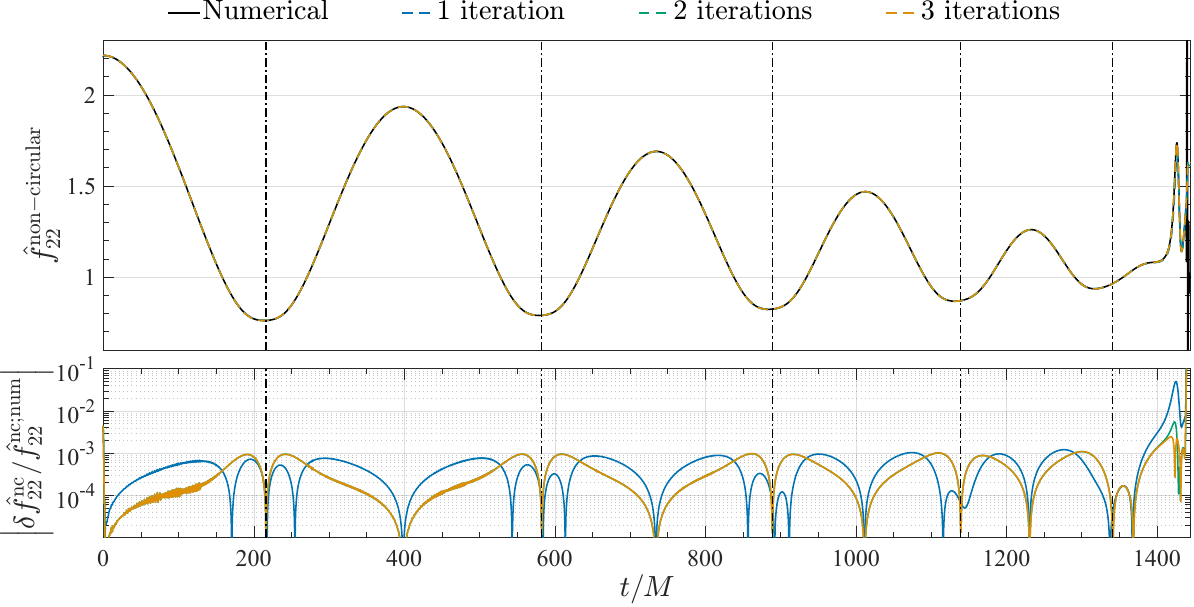}
	\includegraphics[width=0.49\textwidth]{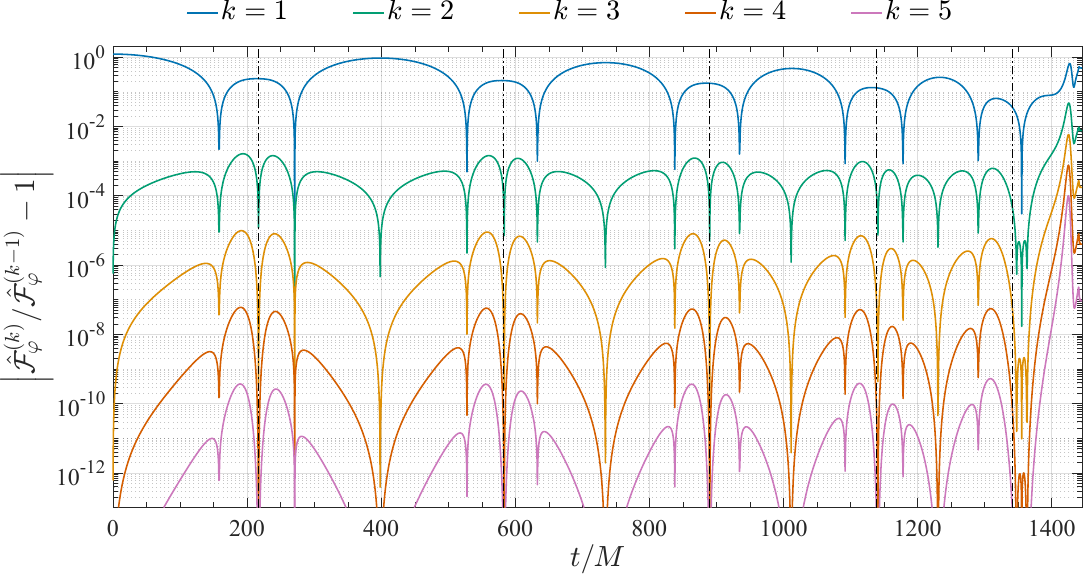}
	\caption{\label{fig:iters_conv}
	Left panel: the first three iterations of the non-circular correcting factor
	$\hat{f}_{22}^{\rm non-circular}$ for the same system parameters considered in Fig.~\ref{fig:iters}. The iterative 
	results are compared with the prefactor calculated using numerical derivatives of $r$ and $\Omega$. Relative differences exceed
	the $10^{-3}$ level only at the very end of the dynamics.
	Right panel: convergence of the complete angular radiation reaction, shown through
	the relative difference between the result of each iteration and the previous one.
	Any iteration beyond the second causes changes below the $0.01\%$ level in the complete flux.}
\end{figure*}

We include in this Appendix a detailed explanation of the implementation of the non-circular radiation
reaction model used in \TEOB~and introduced in~\cite{Chiaramello:2020ehz}, with particular focus on the matter
of the iterative computation of high-order time-derivatives of the dynamical variables.

The \TEOB~radiation reaction model for circularized binaries is based on the special resummation and 
factorization scheme employed for the multipolar waveform~\cite{Damour:2008gu,Damour:2014sva},
as recalled in Sec.~\ref{sec:waveform} of the main text. To more accurately model non-cicularized systems,
the dominant $(2,2)$ contribution to the flux of Eq.~\eqref{eq:sumflm} is dressed with a Newtonian prefactor
that crucially retains explicit time-derivatives of the dynamical variables $r, \varphi$:
\begin{align}
\label{eq:Fnewt}
&\hat{f}_{22}^{\rm non-circular} (r,\Omega;\dot{r}, \dot{\Omega}, \ldots)=1+\dfrac{3}{4}\dfrac{\ddot{r}^2}{r^2\Omega^4} - \dfrac{\ddot{\Omega}}{4\Omega^3}
+\dfrac{3\dot{r}\dot{\Omega}}{r\Omega^3} \nonumber\\
& +\dfrac{4\dot{r}^2}{r^2\Omega^2}
+\dfrac{\ddot{\Omega}\dot{r}^2}{8 r^2\Omega^5} + \dfrac{3}{4}\dfrac{\dot{r}^3\dot{\Omega}}{ r^3\Omega^5}
+\dfrac{3}{4}\dfrac{\dot{r}^4}{r^4\Omega^4} + \dfrac{3}{4}\dfrac{\dot{\Omega}^2}{\Omega^4}\\
&-\dddot{r}\left(\dfrac{\dot{r}}{2r^2\Omega^4}+\dfrac{\dot{\Omega}}{8r\Omega^5}\right)%\nonumber\\
+\ddot{r}\left(-\dfrac{2}{r\Omega^2}+\dfrac{\ddot{\Omega}}{8r \Omega^5}+\dfrac{3}{8}\dfrac{\dot{r}\dot{\Omega}}{r^2\Omega^5}\right)\nonumber.
\end{align}
In principle, these derivatives can be straightforwardly 
determined (irrespective of their algebraic complexity) by consecutive differentiation of the equations of motion.
For instance, consider the case of the orbital frequency. Following the notation of Ref.~\cite{Nagar:2021xnh},
Hamilton's equation for $\varphi$ reads:
\begin{equation}
	\Omega = \dot{\varphi} = \dfrac{\partial \hat{H}_{\rm EOB}}{\partial p_\varphi} = 
	         \frac{p_{\varphi} A u_c^2}{\nu \hat{H}_{\rm EOB} \hat{H}_{\rm eff}^{\rm orb}} + \frac{\Tilde{G} (r_c, p_{r_*}, p_\varphi)}{\nu \hat{H}_{\rm EOB}}
\end{equation}
By further differentiating this expression, one finds:
\begin{widetext}
\begin{align}
	\dot{\Omega} =& \frac{1}{\nu \hat{H}_{\rm EOB} \hat{H}_{\rm eff}^{\rm orb}} \Biggl[ \left( A u_c^2 \right)' \pph \dot{r} + A u_c^2 \left( \dot{p}_{\varphi} - p_{\varphi} \frac{1}{\nu \hat{H}_{\rm EOB} \hat{H}_{\rm eff}^{\rm orb}} \dfrac{d}{dt} \left(\nu \hat{H}_{\rm EOB} \hat{H}_{\rm eff}^{\rm orb}\right) \right) \Biggr] \nonumber \\
	             +& \frac{1}{\nu \hat{H}_{\rm EOB}} \left[ \Tilde{G}' \dot{r} + \dpd{\Tilde{G}}{p_{r_*}} \dot{p}_{r_*} - \Tilde{G} \frac{\dot{\hat{H}}_{\rm EOB}}{\hat{H}_{\rm EOB}} \right], \label{eq:omgdot}
\end{align}
\end{widetext}
where a prime denotes differentiation with respect to $r$. All other time-derivatives of the dynamical variables can be 
computed in a similar fashion.
Notice, however, that the above equation contains time-derivatives of $p_\varphi$ and of the energy $\hat{H}_{\rm EOB}$.
These terms would vanish in the conservative problem. Here, however, they are non-zero and are entirely due to the 
radiation reaction terms:
\begin{align}
	\dot{p}_\varphi &= \hat{\mathcal{F}}_\varphi \ , \\
	\dot{\hat{H}}_{\rm EOB} &= \dot{r} \hat{\mathcal{F}}_r + \Omega \hat{\mathcal{F}}_\varphi \ ,
\end{align}
where the equation for $\dot{\hat{H}}_{\rm EOB}$ is a consequence of the structure of Hamilton's equations modified
by the dissipative terms; cf. Eq.~(2.4) of Ref.~\cite{Bini:2012ji}.
Since these expressions enter the $(2,2)$ prefactor in the angular flux itself, 
the equation defining $\hat{\mathcal{F}}_\varphi$ is actually a non-trivial nonlinear one;
time derivatives of $\hat{\mathcal{F}}_\varphi$ show up as well in, e.g., $\dddot{r}, \ddot{\Omega}$.

In order to more simply and efficiently compute the full angular radiation reaction, we resort to an 
approximate iterative method that generalizes the one introduced in Ref.~\cite{Damour:2012ky} to calculate 
$\ddot{r}$ for the determination of the non-quasi-circular (NQC) corrections to the waveform.

First approximations to the necessary time derivatives are computed by using as a ``$0^{\rm th}$ order'' value $\Fphi^{(0)}$ 
the circularized prescription, excluding the $(2,2)$ Newtonian prefactor; with these, a first solution for 
the full $\Fphi$ is calculated.

Denoting generically by $z$ the time derivatives of the dynamical variables $r, \Omega$, we have:
\begin{align}
z^{(0)} &\equiv z(r, \prs, \pph; \Fphi^{(0)} \equiv \Fphi^{\rm circ.}) \ , \\
\Fphi^{(1)} &\equiv -\frac{32}{5} \nu r_{\omega}^4 \Omega^5 \hat{f} (\Omega; z^{(0)}) \ ,
\end{align}
where we highlighted in the Newton-normalized flux function $\hat{f}$ dependence on the $z$ through the 
Newtonian prefactor of Eq.~\eqref{eq:Fnewt}.
Better approximations of the $z$ are then computed using $\Fphi^{(1)}$ as angular radiation reaction. With them, a 
second-order solution for $\Fphi$ is found:
\begin{align}
z^{(1)} &\equiv z(r, \prs, \pph; \Fphi^{(1)}) \\
\Fphi^{(2)} &\equiv -\frac{32}{5} \nu r_{\omega}^4 \Omega^5 \hat{f} (\Omega; z^{(1)}).
\end{align}
This procedure can then be iterated for better approximations to the correct solution, with convergence tests 
carried out by comparing the iterated time-derivatives with those found through numerical differentiation 
of the output of the integration of the dynamics.
We neglect any time-derivative of the radiation reaction components
themselves, which appear in the third-order derivatives, $\dddot{r}$ and $\ddot{\Omega}$, as well as in $\ddot{p}_{r_*}$.

Performing the aforementioned tests in the comparable mass case, where radiation reaction is strongest, 
we found that two iterations (i.e., going up to $\Fphi^{(2)}$) should be enough for our purposes.
In all cases, looking at either the actual time-derivatives of the dynamical variables, the Newtonian
prefactor, or the flux itself, the iteration converges quite fast, with negligible variations already above three
loops. In particular, the iterated second-order time-derivatives ($\ddot{r}, \dot{\Omega}$) excellently
reproduce the numerical exact ones from the second iteration onward (with a relative error $< 10^{-4}$ during the
inspiral, only exceeding that threshold in the last moments before merger).

When looking at $\ddot{p}_{r_*}, \dddot{r}$ and $\ddot{\Omega}$, the neglected terms in their
analytical expressions prevent similarly precise recovery of their exact values; indeed, differences between
consecutive iterations quickly fall below the weight of the missing terms. As a consequence, even during the 
inspiral, relative differences between the exact numerical derivatives and those found through two iterations
can peak above $\sim 10^{-1}$. Of those mentioned above, $\dddot{r}$ seems to be the one most suffering from
the neglected terms, especially at the end of the inspiral and during plunge.

However, these inaccuracies still have little impact on the Newtonian prefactor, and as such on the flux.
At two iterations, the analytical $\hat{f}_{22}^{\rm non-circular}$ reproduces the correcting factor as
computed using numerical derivatives of the dynamical variables to within $10^{-3}$ relative accuracy throughout
the inspiral. This is true in all cases but those with very high, positive spins and/or high eccentricity,
where inaccuracies still don't exceed $\sim 2 \cdot 10^{-2}$ in relative terms.
Fig.~\ref{fig:iters_conv} shows in an example system how any iteration beyond the second produces extremely small 
corrections to the complete angular radiation reaction $\mathcal{F}_\varphi$, below $0.01\%$ throughout the inspiral,
justifying our choice of stopping at two iterations.

In conclusion, the iterative procedure outlined in this Appendix converges fast, producing very accurate results for 
some of the required time-derivatives ($\ddot{r}, \dot{\Omega}$). Relatively high inaccuracies do remain in 
some cases due to neglected terms in the analytical expressions containing derivatives of the radiation reaction, 
but their effect on the non-circular correcting factor and the angular flux remains small, and mostly confined to the very 
late stages of the binary evolution, at or around plunge, where the conservative dynamics dominates.

\section{Systematics in the determination of $a_6^c$}
\label{app:newa6}

%=====================
% Determining a6c
%=====================
\begin{table*}[t]
	\caption{\label{tab:a6cs0}Informing the orbital sector of the model using NR data to obtain
	different representations of the effective 5PN parameter $a_6^c$ entering the $A$ potential.
	We use either $4$ or $7$ NR datasets to fit the global functions. The two models thus obtained are dubbed 
	$a_{6, {\rm default}}^c$ and $a_{6, {\rm fine-tuned}}^c$ respectively. $a_{6, {\rm default}}^c$ is the default choice
	in \TEOBResumS, and is the one considered in the main text.
	In the case of $a_{6, {\rm fine-tuned}}^c$, the first-guess values of $a_6^c$ are chosen so as to further reduce the phase
	difference accumulated up to the merger time.
	From left to right, the columns report: an ordering index; the name of the SXS simulation used; the mass
	ratio $q$ and the symmetric mass ratio $\nu=q/(1+q)^2$; 
	$\delta\phi^{\rm NR}_{\rm mrg}$ is an estimate of the NR phasing error at merger time, equal to the $\ell=m=2$ phase difference 
	between the highest and second highest resolution available accumulated between $t=600M$ and the 
	peak amplitude of the highest resolution waveform (see also Table~II of Ref.~\cite{Nagar:2019wds});
	the first-guess values of $a_6^c$ for the two determinations;
	the quantities $\Delta\phi^{\rm EOBNR}_{22, \rm mrg}$ are the phase differences 
	$\Delta\phi^{\rm EOBNR}_{22}\equiv \phi^{\rm EOB}_{22}-\phi_{22}^{\rm NR}$ 
	computed at the NR merger (i.e., the peak of the $\ell=m=2$ waveform amplitude) for the corresponding values of $a_6^c$;
	finally, for each $a_6^c$ model we have the values of the maximal EOB/NR unfaithfulness $\bar{\cal F}_{\rm EOBNR}^{\rm max}$, from Eq.~\eqref{eq:barF}.}
	  \begin{center}
		\begin{ruledtabular}
   \begin{tabular}{c c c c c |c c c || c c c } 
   $\#$ & name & $q$ & $\nu$ & $\delta\phi^{\rm NR}_{\rm mrg}$ & $a_{6,{\rm default}}^c$ & $\Delta\phi^{\rm EOBNR}_{22} |_{\rm mrg}$ 
     & $\bar{\cal F}_{\rm EOBNR}^{\rm max}$ & 
     $a_{6,{\rm fine-tuned}}^{c}$  & $\Delta\phi^{\rm EOBNR}_{22} |_{\rm mrg}$ & $\bar{\cal F}_{\rm EOBNR}^{\rm max}$\\
   \hline
   \hline
   1      & SXS:BBH:0180 & 1    & $0.25$       & $-0.42$   & $-31.7$  & $+0.0265$  & 0.00012101  & $-31.7$   & $+0.0265$  & 0.00013565 \\
   2      & SXS:BBH:0169 & 2    & $0.\bar{2}$  & $-0.027$  & $-25.6$  & $+0.1255$  & 0.00020099  & $-24.8$   & $-0.0214$  & 0.00024516  \\
   3      & SXS:BBH:0259 & 2.5  & $0.2041$     & $-0.0080$ & \dots    & $\dots$    & 0.00041247  & $-21.3$   & $-0.0089$  & 0.00051478  \\
   4      & SXS:BBH:0168 & 3    & $0.1875$     & $-0.0870$ & $-17.5$  & $+0.0616$  & 0.00043662  & $-17.1$   & $-0.0095$  & 0.00054175  \\
   5      & SXS:BBH:0294 & 3.5  & $0.1728$     & $+1.32$   & $\dots$  & $\dots$    & 0.00025244  & $-13.2$   & $-0.042$   & 0.00031963  \\
   6      & SXS:BBH:0295 & 4.5  & $0.1488$     & $+0.2397$ & $\dots$  & $\dots$    & 0.00021093  & $-7.1$    & $-0.034$   & 0.00029180  \\
   7      & SXS:BBH:0166 & 6    & $0.1225$     & $\dots$   & $-1.0$   & $+0.219$   & 0.00017874  & $+1.0$    & $-0.034$   & 0.00025490  \\
   8      & SXS:BBH:0298 & 7    & $0.1094$     & $-0.0775$ & $\dots$  & $\dots$    & 0.00019141  & $\dots$   & $\dots$    & 0.00026358  \\
      \end{tabular}
\end{ruledtabular}
\end{center}
\end{table*}

\begin{figure*}[t]
	\center	
	\includegraphics[width=0.32\textwidth]{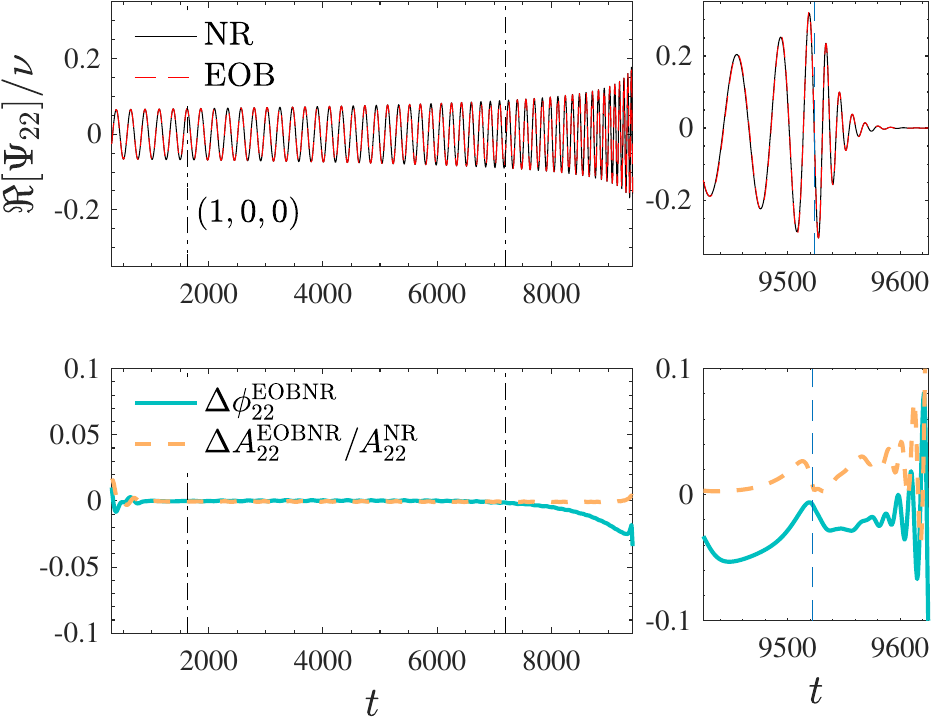}
	\includegraphics[width=0.33\textwidth]{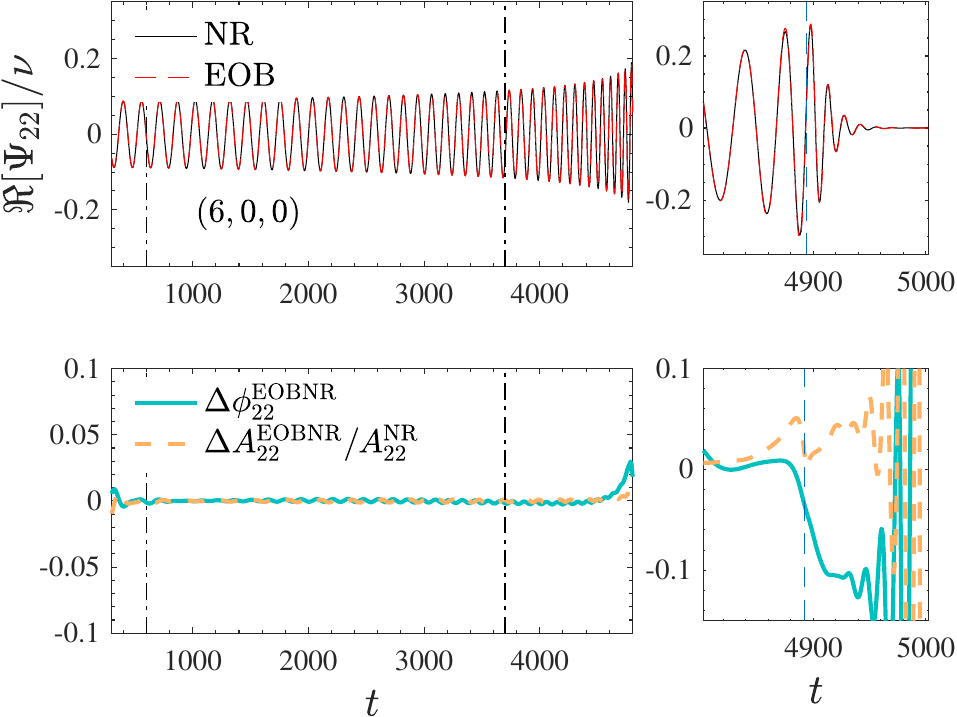}	
	\includegraphics[width=0.31\textwidth]{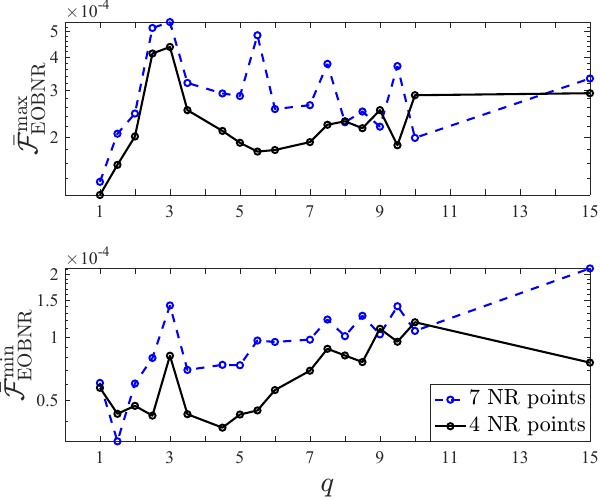}
	\caption{\label{fig:barFs0_finetuned}
	Evaluating the performance of the fine-tuned $a_6^c(\nu)$ from Eq.~\eqref{eq:a6c_ft}
	versus the default choice, given by Eq.~\eqref{eq:a6c}, illustrated in Fig.~\ref{fig:barFs0}. 
	Note the reduced accumulated phase difference for $q=1$ and $q=6$ with respect to the top panels of Fig.~\ref{fig:barFs0}.
	The rightmost panel of the figure compares $\bar{\F}_{\rm EOBNR}^{\rm max}$ and $\bar{\F}_{\rm EOBNR}^{\rm min}$  obtained
	with Eq.~\eqref{eq:a6c} (using 4 NR-informed points) or Eq.~\eqref{eq:a6c_ft} (using 7 NR-informed points).
	%The corresponding median values are $3.07\times 10^{-4}$ and $2.77\times 10^{-4}$ respectively.
	}
\end{figure*}

In this Appendix we expand upon the procedure we follow for the determination of the NR-informed 
function $a_6^c(\nu)$, exploring a different choice of NR datasets used. 
In Sec.~\ref{sec:nrinfo} we presented an initial estimate, that we rename here $a^c_{6, \rm default}$,
fitted from the 4 NR datasets reported in the middle section of Table~\ref{tab:a6cs0}. The phase difference
at merger for these datasets is of order $\sim 0.1$~rad (notably, this is at the level of the NR uncertainty in some cases;
compare with column 5 of Table~\ref{tab:a6cs0}), corresponding to a maximal EOB/NR unfaithfulness 
(on average) of $\sim 3 \times 10^{-4}$. We undertake here a new determination of this coefficient, 
resulting in the function $a^c_{6, \rm fine-tuned} (\nu)$, starting from the NR waveforms in column 
9 of Table~\ref{tab:a6cs0}; we consider now 7 datasets, for a more accurate global fit 
against $\nu$. In doing this, we are also interested in exploring
the model's sensitivity to small ($\sim 10 \%$) variations in $a_6^c$ that are however 
enough to bring the accumulated dephasing down to $\sim 0.01$~rad for the calibration datasets.

% We present in what follows two different NR-informed estimates of $a_6^c(\nu)$.
% The first one is more conservative since: (i) we use only 4 NR simulations to determine it; 
% (ii) we allow for phase differences accumulated up to merger that are $\sim 0.1$~rad, i.e. of the 
% order of the aforementioned nominal NR uncertainty in some cases.
% In the second case we explore how sensitive the model is to variations in $a_6^c$ that are small
% (say order $10\%$) but sufficient to reduce the accumulated phase difference at merger down 
% to  $\sim 0.01$~rad. In this case we consider 7 datasets instead of just $4$ so to have a 
% more accurate global fit of $a_6^c$ versus $\nu$.
% This information is collected in Table~\ref{tab:a6cs0}. The third column reports the uncertainty estimate
% obtained taking the difference between the highest and second highest resolutions available for those
% datasets (see also Ref.~\cite{Nagar:2019wds}). The table highlights that the error estimate obtained in
% this way is not uniform all over the datasets, but roughly we have $\langle \delta\phi_{\rm mrg}^{\rm NR}\rangle\sim 0.1$~rad.
% Let us focus first on the determination of $a_6^c$ using only 4 datasets, with the corresponding values of
% $a_6^c$, dubbed $a_{6,{\rm default}}^c$ in the 6th column of Table~\ref{tab:a6cs0}.

The behavior of the newly chosen values of $a_6^c$ versus $\nu$ is always quasi-linear, but
they are better fitted by a cubic function of the form
\be
\label{eq:a6c_ft}
a_{6,{\rm fine-tuned}}^c(\nu) = -3817.96\nu^3+2479.56\nu^2-766.67\nu+64.71 \  .
\ee
At the level of time-domain phasing difference, $a^c_{6,{\rm fine-tuned}}(\nu)$ yields much closer 
agreement with all nonspinning SXS datasets available, possibly even below the NR uncertainty,
since the accumulated phase difference up merger can be lowered to below $0.1$~rad even after 
several tens of orbits. For this reason we preferred to use the $a_6^c$ function of Eq.~\eqref{eq:a6c} 
as our default choice to construct the spin-aligned model in the main text. 

In Fig.~\ref{fig:barFs0_finetuned} we show the EOB/NR time-domain phasings for $q=1$ and $q=6$, to
be compared to the corresponding ones in Fig.~\ref{fig:barFs0}. Although we didn't change the input
value $a_6^c=-31.7$ for $q=1$, the small differences in the fit yield changes in the phase differences
that eventually determine modifications in $\bar{\cal F}_{\rm EOBNR}$.
In the rightmost panel of the figure we contrast $\bar{\F}_{\rm EOBNR}^{\rm max,min}$ obtained using 
Eq.~\eqref{eq:a6c} (black line) with the one from our new choice, Eq.~\eqref{eq:a6c_ft} (blue line). 
We see that, despite the accumulated $\Delta\phi^{\rm EOBNR}_{22}$  is now smaller, the
global behavior in terms of unfaithfulness is less good than before. Without a more stringent estimate 
of the NR phasing error in the time-domain it seems impossible, at this stage, to choose which 
version of $a_6^c(\nu)$ yields the most faithful representation of the NR data\footnote{Let us also note 
that the SXS collaboration typically gives uncertainties in terms of unfaithfulness with a flat-noise curve,
pointing to $10^{-4}$ as the average uncertainty of the catalog, see Fig.~9 of Ref.~\cite{Boyle:2019kee}.} 
Looking at the rightmost panel of Fig.~\ref{fig:barFs0_finetuned} we conclude that variations of $a_6^c$ 
of order $1$ (or less) entail variations of $\Delta\phi_{22}^{\rm EOBNR}$ at merger of order 0.1~rad 
that eventually determine modifications on $\bar{\F}_{\rm EOBNR}^{\rm max}$ of  $\sim 2\times 10^{-4}$. 
It thus seems that the value $2\times 10^{-4}$ might be considered as a reasonably conservative 
estimate of the NR uncertainty on the nonspinning SXS data considered here 
(extrapolated to infinity and not CCE evolved~\cite{Mitman:2020bjf}) and it might 
be better to not attempt to reduce the EOB/NR disagreement further.
In any case, we explored to which extent the model is flexible and whether it can be made
close to NR at will. Focusing on the most significative $q=1$ case, we found that tuning only 
$a_6^c$ it doesn't seem possible to reduce $\bar{\F}_{\rm EOBNR}$ below $1\times 10^{-4}$.
This might be probably achieved also additionally tuning the parameter\footnote{A procedure analogous to this
is implemented in the {\tt SEOBNRv5HM} model~\cite{Pompili:2023tna}, that shows, on average, a smaller
EOB/NR phasing disagreement around merger.} $\Delta t_{\rm NQC}$, that is fixed for simplicity to $\Delta t_{\rm NQC}=1$.
Although our reasoning is here limited to only all available nonspinning simulations, it should also hold
for the spin-aligned ones, since a priori larger numerical uncertainties are expected in the presence of spin.
We believe that any additional improvement in the construction of an EOB model by comparison with 
numerical data in the nonspinning case cannot come without different NR simulations that are at least 
SXS quality, but are obtained with a completely different code and infrastructure, see e.g. Ref.~\cite{Rashti:2024yoc}
for recent attempts in this direction.

\section{Optimized EOBNR unfaithfulness for eccentric binaries}
\label{sec:opt_dali}

\begin{figure}[t]
	\center	
	\includegraphics[width=0.49\textwidth]{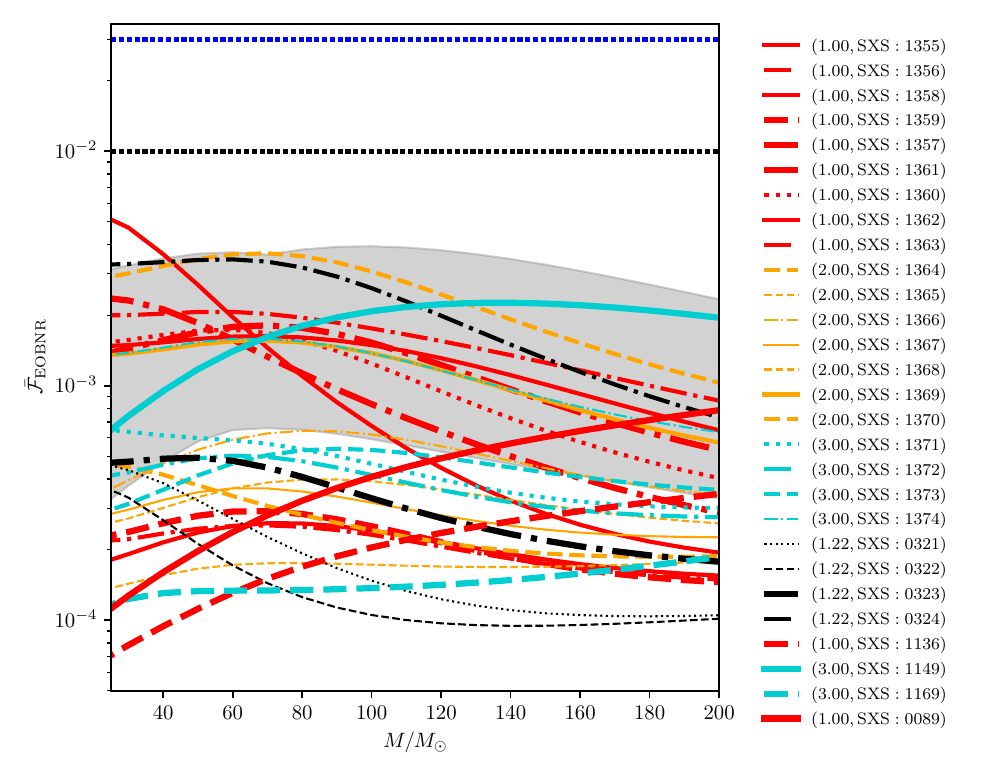}
	\caption{\label{fig:optmm_ecc}EOB/NR unfaithfulness between the new version of the EOB model and the $28$ publicly available eccentric
	SXS waveforms. The unfaithfulness is computed using the optimized initial conditions ($e_0$ and $f_0$) for the EOB model,
	determined by minimizing $\bar{\cal F}_{\rm EOBNR}$ at a reference mass of $M=25 M_{\odot}$. We find a considerable improvement
	with respect to both the previous version of the model (which spans mismatches in the range within the grey bands, see Fig.~13 of Ref.~\cite{Gamboa:2024hli}) 
	and the non-optimized mismatches (Fig.~\ref{fig:barF_ecc}): no configuration exceeds the $1\%$ threshold, and most of them are now below $0.1\%$.}
\end{figure}

The EOBNR model unfaithfulness reported in Fig.~\ref{fig:barF_ecc} is computed using
initial conditions (that is, initial eccentricity $e_0$ and frequency $f_0$) estimated by aligning the
EOB and NR waveforms in the time domain, and minimizing their phase difference over the inspiral.
As previously discussed, this procedure doesn't necessarily ensure an optimization of the EOBNR unfaithfulness.

In this appendix we therefore complement the result shown above with the EOBNR unfaithfulness obtained via
numerical optimization of the initial conditions for the EOB model. We use a loss function $\mathcal{L}(e_0, f_0)$
given by $\bar{\mathcal F}_{\rm EOBNR}$ at a reference mass of $M = 25 M_{\odot}$:
\begin{equation}
	\mathcal{L}(e_0, f_0) = 1 - \max_{t_0,\phi_0}\dfrac{\langle h_{\rm NR},h_{\rm EOB}(e_0, f_0)\rangle}{||h_{\rm NR}||||h_{\rm EOB}(e_0, f_0)||} \, .
\end{equation}
This choice of reference mass is motivated by the desire to give increased weight to the inspiral phase, where the
effect of eccentricity is most pronounced. All other intrinsic parameters are fixed to the NR ones. 
The minimization is then performed using a dual annealing algorithm, as implemented in the {\tt scipy} 
library~\cite{scipy}. For each configuration, we allow the algorithm to explore
the parameter space $e_0 \in [0, 0.7]$ and $f_0 \in [0.002, 0.007]$.

Results are presented in Fig.~\ref{fig:optmm_ecc}. Thanks to the optimization procedure, no configuration 
now exceeds the $1\%$ mismatch threshold at any mass value. Compared to the previous version of the model -- which spans
a mismatch range of $3\times 10^{-4}$ to $4\times  10^{-3}$ over the same configurations and masses, grey band 
in the plot (see also Fig.~13 of Ref.~\cite{Gamboa:2024hli} )-- the updated model demonstrates greater accuracy
in the description of eccentric binaries, with most configurations having $\bar{\mathcal{F}}_{\rm EOBNR} \leq 0.1\%$.

\section{NR datasets}
\label{app:tables}

We include in this Appendix tables detailing the restricted NR datasets considered in this work.
Table~\ref{tab:nonspinning_datasets} lists the nonspinning simulations used in Fig.~\figref{fig:barFs0}
and Table~\ref{tab:model_performance_spin_ns}.
Tables~\ref{tab:c3_eqmass} and~\ref{tab:c3_uneqmass} contain the properties of the simulations used
for the calibration of $c_3$, indicating for each also the corrisponding first-guess value of the coefficient,
used to construct the global fit of Eqs.~\eqref{eq:c3fit}-\eqref{eq:c3fit_pieces}.

%========================
% Equal-mass, equal-spin values
%========================
\begin{table}[t]
	\caption{\label{tab:nonspinning_datasets}Nonspinning SXS datasets used in the main text. The values
	of the maximum of the EOB/NR unfaithfulness with the advanced LIGO power spectral density are
	also reported as complement to the information collected in the main text. The median value is $2.13\times 10^{-4}$.}
	\begin{center}
  \begin{ruledtabular}
	\begin{tabular}{cccc}
	  ID & $q$ & $\nu$ & ${\bar{\cal F}}_{\rm EOBNR}^{\rm max}$ \\ 	  
	  \hline
    SXS:BBH:0180   & 1   & 0.25        & 0.000121 \\
	SXS:BBH:0007   & 1.5 & 0.24        & 0.000157 \\
	SXS:BBH:0169   & 2   & $0.\bar{2}$ & 0.000201 \\
	SXS:BBH:0259   & 2.5 & 0.204       & 0.000412 \\
	SXS:BBH:0168   & 3   & 0.1875      & 0.000437 \\
	SXS:BBH:0294   & 3.5 & 0.1728      & 0.000252 \\
	SXS:BBH:0295   & 4.5 & 0.1488      & 0.000211 \\
	SXS:BBH:0056   & 5   & 0.1389      & 0.000190 \\
	SXS:BBH:0296   & 5.5 & 0.1302      & 0.000176 \\
	SXS:BBH:0166   & 6   & 0.1224      & 0.000179 \\
	SXS:BBH:0298   & 7   & 0.1094      & 0.000191 \\
	SXS:BBH:0299   & 7.5 & 0.1038      & 0.000222 \\
	SXS:BBH:0063   & 8   & 0.0988      & 0.000229 \\
	SXS:BBH:0300   & 8.5 & 0.0942      & 0.000216 \\
	SXS:BBH:0301   & 9   & 0.09        & 0.000253 \\
	SXS:BBH:0302   & 9.5 & 0.0862      & 0.000186 \\
	SXS:BBH:0303   & 10  & 0.0826      & 0.000287 \\
	SXS:BBH:2477   & 15  & 0.0586      & 0.000292 
 \end{tabular}
  \end{ruledtabular}
  \end{center}
  \end{table}
 
%#1 q=1 a0=0 chi=s0 max(F)=0.00023198
%#2 q=1.5 a0=0 chi=s0 max(F)=0.00020617
%#3 q=2 a0=0 chi=s0 max(F)=0.00050811
%#4 q=2.5 a0=0 chi=s0 max(F)=0.00025215
%#5 q=3 a0=0 chi=s0 max(F)=0.00028756
%#6 q=3.5 a0=0 chi=s0 max(F)=0.00031051
%#7 q=4.5 a0=0 chi=s0 max(F)=0.00026724
%#8 q=5 a0=0 chi=s0 max(F)=0.0002432
%#9 q=5.5 a0=0 chi=s0 max(F)=0.00023271
%#10 q=6 a0=0 chi=s0 max(F)=0.00022665
%#11 q=7 a0=0 chi=s0 max(F)=0.00030576
%#12 q=7.5 a0=0 chi=s0 max(F)=0.00035411
%#13 q=8 a0=0 chi=s0 max(F)=0.00024815
%#14 q=8.5 a0=0 chi=s0 max(F)=0.00034334
%#15 q=9 a0=0 chi=s0 max(F)=0.00038786
%#16 q=9.5 a0=0 chi=s0 max(F)=0.00031973
%#17 q=10 a0=0 chi=s0 max(F)=0.00032386
%#18 q=15 a0=0 chi=s0 max(F)=0.00041326

%========================
% Equal-mass, equal-spin values
%========================
\begin{table}[t]
	\caption{\label{tab:c3_eqmass}First-guess values for $c_3$ for equal-mass, equal-spin configurations.
	They are  used to determine $c_3^{=}$ in Eq.~\eqref{eq:c3fit}.}
\begin{center}
\begin{ruledtabular}
\begin{tabular}{lllc|ccc}
$\#$ & ID & $(q,\chi_1,\chi_2)$ & $\tilde{a}_0$ &$c_3^{\tt newlogs}$ & \\
\hline
1  & BBH:1137 & $(   1, -0.   9692, -0.   9692)$ & $-0.9692$   & 92.5  \\ 
2  & BBH:0156 & $(   1, -0.9498, -0.9498)$       & $-0.95$     & 91.0  \\  
3  & BBH:2086 & $(   1, -0.80, -0.   80)$        & $-0.80$     & 83.3  \\ 
4  & BBH:2089 & $(   1, -0.60, -0.   60)$        & $-0.60$     & 72.0  \\ 
5  & BBH:0149 & $(   1, -0.20, -0.   20)$        & $-0.20$     & 53.2  \\
6  & BBH:0150 & $(   1, +0.   20, +0.   20)$     & $+0.20$     & 35.0  \\ 
7  & BBH:0170 & $(   1, +0.   4365, +0.   4365)$ & $+0.20$     & 27.8  \\ 
8  & BBH:2102 & $(   1, +0.   60, +0.   60)$     & $+0.60$     & 23.3  \\ 
9  & BBH:2104 & $(   1, +0.   80, +0.   80)$     & $+0.80$     & 18.5  \\  
10 & BBH:0160 & $(   1, +0.8997, +0.8997)$       & $+0.8997$   & 16.0  \\ 
11 & BBH:0157 & $(   1, +0.949586, +0.949586)$   & $+0.949586$ & 14.9  \\ 
12 & BBH:0177 & $(   1, +0.989253, +0.989253)$   & $+0.989253$ & 14.2    
\end{tabular}
\end{ruledtabular}
\end{center}
\end{table}
 %=====================================
 % The values of c3: first-guess & fits
 %=====================================
\begin{table}[t]
\caption{\label{tab:c3_uneqmass}First-guess values for $c_3$ for the unequal-spin and
	unequal-mass configurations. They are  used to determine $c_3^{\neq}$ in Eq.~\eqref{eq:c3fit}.
	Note that configurations with close values of $\tilde{a}_0$ consistently share similar values of $c_{3}^{\tt newlogs}$.}
\begin{center}
\begin{ruledtabular}
\begin{tabular}{lllc|ccc}
$\#$ & ID & $(q,\chi_1,\chi_2)$ & $\tilde{a}_0$ & $c_3^{\tt newlogs}$  \\
\hline
13 & BBH:0004 & $(   1.0, -0.   50,  0.    0)$ & $-0.25$   & 56    \\ 
14 & BBH:0016 & $( 1.5, -0.   50,  0.    0)$   & $-0.30$   & 56.6  \\ 
15 & BBH:0036 & $(   3.0, -0.   50,  0.    0)$ & $-0.38$   & 59.4  \\ 
16 & BBH:2139 & $(   3.0, -0.   50, -0.   50)$ & $-0.50$   & 66.5  \\ 
17 & BBH:0111 & $(   5.0, -0.   50,  0.    0)$ & $-0.42$   & 56    \\ 
18 & BBH:1428 & $(5.516,-0.80,-0.70)$          & $-0.784$  & 80.7  \\
19 & BBH:1419 & $(8.0,-0.80,-0.80)$            & $-0.80$   & 82.0  \\
20 & BBH:1375 & $(   8.0, -0.   90,  0.    0)$ & $-0.80$   & 67    \\ 
21 & BBH:0114 & $(   8.0, -0.   50,  0.    0)$ & $-0.44$   & 61    \\ 
\hline
22 & BBH:0005 & $(   1, +0.   50,  0.    0)$   & $+0.25$   & 34.8  \\ 
23 & BBH:2105 & $(   1, +0.   90,  0.    0)$   & $+0.45$   & 28.1  \\ 
24 & BBH:2106 & $(   1, +0.   90, +0.   50)$   & $+0.70$   & 20.9  \\ 
25 & BBH:1146 & $( 1.5, +0.   95, +0.   95)$   & $+0.95$   & 15.8  \\ 
26 & BBH:0552 & $(1.75,+0.80,-0.40)$           & $+0.36$   & 29.6  \\
27 & BBH:1466 & $(1.90,+0.70,-0.80)$           & $+0.18$   & 33.7  \\
28 & BBH:2129 & $(   2, +0.   60,  0.    0)$   & $+0.40$   & 29.7  \\ 
29 & BBH:2130 & $(   2, +0.   60, +0.   60)$   & $+0.60$   & 24.1  \\ 
30 & BBH:2131 & $(   2, +0.   85, +0.   85)$   & $+0.85$   & 18.0  \\ 
31 & BBH:1453 & $(2.352,+0.80,-0.78)$          & $+0.328$  & 30.8  \\
32 & BBH:0292 & $(3,+0.73,-0.85)$              & $+0.335$  & 30.5  \\
33 & BBH:0174 & $(   3, +0.   50,  0.    0)$   & $+0.37$   & 29.1  \\ 
34 & BBH:2158 & $(   3, +0.   50, +0.   50)$   & $+0.50$   & 27.6  \\ 
35 & BBH:2163 & $(   3, +0.   60, +0.   60)$   & $+0.60$   & 24.9  \\ 
36 & BBH:0293 & $(   3, +0.   85, +0.   85)$   & $+0.85$   & 18.2  \\ 
37 & BBH:1447 & $(3.16, +0.7398, +0.   80)$    & $+0.75$   & 20.5  \\ 
38 & BBH:1452 & $(3.641,+0.80,-0.43)$          & $+0.534$  & 27.1  \\
39 & BBH:2014 & $(   4, +0.   80, +0.   40)$   & $+0.72$   & 22.0  \\ 
40 & BBH:1434 & $(4.368, +0.7977, +0.7959)$    & $+0.80$   & 19.5  \\ 
41 & BBH:0110 & $(   5, +0.   50,  0.    0)$   & $+0.42$   & 30.0  \\ 
42 & BBH:1440 & $(5.64,+0.77,+0.31)$           & $+0.70$   & 23.5  \\
43 & BBH:1432 & $(5.84, +0.6577, +0. 793)$     & $+0.68$   & 23.7  \\ 
44 & BBH:1437 & $(6.038,+0.80,+0.1475)$        & $+0.707$  & 23.6  \\
45 & BBH:0065 & $(   8, +0.   50,  0.    0)$   & $+0.44$   & 26.8  \\ 
46 & BBH:1426 & $(   8, +0.4838, +0.7484)$     & $+0.51$   & 27.8  
\end{tabular}
\end{ruledtabular}
\end{center}
\end{table}

\bibliography{refs20250122.bib,local.bib}

\end{document}